\documentclass[a4paper,12pt]{article}
\pdfoutput=1
\usepackage{jcappub}
\usepackage{amsthm,graphicx}
\usepackage{epsfig}
\usepackage{latexsym, amssymb} 
\usepackage{amsmath}

\def\beq{\begin{equation}}
\def\eeq{\end{equation}}
\def\br{\begin{eqnarray}}
\def\er{\end{eqnarray}}
\def\benu{\begin{enumerate}}
\def\efnu{\end{enumerate}}
\def\nn{\nonumber}

\def\l{\left}
\def\r{\right}

\def\cl{{\cal C}_{\ell}}
\def\psk{{ P}_{k}}
\def\psk{{P}_{\rm S}(k)}

\begin{document}
\title{Primordial power spectrum from Planck}
 \author[a]{Dhiraj Kumar Hazra} 
 \author[a,b]{Arman Shafieloo}
 \author[c]{Tarun Souradeep}  
\affiliation[a]{Asia Pacific Center for Theoretical Physics, Pohang, Gyeongbuk 790-784, Korea}
\affiliation[b]{Department of Physics, POSTECH, Pohang, Gyeongbuk 790-784, Korea}
\affiliation[c]{Inter-University Centre for Astronomy and Astrophysics, Post Bag 4, Ganeshkhind, Pune 411~007, India}

\emailAdd{dhiraj@apctp.org, arman@apctp.org, tarun@iucaa.ernet.in}

\abstract 
{Using modified Richardson-Lucy algorithm we reconstruct the primordial power spectrum (PPS) from Planck Cosmic Microwave Background (CMB) temperature anisotropy data. In our analysis we use different combinations of angular power spectra from Planck to reconstruct the shape of the primordial power spectrum and locate possible features. Performing an extensive error analysis we found the dip near $\ell\sim750-850$ represents the most prominent feature in the data. Feature near $\ell\sim1800-2000$ is detectable with high confidence only in 217 GHz spectrum and is apparently consequence of a small systematic as described in the revised Planck 2013 papers. Fixing the background cosmological parameters and the foreground nuisance parameters to their best fit baseline values, we report that the best fit power law primordial power spectrum is consistent with the reconstructed form of the PPS at 2$\sigma$ C.L. of the estimated errors (apart from the local features mentioned above). As a consistency test, we found the reconstructed primordial power spectrum from Planck temperature data can also substantially improve the fit to WMAP-9 angular power spectrum data (with respect to power-law form of the PPS) allowing an overall amplitude shift of $\sim2.5\%$. In this context low-$\ell$ and 100 GHz spectrum from Planck which have proper overlap in the multipole range with WMAP data found to be completely consistent with WMAP-9 (allowing amplitude shift). As another important result of our analysis we do report the evidence of gravitational lensing through the reconstruction analysis. Finally we present two smooth form of the PPS containing only the important features. These smooth forms of PPS can provide significant improvements in fitting the data (with respect to the power law PPS) and can be helpful to give hints for inflationary model building. 
}

\maketitle

\section{Introduction and the road-map}
Planck~\cite{Planck}, the most recent full sky Cosmic Microwave Background (CMB) survey has been able to provide us with the most precise
information about the early Universe. However, few underlying assumptions about the Universe largely dominate our understanding of the 
history of the Universe. The angular power spectrum from CMB temperature fluctuations contains a convolved signal of the shape of initial fluctuation, {\it i.e.} the primordial power spectrum (PPS) and 
the transition of this fluctuation through different phases of the Universe. To extract these information we usually make assumptions in two sectors, namely
the formation and evolution of the perturbations/fluctuations. The first assumption we make in primordial formation sector where 
the primordial power spectrum is assumed to be nearly scale invariant, following the theory of slow-roll inflation. 
In the evolution sector, we model the evolution of the initial perturbations.

Favored deviations from our standard assumptions may hint towards the necessity of a new model of the Universe. 
Hunting down these deviations requires extensive search in both the sectors mentioned, individually or jointly. The nature of the 
search can be model specific or model independent. Model specific search, though being effective, is limited in terms 
of its flexibilities. On the other hand, model independent reconstruction of the phenomenology directly from the data can 
immediately pin point all the places where we might need deviations. 

The aim of this paper is to reconstruct the PPS from Planck data. We list few issues that can be addressed with the PPS 
reconstruction.

\begin{itemize}
 \item {\bf Features :} Features in the PPS~\cite{features-all} have been widely discussed in literature to address the deviations from power 
 law PPS and thereby categorize a class of inflationary models that deal with departures from slow-roll hypothesis. 
 The location of possible features can be addressed by direct reconstruction that is one of the main aims of this 
 paper. Moreover, Planck had observed the CMB in 9 different frequency channels and provided 5 angular power spectra for 
 parameter estimation. It is interesting to examine whether features found in different spectra are consistent to each other. 
 
 \item {\bf Falsifying power law PPS :} Certain deviations from power law PPS can address the data 
 better with additional degrees of freedom. However it does not always mean that these features indicate any physical effect. Statistical fluctuations 
 and noise or systematics in the data can also lead to certain features. Reconstruction of PPS for a large number of realizations 
 of the data can address this issue and in our work~\cite{Hazra:2013xva} with WMAP-9 data~\cite{Hinshaw:2012fq} we have shown that within the uncertainties in the WMAP data
 the power law performs perfectly well. Planck constrains the PPS with much better precision than WMAP-9 which invites a re-analysis of the falsification of 
 power law PPS with Planck data. Moreover, Planck analysis has reported features near multipoles $\ell\simeq 20-30$ and near $\ell\simeq 1800$~\footnote{
 $\ell=1800$ feature is acknowledged by the Planck collaboration to be a systematic caused by 4K cooler-bolometer read--out electromagnetic interference~\cite{Planck:inflation}.}. 
 In our analysis we provide the significance of these features in different spectra provided by Planck.
 
 \item {\bf Gravitational lensing :} With the Planck data the effect of gravitational lensing is confirmed with 25$\sigma$ 
 confidence~\cite{Planck:lensing}. The convolved information of the initial fluctuations and its evolution reach us after getting 
 lensed by matter distributed across the Universe. Due to this convolution we can expect a degeneracy between the background cosmology, the gravitational 
 lensing and the shape of the PPS. 
 In this paper we address the degeneracy between the effects of gravitational lensing and the PPS in detail. Despite 
 of the degeneracy it is interesting to examine whether the lensing effect can be captured/indicated through the reconstruction.   

 \item {\bf Consistency with WMAP-9:} Resolution of Planck is significantly better than WMAP and hence cosmological scales
 probed by Planck includes WMAP probed scales. This overlap enables us to check how well the reconstructed PPS 
 from Planck can fit the WMAP dataset. We explore the consistency in PPS obtained from different combinations 
 of angular power spectra provided by Planck. 
 
 \item {\bf Smooth primordial power spectrum :} Having investigating all the above issues, we present a smooth 
 PPS which can be described by a simple form. We expect the smooth PPS to contain 
 {\it only} significant features in the data, which might be coming from some underlying physical effects.  
 
 \end{itemize}
 
This paper is organized as follows. In section~\ref{sec:formalism} we discuss the main algorithm of the reconstruction. 
Following that, in section~\ref{sec:results} we present the results of our analysis
and in section~\ref{sec:discussion} we end with concluding remarks.

%  Over the last year the consistency between WMAP and Planck has become an 
%  important topic the results from 2 observations differ in 

%  As the CMB signal provides the informations power spectrum of initial fluctuations 
%  and their evolution in a convolved form, there exist a large degeneracy between the background and primordial cosmology. With WMAP-9 data 
%  using the reconstruction we have been able to understand the degeneracy~\cite{Hazra:2013eva}. However, with Planck, the study of  
%  complete degeneracy is becomes complicated as the foregrounds in the CMB map increase the parameter space of our analysis. In this paper,
%  therefore we address a specific degeneracy. 

% For example, in a recent work~\cite{Hazra-Shafieloo-14}, in a model independent way 
% we have shown that Planck data strongly favors a specific and significant damping in the CMB power spectrum.  

% Here, by theory we mean only the shape of the PPS. A model independent reconstruction of the PPS 
% is certainly based on the theory of the evolution of the universe which is model dependent. Fixing this model one can hunt for the 
% free form PPS.      

%%%%%%%%%%%%%%%%%%%%%%%%%%%%%%%%%%%%%%%%%%%%%%%%%%%%%%%%%%%%%%%%%%%%%%%%%%%%%%%

\section{Reconstruction algorithm}~\label{sec:formalism}

Reconstruction of the PPS directly from the data can be achieved through various 
methods~\cite{Shafieloo:2003gf,Shafieloo:2007tk,Hazra:2013xva,Hazra:2013eva,reconstruction-all,Hazra:2013nca}~\footnote{Also see~\cite{Paykari:2014zua}, 
though we do not agree with some claims made in this paper regarding the MRL algorithm.}. In this analysis we shall 
re-use the Richardson-Lucy (RL)~\cite{richardson,lucy,baugh1,baugh2} deconvolution algorithm. The error-sensitive IRL method was 
introduced in~\cite{Shafieloo:2003gf,Shafieloo:2007tk} for the binned 
data. We further modified the IRL algorithm to work for WMAP-9 in case of binned and unbinned data in a combined 
analysis~\cite{Hazra:2013xva}, referred to as MRL. However, data from Planck necessitates yet another modification to the MRL algorithm.  

Let us revisit the convolution (Eq.~\ref{eq:clequation}) of PPS (${P}_{k}$) and the radiative transport kernel 
(${G}_{\ell k}$) that generates the angular power spectrum ($\cl^{\rm T}$).  

\beq
\cl^{\rm T}=\sum_{i}{G}_{\ell k_{i}}{{P}_{k_{i}}}~\label{eq:clequation}
\eeq

The radiative transport kernel here contains the information about the assumed background cosmological model. We should mention that 
throughout this paper we have used the Friedmann-Lema\^{\i}tre-Robertson-Walker spatially flat Universe with cosmological constant as 
the dark energy. The dark matter is assumed to be non-relativistic (CDM). Following the standard model the effective number of relativistic 
species is fixed to be 3.046. For neutrinos we follow standard mass hierarchy and mass of the single massive neutrino eigenstate is 
fixed to be $m_{\nu}=0.06$ eV. Note our assumptions simply follow the {\it baseline} model of the Planck analysis~\cite{Planck:cparam}. The radiative
transport matrix ${G}_{\ell k}$ is calculated for the Planck best fit values of the parameters $\Omega_{\rm b}h^2$ (baryon density), 
$\Omega_{\rm CDM}h^2$ (CDM density), $H_0$ (Hubble parameter) and $\tau$ (reionization optical depth). 

\begin{figure*}[!htb]
\begin{center} 
\resizebox{420pt}{300pt}{\includegraphics{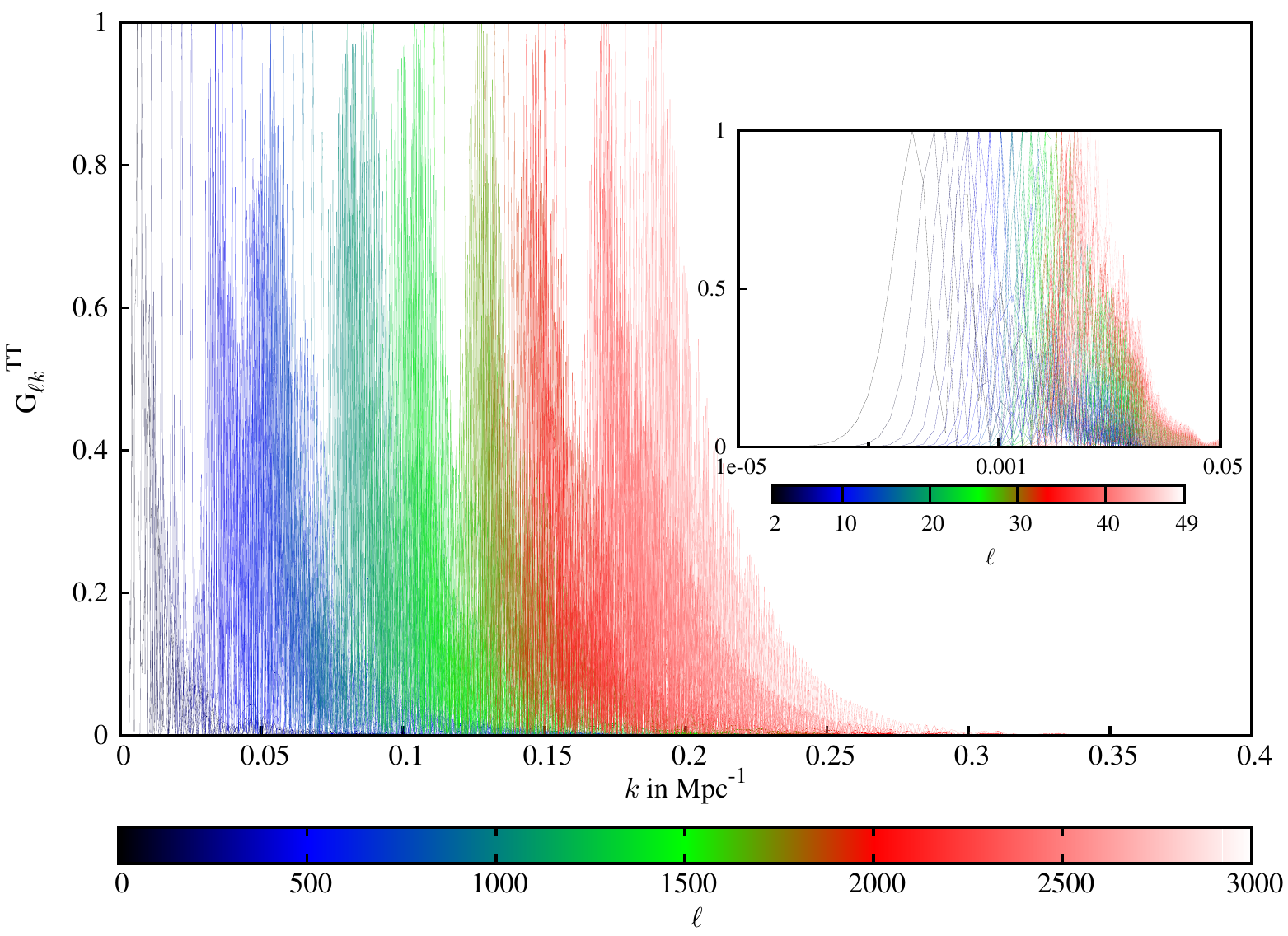}} 
\end{center}
\caption{\footnotesize\label{fig:glk} The radiative transport kernel (appearing in Eq.~\ref{eq:clequation}) as a function of wavenumber is plotted.
The colorbar at bottom represents the multipole moment ($\ell$) ranging from 50 to 2500. The inset contains the plot for $\ell=2-49$ in logarithmic
scale in $k$.}
\end{figure*}

It is of importance that we present the form of the radiative transport kernel pictorially. The angular power spectrum at  
multipole $\ell$ is related to a window of the PPS in wavenumber ($k$) space. The particular window in 
$k$ space is given by the distribution of radiative transport kernel which is plotted in Fig.~\ref{fig:glk}. In this figure
we have plotted the normalized ${G}_{\ell k}$ (maximum value normalized to 1) as a function of $k$. The colorbar at 
the bottom represents the multipole from $\ell=50$ to $2500$. In the inset the same is plotted for low-$\ell$ (2-49), in 
logarithmic scale in $k$. Note that the peak-position of the kernel (corresponding to a particular color) refers to the $k$ 
that contributes maximum to the convolution integral of angular power spectrum at a particular $\ell$ 
(corresponding to the same color).

The $\cl^{\rm T}$ appearing in Eq.~\ref{eq:clequation} does not exactly correspond to the angular power spectrum from CMB map that we measure. 
The signal $\cl^{\rm T}$ gets lensed by the underlying matter distribution while reaching to us. Moreover, we should remember that 
at small scales due to point sources, cosmic infrared background (CIB), Sunyaev-Zeldovich (SZ) effect the foregrounds 
dominate the power spectrum. Hence, to reconstruct the PPS from the angular power spectrum obtained from Planck we need to extract the 
$\cl^{\rm T}$ information of Eq.~\ref{eq:clequation} from the $\cl^{\rm Data}$ in all Planck spectra.

In order have a clean angular power spectrum let us begin with identifying and naming individual spectra and their combinations used in this paper. 
Table.~\ref{tab:nomen} lists the multipole range covered by individual spectrum. For simplicity to address our results we follow a nomenclature as has been 
indicated in the table. The largest scale angular power spectrum is denoted by $\alpha$, while the two intermediate scale angular power spectrum are identified by $a$ (100 GHz)
and $b$ (143 GHz). Two smallest scale angular power spectrum are marked as 1 (217 GHz) and 2 ($143{~\rm GHz}\times 217{~\rm GHz}$). We work with 8 different combinations of the power 
spectrum. Starting from the largest scales the different combinations are chosen to cover the multipole range step-by-step with minimal overlap in scales. Hence the possible 
combinations are $\alpha,~\alpha+a,~\alpha+b,~\alpha+a+1,~\alpha+a+2,~\alpha+b+1,~\alpha+b+2$. Finally we choose the 8'th combination to be the entire spectrum, denoted by
$\alpha+a+b+1+2$. For our analysis throughout the paper we shall refer to this nomenclature. 

\begin{table*}[!htb]
\begin{center}
\vspace{6pt}
\begin{tabular}{| l | l | l | l|}
\hline\hline
Our symbol &  Spectra & Multipoles($\ell$)& Scales\\
\hline
$\alpha$ & low-$\ell$ & 2-49 & Largest scales\\
\hline
a & $100{~\rm GHz}\times100{~\rm GHz}$ & 50-1200& Intermediate scales\\
\hline
b & $143{~\rm GHz}\times 143{~\rm GHz}$ & 50-2000& Intermediate scales\\
\hline
1 & $217{~\rm GHz}\times 217{~\rm GHz}$ & 500-2500& Small scales\\
\hline
2 & $143{~\rm GHz}\times 217{~\rm GHz}$ & 500-2500& Small scales\\
\hline\hline
\end{tabular}
\end{center}\caption{~\label{tab:nomen} Our nomenclature of referring to individual Planck spectra. We have divided the 
complete angular power spectra from Planck in largest, intermediate and smallest cosmological scales.}
\end{table*}

We now present the complete algorithm of modified Richardson-Lucy (MRL), {\it particularly} designed for 
the analysis with Planck, though the main algorithm being similar to~\cite{Hazra:2013xva,Hazra:2013eva}. MRL algorithm is an iterative method, where the 
PPS at $i+1$'th iteration, ${P}_{k}^{(i+1)}$, 
is given as a modification to the PPS at $i$'th iteration, ${P}_{k}^{(i)}$ as has been provided in Eq.~\ref{eq:mrl}. 
${\widetilde{G}}_{\ell k}$ is the transport kernel, normalized in each $\ell$. $\cl^{{\rm T}(i)}$ is the theoretical angular power 
spectrum corresponding to the PPS at $i$'th iteration, ${P}_{k}^{(i)}$.

\begin{eqnarray}
{{P}_{k}^{(i+1)}}-{{P}_{k}^{(i)}}&=&{{P}_{k}^{(i)}}\times\sum_{\nu}\Biggl[\sum_{\ell=\ell_{\rm min}^{\nu}}^{\ell_{\rm max}^{\nu}(\le 1900)}
\frac{1}{g_{\nu}(\ell)}{\widetilde{G}}_{\ell k}\Biggl\{\l(\frac{\cl^{\rm {D'_\nu}}-\cl^{{\rm T}(i)}}{\cl^{{\rm T}(i)}}\r)~\tanh^{2}
\l[Q_{\ell} (\cl^{\rm {D'_\nu}}-\cl^{{\rm T}(i)})\r]\Biggr\}_{\rm unbinned}\nn\\
&+&\sum_{\ell=\ell_{\rm min}^{\nu}(>1900)}^{\ell_{\rm max}^{\nu}}
\frac{1}{g'_{\nu}(\ell)}{\widetilde{G'}}_{\ell k}\Biggl\{\l(\frac{\cl^{\rm {D'_\nu}}-\cl^{{\rm T}(i)}}{\cl^{{\rm T}(i)}}\r)~\tanh^{2}
\l[\frac{\cl^{\rm D'_{\nu}}-\cl^{{\rm T}(i)}}{{\sigma_{\ell}^{\rm D_\nu}}}\r]^{2}\Biggr\}_{\rm binned}\Biggr]~\label{eq:mrl}
 \end{eqnarray}

$\cl^{\rm {D'_\nu}}$ is the {\it clean} angular power spectrum ({\it i.e.} the data) from the spectrum $\nu$. Clean angular power spectrum refers to the spectrum obtained after calibrating and subtracting 
the foreground power spectrum from each of the {\it raw} angular power spectrum, $\cl^{\rm {D_\nu}}$ provided by Planck. Foreground and calibration effects are calculated from {\tt CAMspec}~\cite{Planck:likelihood,plc} 
for the best fit foreground and calibration parameters obtained from Planck analysis for the baseline model. Moreover, we calculate the lensed angular power spectrum ($\cl^{\rm Lensed}$) and un-lensed 
power spectrum($\cl^{\rm un-lensed}$)~\footnote{The lensing effect is calculated using {\tt CAMB} assuming curved sky correlation function method.} 
for the best fit baseline model. We define the lensing template as their difference following Eq.~\ref{eq:lens}

\begin{equation}
\cl^{\rm Lens-template}= \cl^{\rm Lensed}-\cl^{\rm un-lensed}~\label{eq:lens}
\end{equation}
Hence, correspond to the $\cl^{\rm T}$ in Eq.~\ref{eq:clequation} and~\ref{eq:mrl}, we use $\cl^{\rm {D'_\nu}}$ which we obtain upon subtracting the lensing template after cleaning the foregrounds
from raw data. Note that throughout the paper we use the same convention apart from subsection~\ref{subsec:lens} where we check the lensing effect with and without subtracting the 
lensing template.  

The term ${g_{\nu}(\ell)}$ in Eq.~\ref{eq:mrl} appears as a degeneracy factor and counts the number of spectra contributing to a specific multipole $\ell$. For instance, 
while analyzing the combination $\alpha+a+b+1+2$, for $\ell=2-49$ only the spectrum $\alpha$ contributes which makes ${g_{\nu}(\ell)}=1$ and at $\ell=500$ where four spectra ($a,b,1$ and $2$), 
contribute ${g_{\nu}(\ell)}$ becomes 4. Note that Eq.~\ref{eq:mrl} breaks the modification to the PPS in 2 parts, namely binned and un-binned similar to~\cite{Hazra:2013xva}. 
Due to noise we get negative $\cl$'s from the data where the signal-to-noise ratio is very low that is theoretically impossible. With negative 
angular power spectrum the MRL algorithm fails to work as it is designed to work for positive definite matrices and hence we need to work with unbinned and binned 
data in a combined analysis as in~\cite{Hazra:2013xva}. We find that $\cl^{\rm {D'_\nu}}$ starts picking up negative data points after $\ell=1900$ (starting 
from $b$). Hence, we fix $\ell=1900$ to be the transition point from un-binned to binned data analysis. After multipole 1900 we bin the {\it clean} data from 
$b,1$ and $2$ with $\ell_{\rm bin}=50$ such that the binned data points beyond that are certainly positive. Similar to the unbinned analysis $g'_{\nu}(\ell)$
represents the degeneracy factor for the binned analysis. ${\widetilde{G'}}_{\ell k}$ is the binned radiative transport kernel with same binning 
width ($\ell_{\rm bin}=50$). The $\tanh$ factors in both the parts of Eq.~\ref{eq:mrl} represent the convergence factors introduced in~\cite{Shafieloo:2003gf,Shafieloo:2007tk}.
The term $Q_{\ell}$ is given by Eq.~\ref{eq:covmat},
\beq
Q_{\ell}=\sum_{\ell'}(C_{\ell'}^{\rm D'_{\nu}}-C_{\ell'}^{{\rm T}(i)}) {\rm COV}^{-1}(\ell,\ell')~\label{eq:covmat}
\eeq
where, ${\rm COV}(\ell,\ell')$ represents the error covariance matrix. Ideally the full 
covariance matrix should be used in the analysis, however, in our analysis we shall only use the diagonal terms of the covariance matrix 
to optimize between computational expense and insignificant improvement in results. For binned analysis after $\ell=1900$ we use 
the ${\sigma_{\ell}^{\rm D_\nu}}$ as the error bars in spectra $\nu$ obtained after computation of the errors for the binned data. Note that due to asymmetric 
errors for low-$\ell$ data in $\alpha$ the algorithm~\ref{eq:mrl} is modified. For $i$'th iteration, if the $\cl^{\rm T}$ is below the 
data, the algorithm selects the error-bar at below to work with and it chooses the upper error-bar if the opposite happens.   

Before continuing to the next section we would like to note a few points. Throughout our analysis we shall only use the publicly available 
Planck likelihood code~\cite{plc}. The high-$\ell$ likelihood is obtained through {\tt CAMspec} and the low-$\ell$ likelihood 
is estimated by {\tt commander} which are available in public domain. For calculation of angular power spectrum we have used {\tt CAMB}
~\cite{cambsite,Lewis:1999bs}. We have developed a new code for MRL with Planck data and used it as an add-on of {\tt CAMB}. { The number of $k$-points 
used for the convolution and the $k_{\rm min}$ and $k_{\rm max}$ depend on the multipoles used for the reconstruction. For example,
when we use $\alpha+a+b+1+2$, we work with $\sim$2700 wavenumbers between $k_{\rm min}=7\times10^{-6}~{\rm Mpc}^{-1}$ and $k_{\rm max}=0.44~{\rm Mpc}^{-1}$. This is 
a conservative bound where the transport kernel becomes negligible to contribute in the convolution integral}. 
After the reconstruction the likelihood is obtained upon adding the same lensing-template which has been subtracted from the data and use the 
same foreground and calibration parameters from which the foreground power spectrum was calculated. While comparing 
with WMAP data we have used the complete WMAP-9 likelihood code supplied by WMAP~\cite{lambdasite}.

%%%%%%%%%%%%%%%%%%%%%%%%%%%%%%%%%%%%%%%%%%%%%%%%%%%%%%%%%%%%%%%%%%%%%%%%%%%%%%
\section{Applications and results}\label{sec:results}
In this section we shall discuss the different applications of the reconstruction as has been pointed out in the introduction. 

\subsection{Consistency with WMAP data}

We begin with by discussing the consistency of the Planck and WMAP-9 data. Checking this consistency has become an important issue since we find the cosmological parameters from WMAP-9 and 
Planck differ significantly~\cite{Hinshaw:2012fq,Planck:cparam} and the Planck angular power spectrum is $\sim2.5\%$ lower~\cite{Planck:likelihood,Hazra:2013oqa} than the WMAP in all the 
scales probed by WMAP. We performed a consistency check between the two data using Crossing statistic~\cite{Hazra:2013oqa} and found that disallowing an amplitude shift, two data 
disagree with each other at worse than 3$\sigma$ confidence. 

In this paper we present a Planck spectrum dependent check of consistency between Planck and WMAP-9 (For other works, see~\cite{Hazra:2013oqa,Kovacs:2013vja}). Using the radiative transport 
kernel obtained from Planck best fit parameters we obtain a PPS using Eq.~\ref{eq:mrl}. In this section we shall use the PPS obtained after 50 iteration. Using that PPS and the 
same kernel we shall check how well we can fit the WMAP-9 data. Of course, through this method we can only check a relative agreement/disagreement of Planck spectra 
with WMAP data. Given a set of cosmological parameters MRL reconstructs a PPS that contains the features and random noise in the data. As we do not expect the noise in Planck data to 
match with WMAP we implement a Gaussian smoothing algorithm following~\cite{Hazra:2013xva}. 
\beq
{{P}_{k}^{\rm Smooth}}=\frac{\sum_{{\tilde k}={\rm k_{\rm min}}}^{\rm k_{\rm max}}{{P}_{\tilde k}^{\rm Raw}
}\times\exp\l[-\l(\frac{\log{\tilde k}-\log k}{\Delta}\r)^2\r]}{\sum_{{\tilde k}={\rm k_{\rm min}}}^{\rm k_{\rm max}}\exp\l[-\l(\frac{\log{\tilde k}-\log k}{\Delta}\r)^2\r]}~\label{eq:gauss}
\eeq
Using Eq.~\ref{eq:gauss} we smooth the raw PPS ${P}_{\tilde k}^{\rm Raw}$ and get a smooth PPS ${{P}_{k}^{\rm Smooth}}$ depending on the smoothing width $\Delta$. For our analysis in this 
section we shall use a constant smoothing width at all scales of interest. As a function of smoothing width $\Delta$ we examine the WMAP-9 likelihood of the PPS obtained from 
different combinations of Planck spectra. To check the amplitude difference we have allowed the overall amplitude of the PPS to vary from $90\%$ to $110\%$. We shall 
define the amplitude shift by factor $A$, which ranges from 0.9 to 1.1.

In Fig.~\ref{fig:vark-1} and Fig.~\ref{fig:vark-2} we plot our results for different combinations of Planck spectra. The plots at the left panel 
contain the $-\ln {\cal L}$ (${\cal L}$ refers to likelihood) from Planck and WMAP-9 as a function of the $\Delta$. The red and blue horizontal straight lines represent the 
$-\ln {\cal L}$ from Planck and WMAP-9 for best fit baseline model respectively.
The red and the blue curves indicate the Planck and WMAP-9 likelihood from the PPS reconstructed from Planck for the corresponding combinations of
spectra. The green curve represents the best likelihood to WMAP-9 from the same reconstructed PPS with allowing an overall amplitude shift.
Note that apart from $\alpha$ in all the cases from $\Delta\simeq0.01$ the PPS fits the Planck data worse than power law PPS. The reason behind 
this is the following. The MRL algorithm fits the noise in the data along with possible features. 
A smoothing of the PPS with higher smoothing width smears out non-local features with higher frequencies resulting in worse fit to the likelihood compared to power 
law spectrum. Thereby we shall give importance to the results obtained till the value of $\Delta$ we get a better fit to the Planck data compared 
to power law model as till that point the combination of smoothing and the MRL work well. Plots at the right panel represent the amplitude 
factor $A$ which provides best fit to the WMAP-9 data, as a function $\Delta$. The red shaded regions represent the band of amplitude factor 
where we get better fit to WMAP-9 compared to power law best fit. Absence of the red band in a combination of Planck spectra indicates that 
even with an overall amplitude shift the 
PPS from that particular combination fails to provide a better fit to WMAP-9 {\it w.r.t.} to power law best fit. 
The green curve represent the best fit, $A$. The blue horizontal line represents $A=1$, no amplitude shift. 

Below we list the results as has been indicated in Fig.~\ref{fig:vark-1} and Fig.~\ref{fig:vark-2} at different scales. 
\begin{itemize}
 \item 

{\bf Largest scales ($\alpha$):} In this particular case, the left plot provides the $-\ln {\cal L}$  from {\tt commander} from Planck (from $\ell=2-49$) 
and low-$\ell$ likelihood from WMAP-9(from $\ell=2-32$). Note that the reconstructed PPS from $\alpha$ is able to fit the WMAP-9 
low-$\ell$ likelihood without an overall amplitude shift. Moreover the broad red band in the right panel indicates that the reconstructed PPS from Planck 
can fit WMAP-9 data better than power law even with overall $10\%$ amplitude shift. This reflects that low-$\ell$ features ranging from $\ell=2-32$ are similar
in both the surveys. However, the green line, that represents the best fit, indicates approximately $4\%$ increment in power provides the best fit to WMAP-9. 
 \item 

{\bf Largest + intermediate scales:}

 The reconstructed PPS from $\alpha+a$ provides a better fit to WMAP-9 complete datasets without any amplitude shift. Although the best fit to WMAP-9 is 
 obtained by increasing the power around $2.5\%$, as indicated by the green line. This fact indicates that the features in WMAP-9 and in 
 Planck 100 GHz spectrum (combined with with low-$\ell$) agree. Moreover there is a clear 
 mismatch in amplitude between the two and the WMAP-9 data is significantly higher in amplitude (the amplitude band does not extend below 1).   

 For $\alpha+b$, we notice that even allowing the amplitude shift we are unable to fit the WMAP-9 data better than power law with 
 the reconstructed PPS obtained from the 143 GHz (with low-$\ell$) spectrum. We expect to revisit this issue with 
 taking into account the correlation between multipoles and having detailed analysis~\cite{Hazra-Shafieloo-Hope}. 
 The best fit green line in the right panel here indicates amplitude difference 
 similar to the case of $\alpha+a$.

 \item
 {\bf Largest + intermediate + smallest scales~\footnote{Here by smallest scales we refer to the smallest scale probed by Planck {\it i.e.} till $\ell=2500$}:} 

 The primordial spectra reconstructed from $\alpha+a+1$ and $\alpha+a+2$ show results similar to the case of $\alpha+a$. We find that the inclusion of 1 
 and 2 spectra does not change the results (obtained for $\alpha+a$) significantly. This is an expected result since 1 and 2 have less overlap in 
 multipoles with WMAP-9 compared to $a$ and $b$. 
 However, the improvement in fit to WMAP-9 over power law degrades marginally without an amplitude shift. As before, 
 the best fit is obtained with $2.5\%$ shift in amplitude.

%%%%%%%%%%%%%%%%%%%%%%%%%%%%%%%%%%%%%%%%%%%%%%%%%%%%%%%%%%%%%%%%%%%%%%%%%%%%%%%
\begin{figure*}[!htb]
\vskip -75pt
\begin{center} 
\resizebox{210pt}{160pt}{\includegraphics{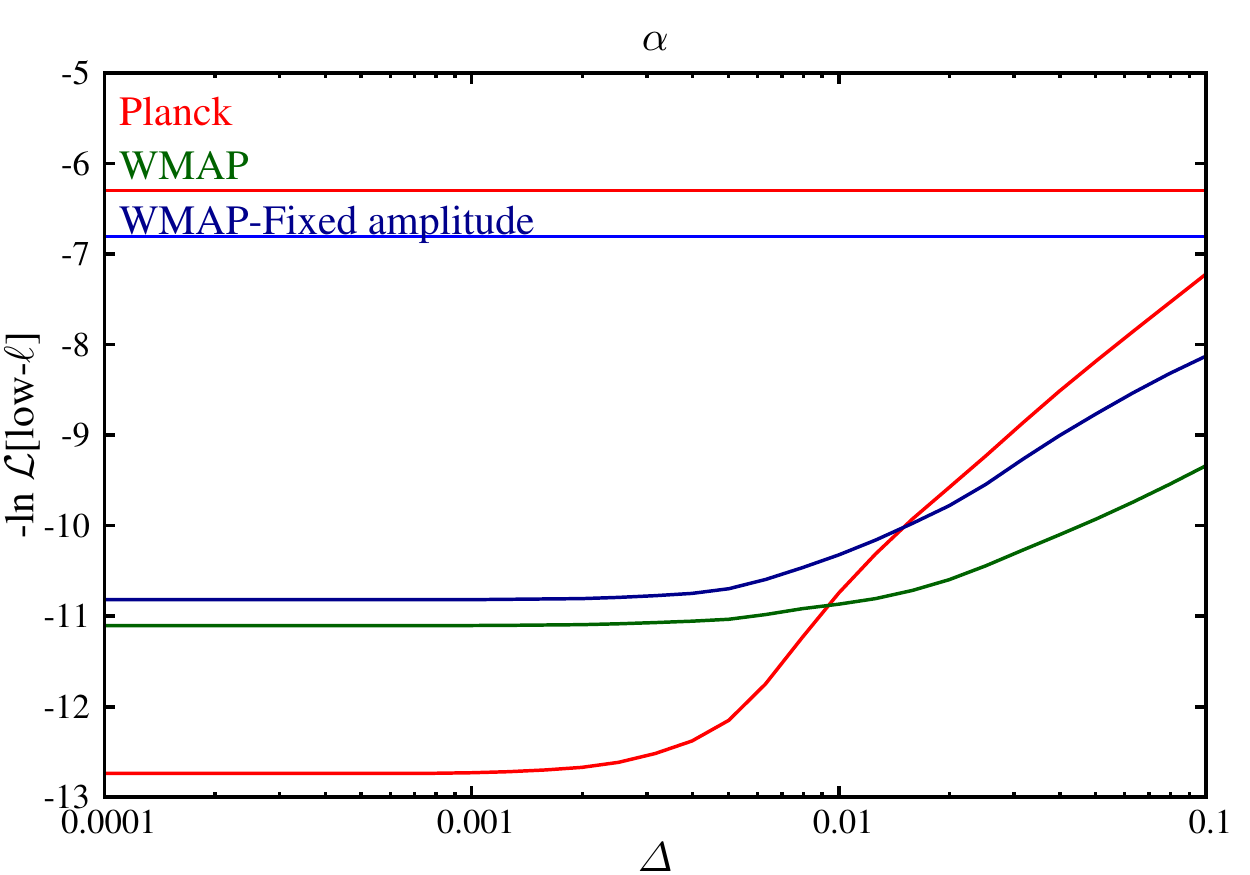}} 
\resizebox{210pt}{160pt}{\includegraphics{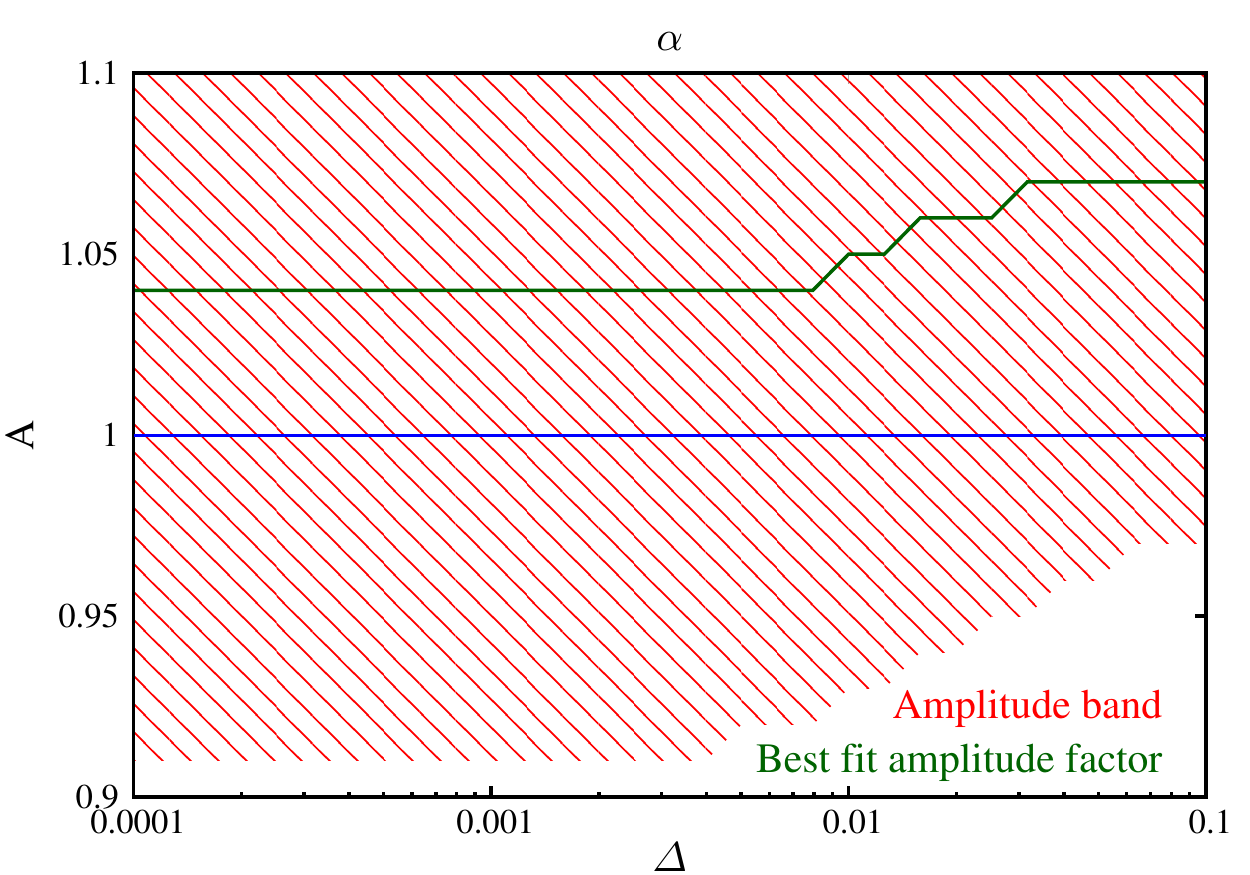}} 

\resizebox{210pt}{160pt}{\includegraphics{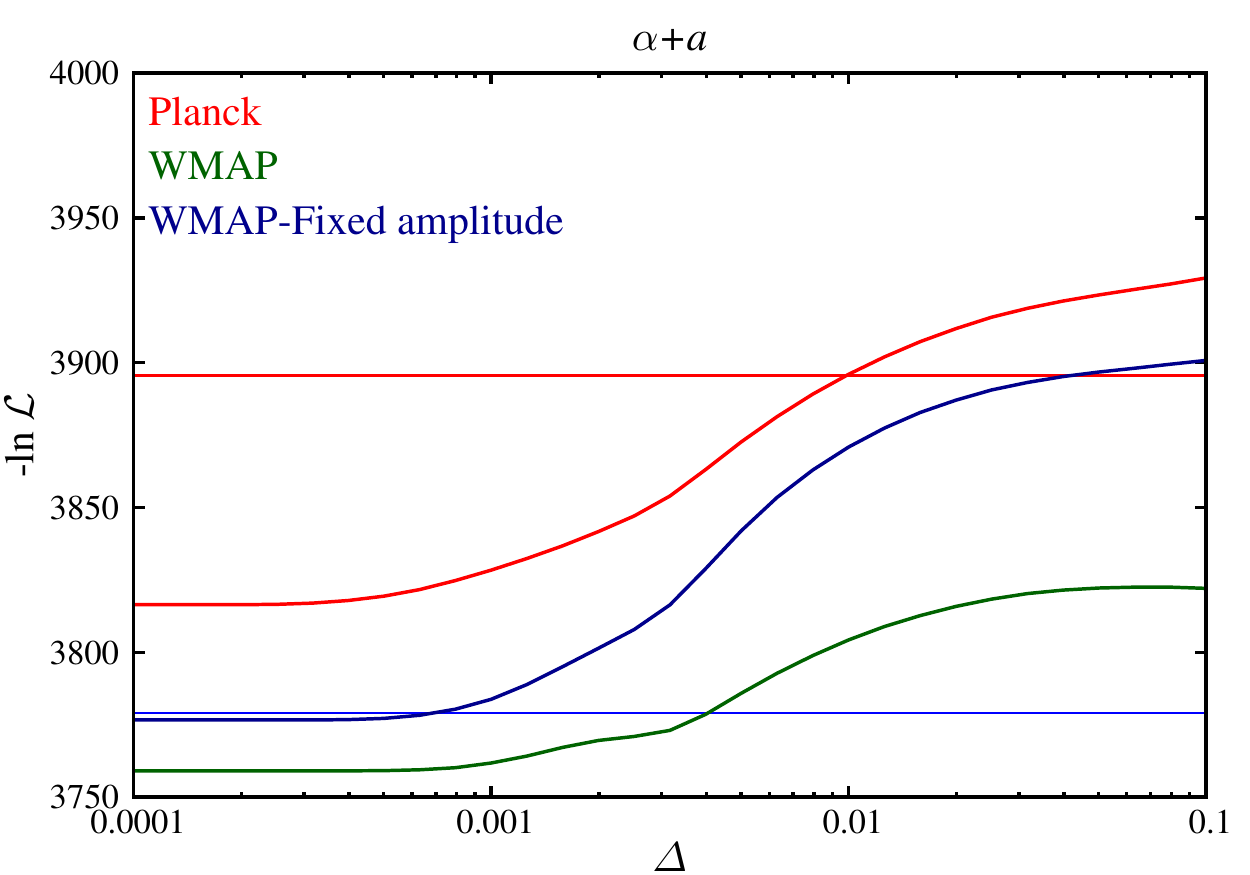}} 
\resizebox{210pt}{160pt}{\includegraphics{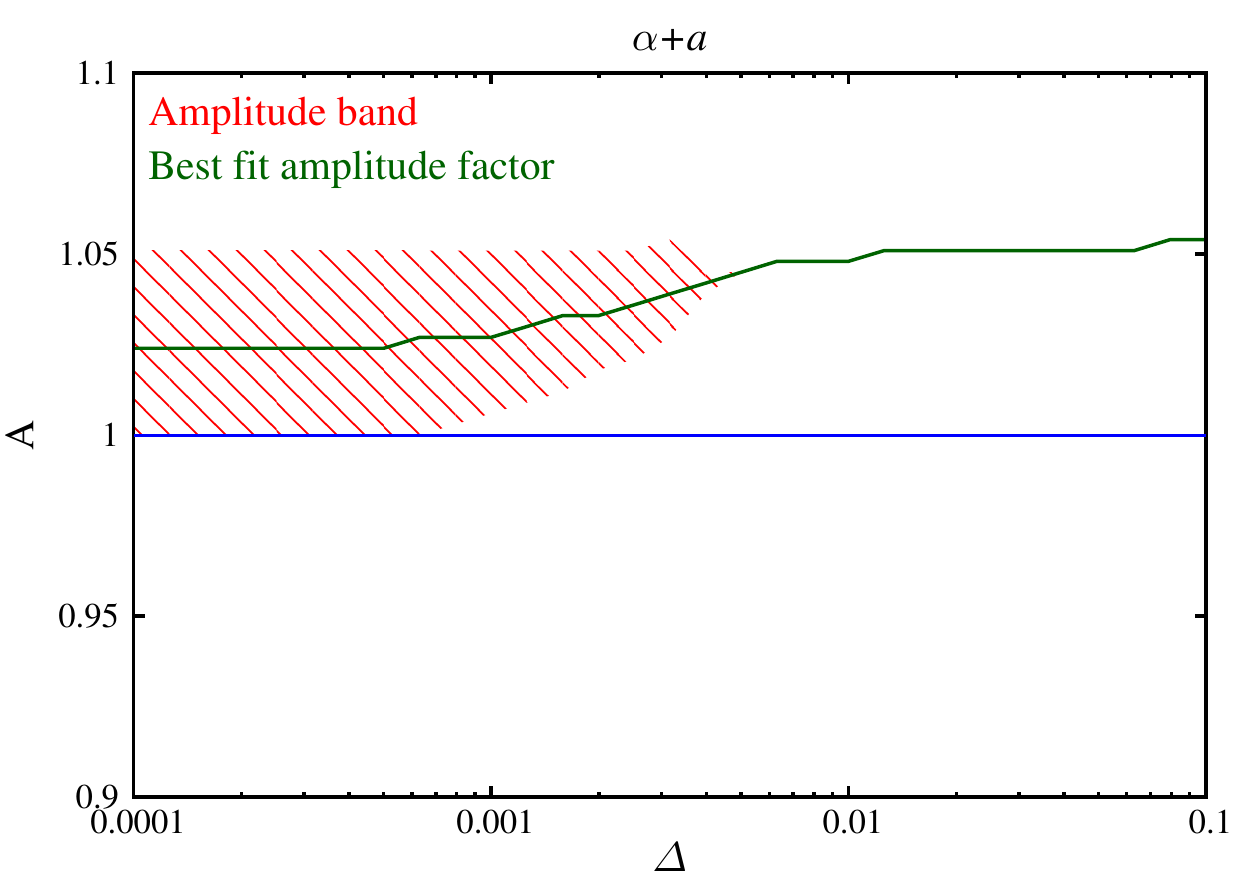}}

\resizebox{210pt}{160pt}{\includegraphics{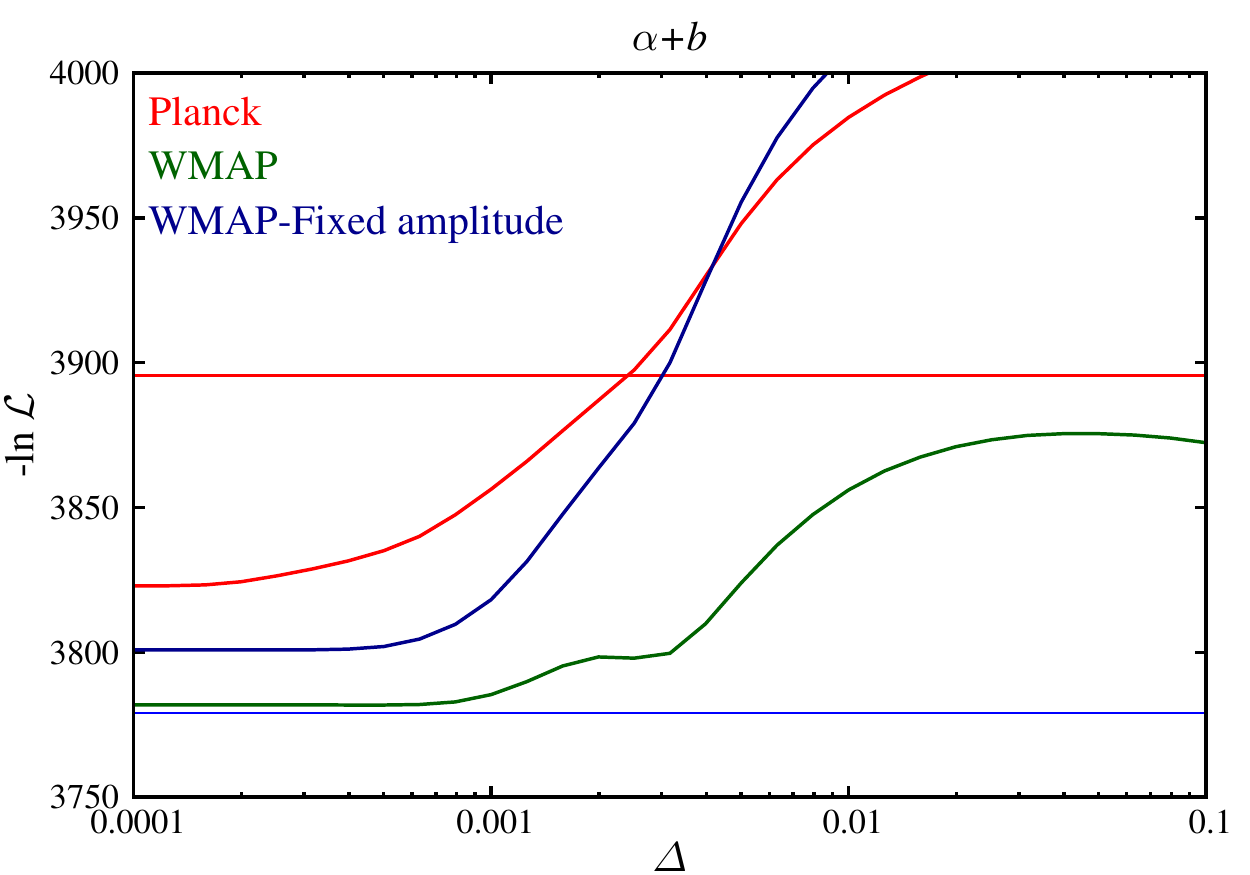}} 
\resizebox{210pt}{160pt}{\includegraphics{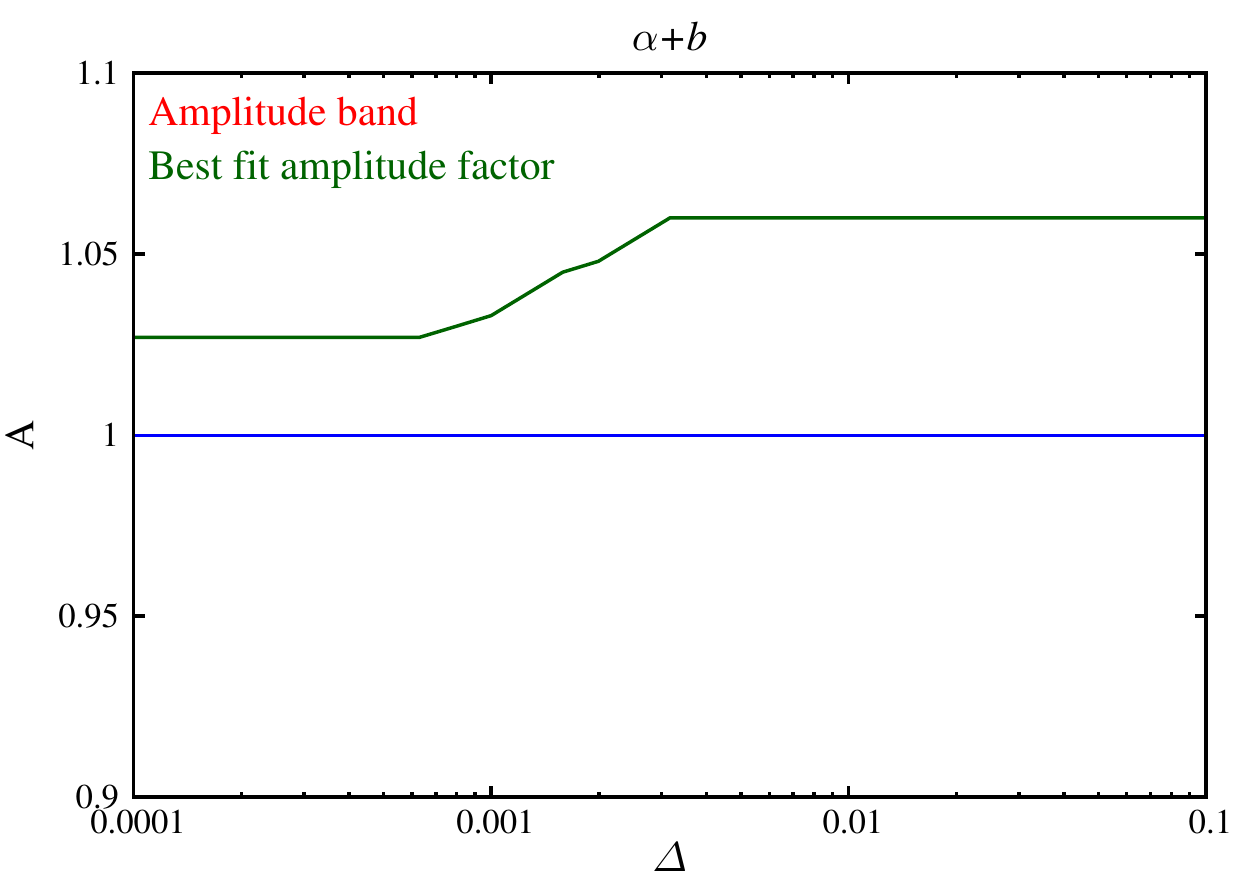}} 

\resizebox{210pt}{160pt}{\includegraphics{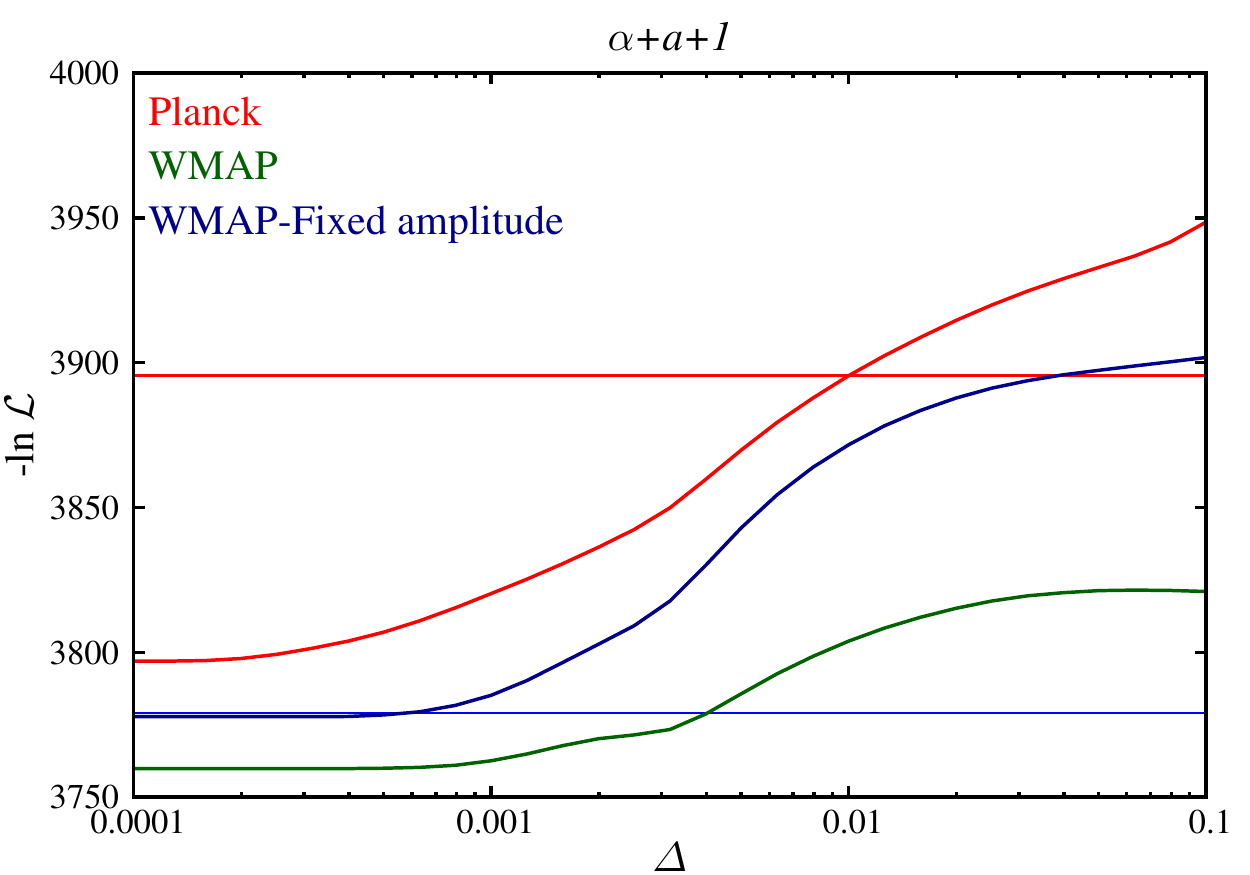}} 
\resizebox{210pt}{160pt}{\includegraphics{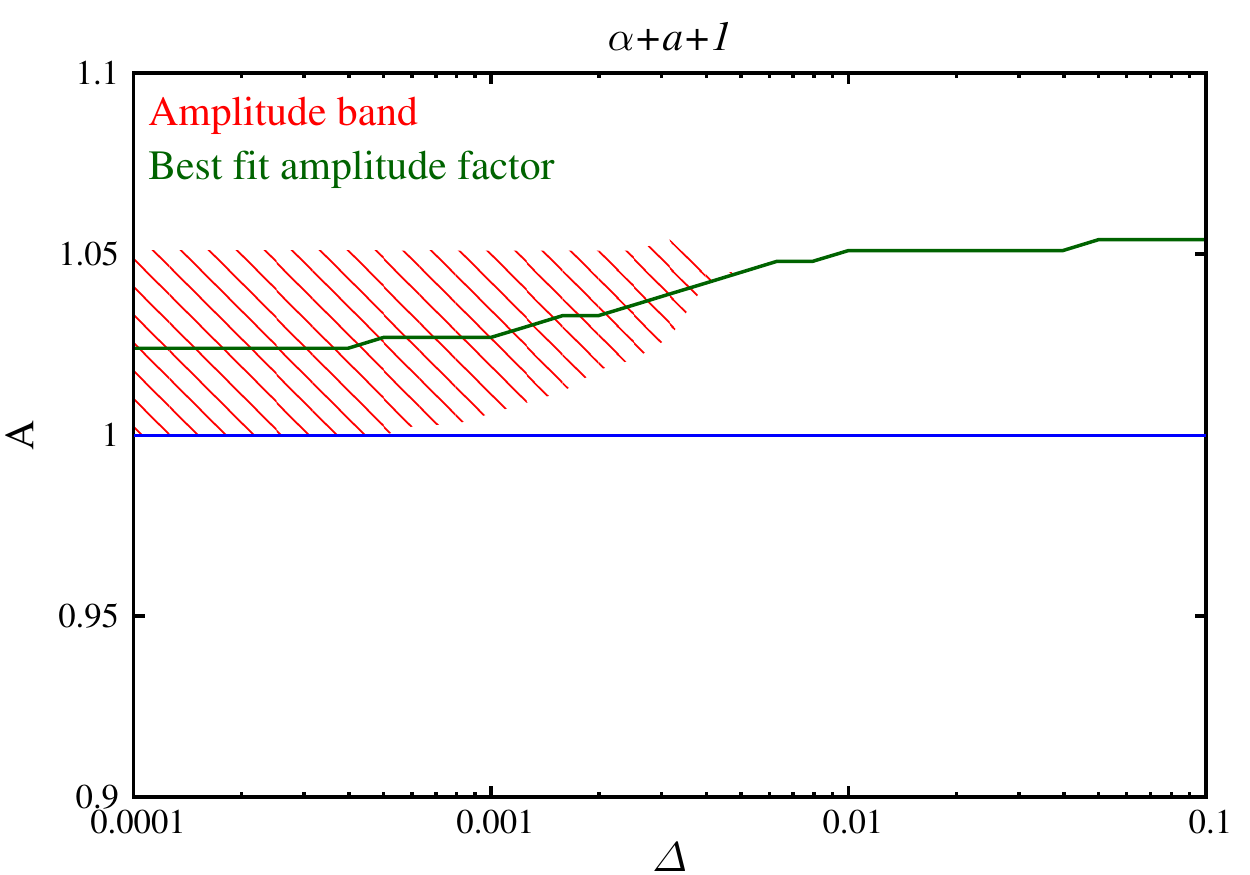}}

\end{center}
\caption{\footnotesize\label{fig:vark-1} [Left] The $-\ln {\cal L}$ obtained from Planck (in red) and WMAP-9 (in green and blue) 
for reconstructed PPS as a function of smoothing width. Red and blue straight lines represent the power law best fit $-\ln {\cal L}$
from Planck and WMAP-9 respectively. Green line represents the best fit WMAP-9 $-\ln {\cal L}$ obtained 
upon comparing the reconstructed PPS from Planck, allowing an overall amplitude shift. The blue line corresponds to WMAP-9 $-\ln {\cal L}$
obtained from the reconstructed PPS from Planck without an overall amplitude shift. [Right] The region of amplitude factor, $A$, 
to the PPS where we get better fit to WMAP-9 data from reconstructed PPS compared to power law PPS (red shaded area) and the 
value of $A$ which provides the best fit to WMAP-9 data(green line).}
\end{figure*}
 
\begin{figure*}[!htb]
\vskip -75pt
\begin{center}

\resizebox{210pt}{160pt}{\includegraphics{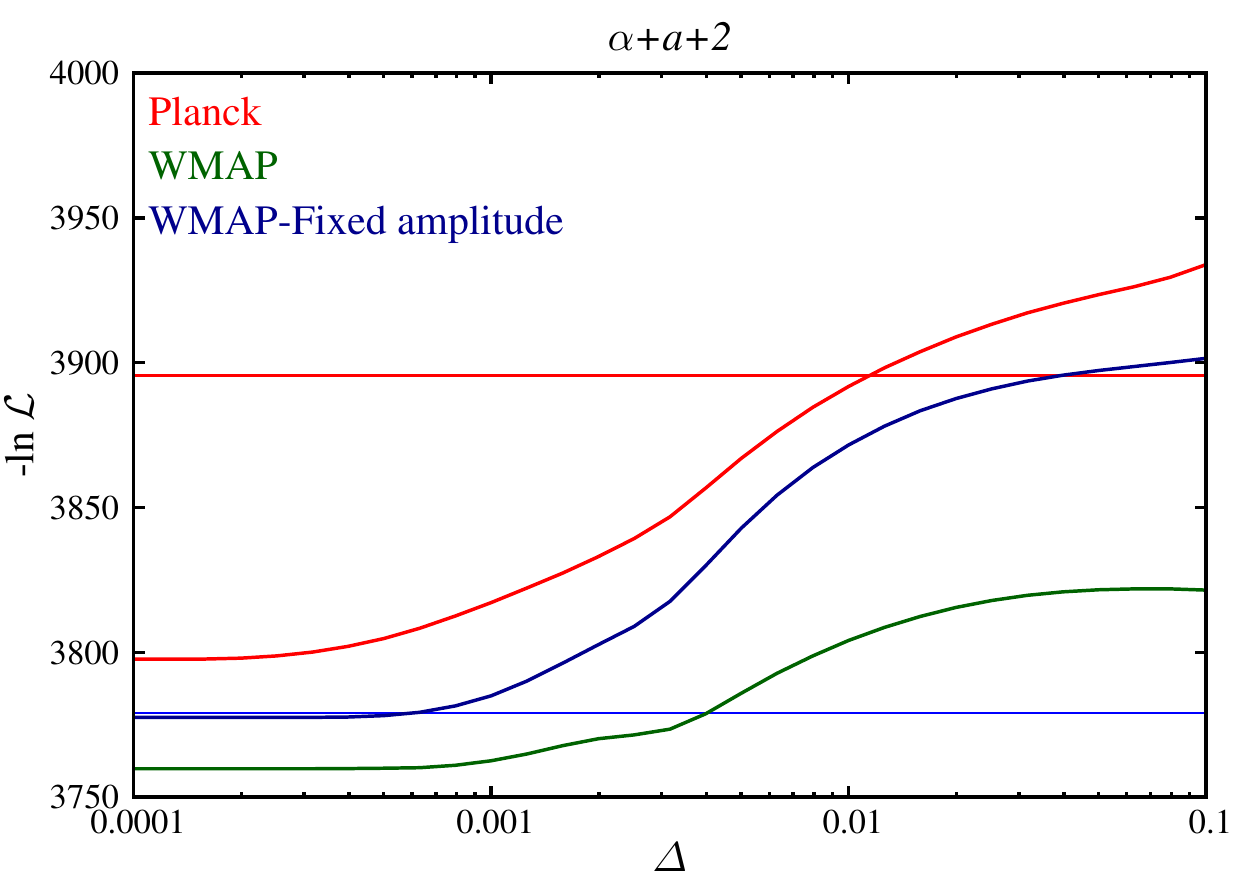}} 
\resizebox{210pt}{160pt}{\includegraphics{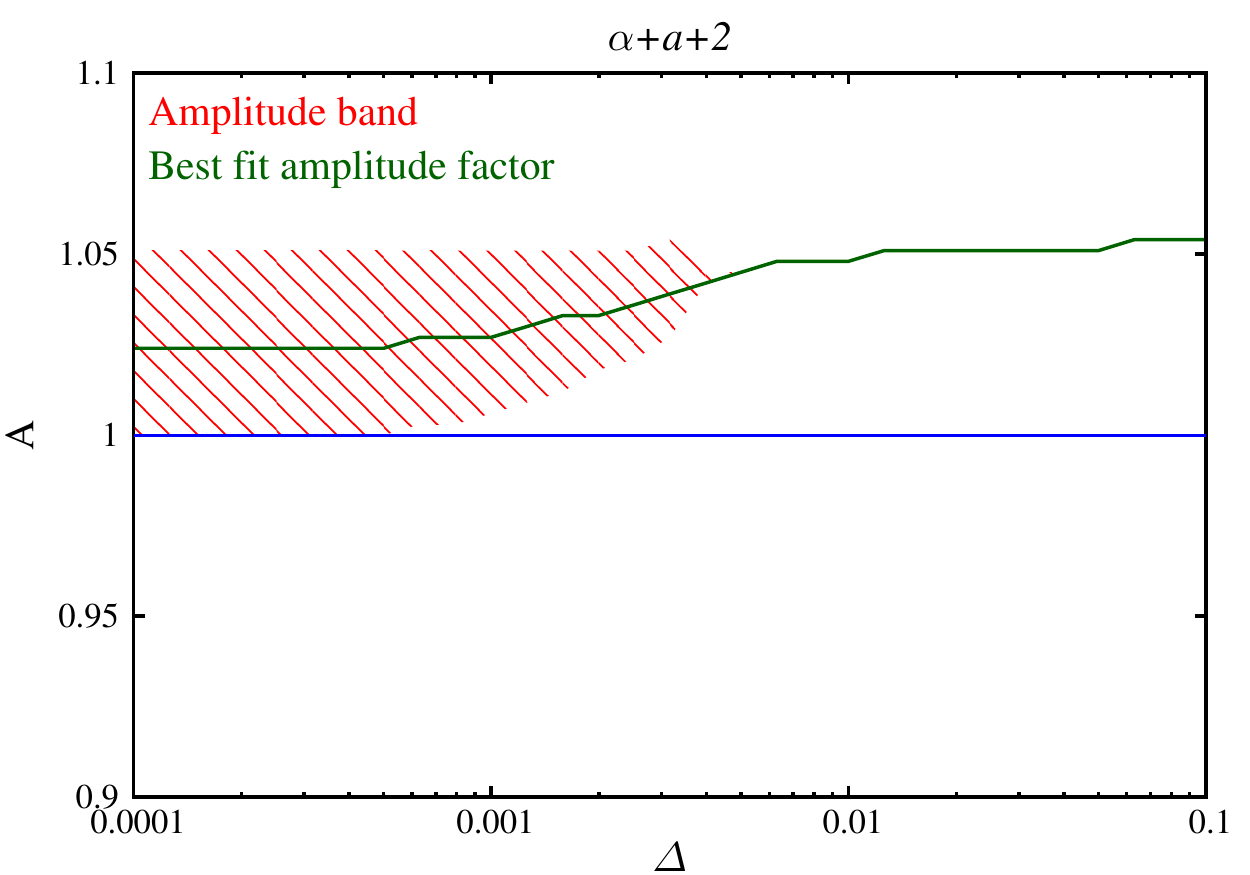}} 

\resizebox{210pt}{160pt}{\includegraphics{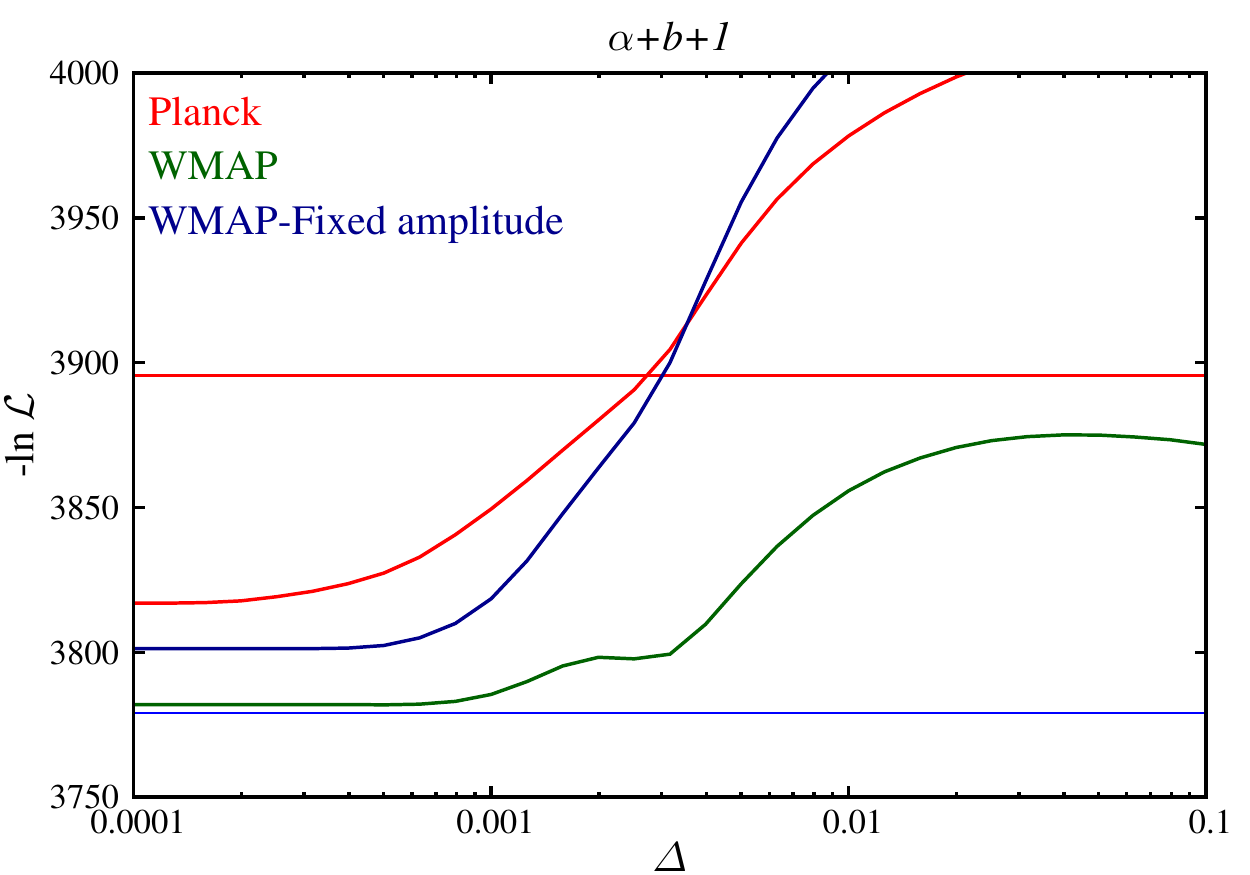}} 
\resizebox{210pt}{160pt}{\includegraphics{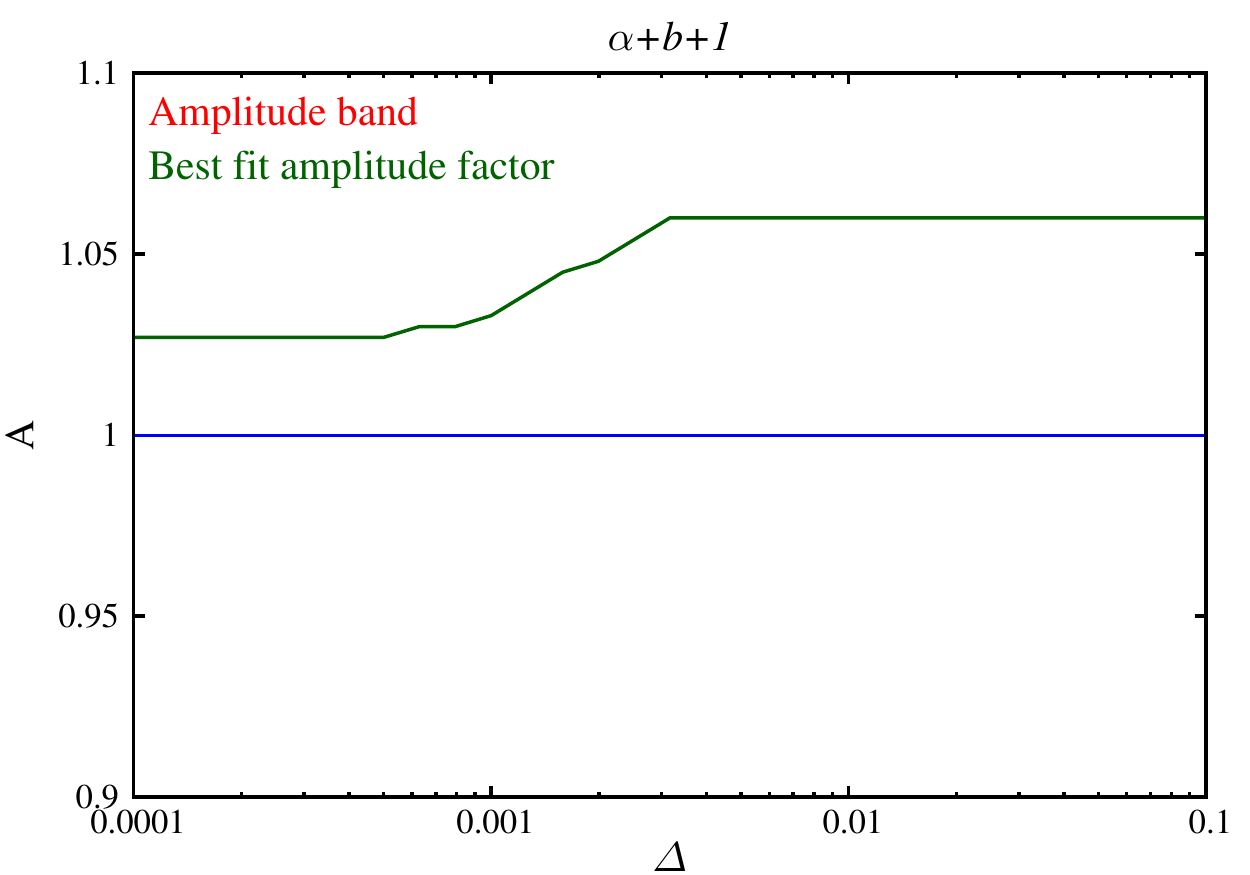}} 

\resizebox{210pt}{160pt}{\includegraphics{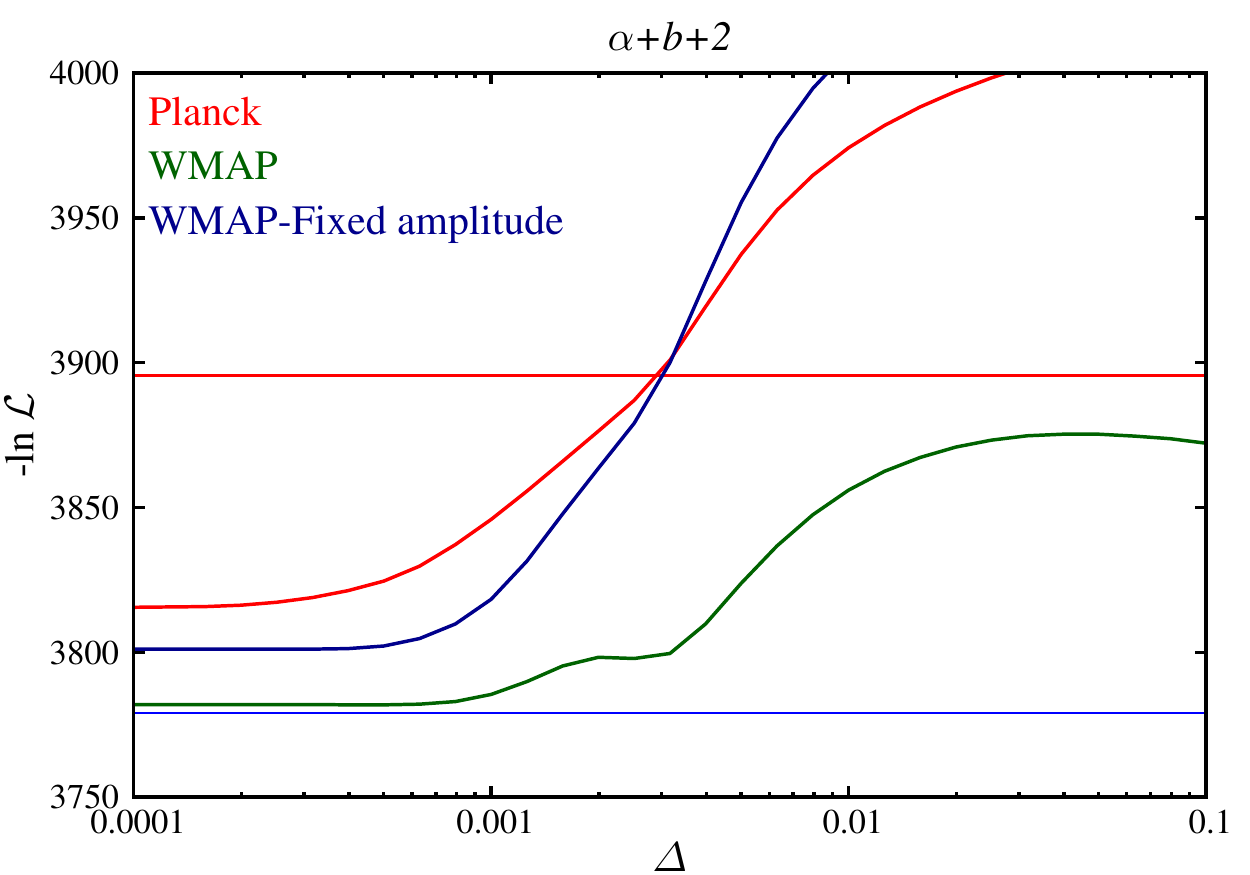}} 
\resizebox{210pt}{160pt}{\includegraphics{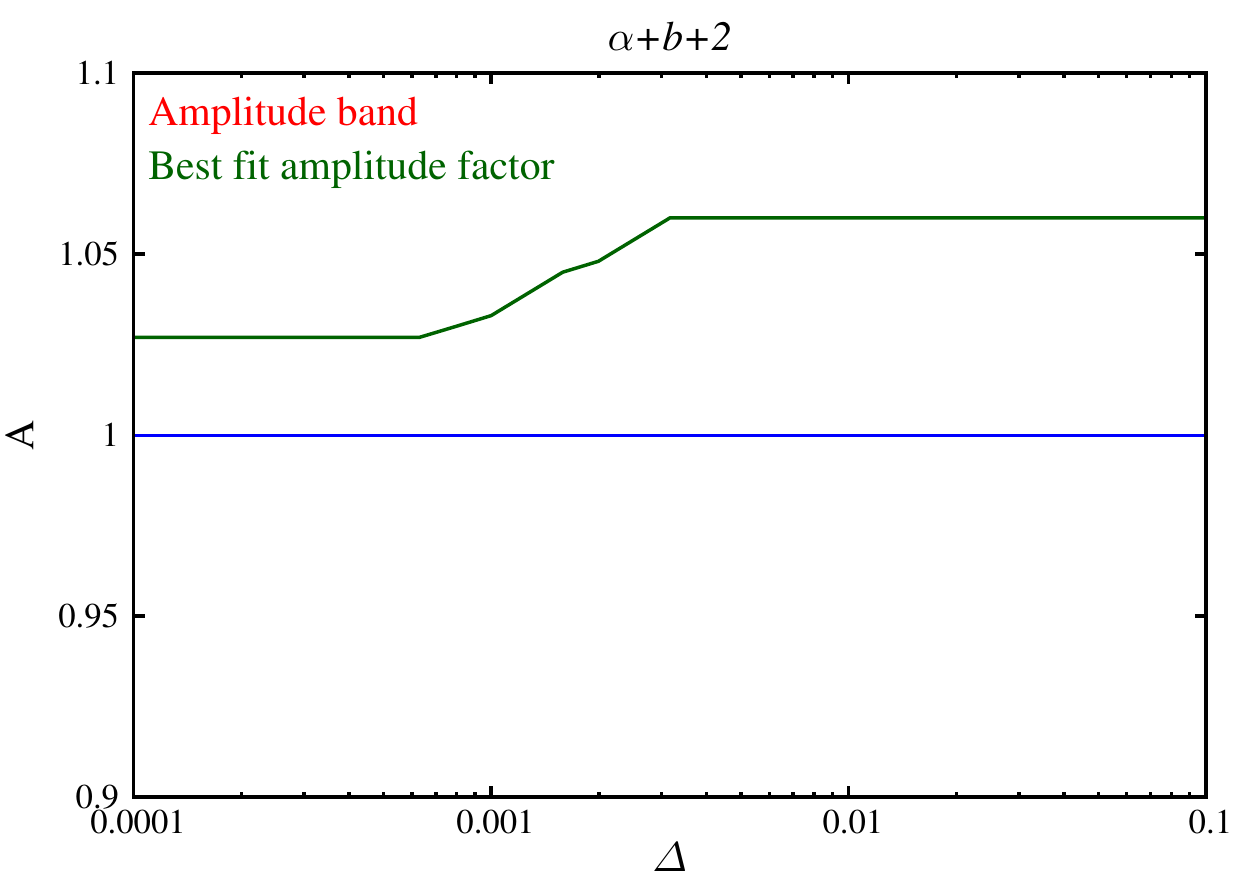}} 

\resizebox{210pt}{160pt}{\includegraphics{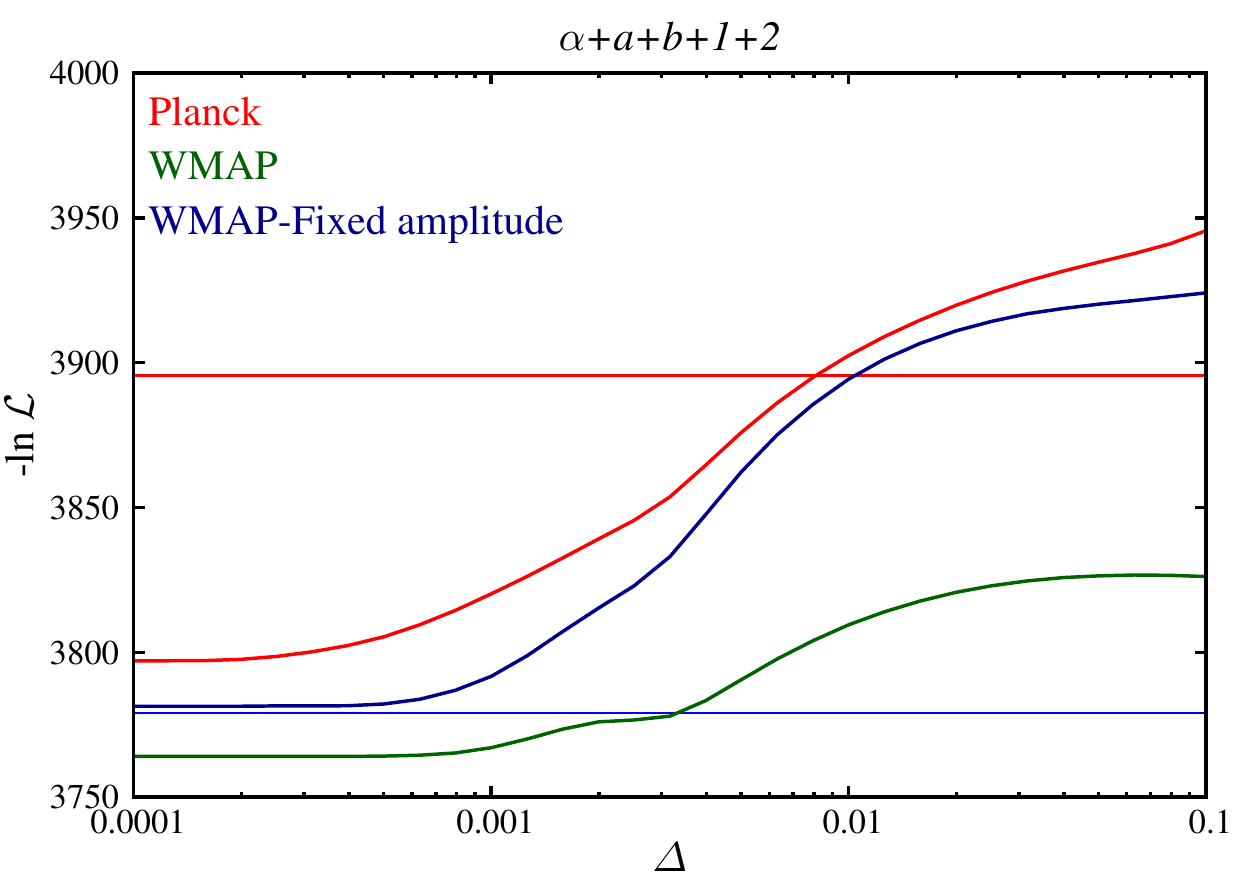}} 
\resizebox{210pt}{160pt}{\includegraphics{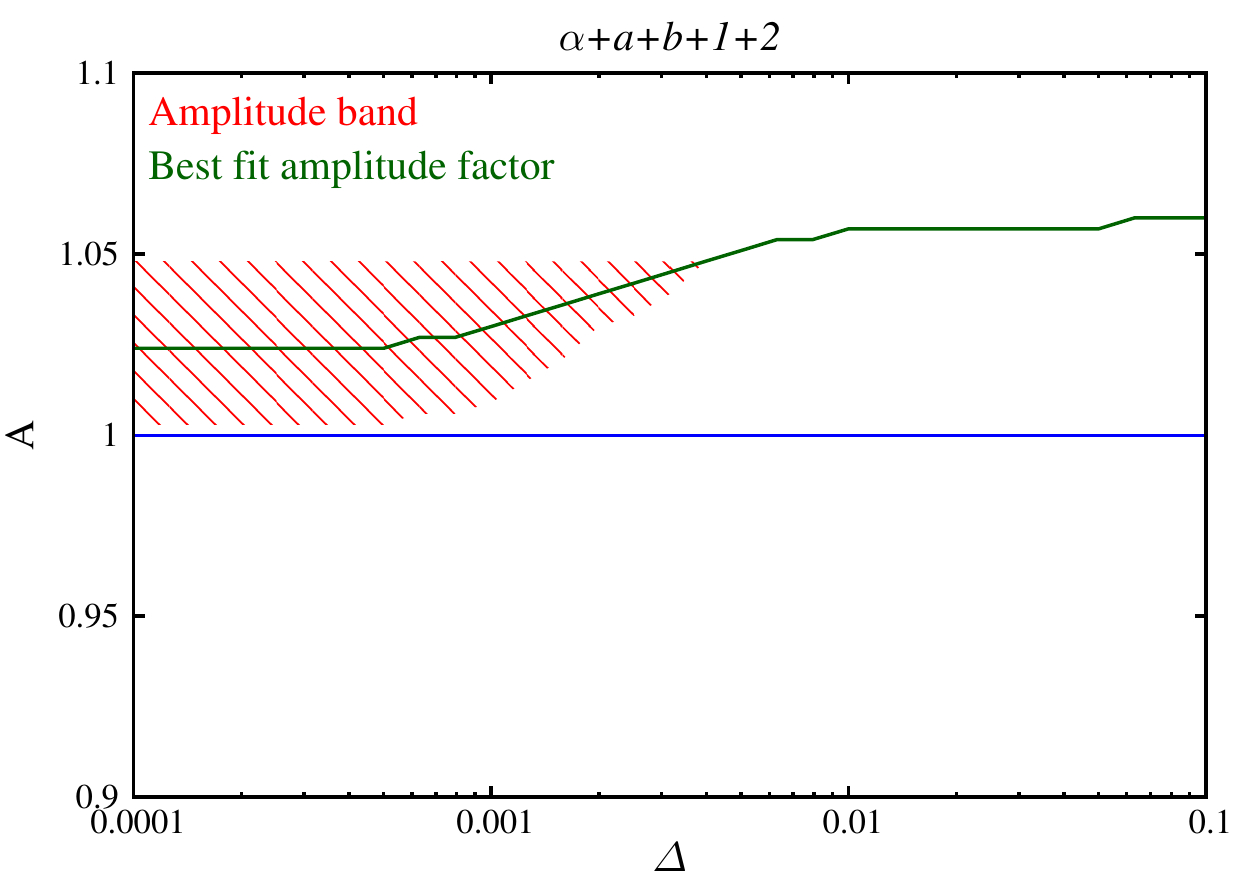}} 
\end{center}
\caption{\footnotesize\label{fig:vark-2} [Left] The $-\ln {\cal L}$ obtained from Planck (in red) and WMAP-9 (in green and blue) 
for reconstructed PPS as a function of smoothing width. Red and blue straight lines represent the power law best fit $-\ln {\cal L}$
from Planck and WMAP-9 respectively. Green line represents the best fit WMAP-9 $-\ln {\cal L}$ obtained 
upon comparing the reconstructed PPS from Planck, allowing an overall amplitude shift. The blue line corresponds to WMAP-9 $-\ln {\cal L}$
obtained from the reconstructed PPS from Planck without an overall amplitude shift. [Right] The region of amplitude factor, $A$, 
to the PPS where we get better fit to WMAP-9 data from reconstructed PPS compared to power law PPS (red shaded area) and the 
value of $A$ which provides the best fit to WMAP-9 data(green line).}
\end{figure*}
 Results from $\alpha+b+1$ and $\alpha+b+2$ is dominated by spectrum $b$ since it has the larger overlap with WMAP-9 probed scales 
 ($\ell=2-1200$) than 1 and 2. We find similar results as in $\alpha+b$.

 \item{\bf Using the complete Planck spectra:} Upon using the complete Planck spectra, we find that 
 $\alpha+a+b+1+2$ shows behavior similar to $\alpha+a$, $\alpha+a+1$ and $\alpha+a+2$. The inclusion of $b$ degrades the result slightly 
 since we find that PPS from $\alpha+a+b+1+2$ fails to fit the WMAP-9 data better than power law without an overall amplitude shift. 
 Note that the amplitude band is now strictly greater than unity. 
 The best fit $A$ is found to be approximately $2.4\%-2.5\%$. 
 \end{itemize}

 From all the analyses above, we find $\alpha$ and $a$ (which are the most relevant spectra to be compared with WMAP) are consistent with WMAP-9 data.
 Combining all spectra we find WMAP-9 and Planck are in 
 agreement allowing $2.4\%-2.5\%$ amplitude shift. Although the nature of 
 investigation is different, this result is consistent with the conclusion of ~\cite{Hazra:2013oqa}.

\clearpage
%%%%%%%%%%%%%%%%%%%%%%%%%%%%%%%%%%%%%%%%%%%%%%%%%%%%%%%%%%%%%%%%%%%%%%%%%%%%%%%
\subsection{Reconstruction indicates lensing}~\label{subsec:lens}
Gravitational lensing distorts the CMB spectrum, specifically lensing can increase/decrease the power of the CMB acoustic peaks. With the detection of small scale CMB power 
spectrum, Planck has confirmed effect of lensing by $25\sigma$~\cite{Planck:lensing} assuming a power law PPS. With our reconstruction 
we demonstrate signatures of lensing in the angular power spectra. We follow the following procedures : 

\begin{enumerate}
 \item Using the MRL we reconstruct 2 PPS. For the first PPS we consider the lensing effect, {\it i.e.} we assume the 
 detected CMB angular power spectrum is lensed. After foreground subtraction from 
 the raw data we also subtract the lensing template (Eq.~\ref{eq:lens}) from each spectra and obtain $\cl^{\rm {D'_\nu}}$. Performing 
 the reconstruction with the resulting $\cl^{\rm {D'_\nu}}$ we get the PPS as $\psk^{\rm Lens}$. The second PPS ($\psk^{\rm No-Lens}$)
 is reconstructed using $\cl^{\rm {D'_\nu}}$ obtained after foreground subtraction but without subtracting the lensing template. 
 Hence, for the second case we assume that the data indicates no lensing effect and argue that $\cl^{\rm T}$ (Eq.~\ref{eq:clequation}) 
 along with the foregrounds are enough to address the CMB data obtained.
 
 \item We define the difference in the PPS : $\Delta\psk \equiv \psk^{\rm Lens}-\psk^{\rm No-Lens}$.
 
 \item We obtain 2 reconstructed $\cl^{\rm T}$ from both the PPS. 
 We add the lensing template back to the first $\cl^{\rm T}$ obtained only and tag it as $\cl^{\rm Lens}$. The second power 
 spectrum is named as $\cl^{\rm No-Lens}$
 
 \item We define the difference in the angular power spectrum as, $\Delta\cl^{\rm TT} \equiv \cl^{\rm Lens}-\cl^{\rm No-Lens}$.
 
 \item We examine the $-\ln {\cal L}$ as a function of MRL iterations for $i\le500$ in different combinations of Planck spectra.
 
\end{enumerate}

 If the lensing effect is significant, we can expect the first PPS can provide a better likelihood compared to the second one. Moreover, the second PPS can be expected to contain oscillations 
 since the MRL will enforce the lensing effect to be encoded in the PPS. Hence, we can expect periodic oscillations in $\Delta\psk$. If the lensing effect
 is completely captured by the oscillations in the second PPS, we expect $\Delta\cl^{\rm TT}$ to be {\it zero} at all scales and both the angular 
 power spectra provide similar likelihood to the data.
 
\begin{figure*}[!htb]
\begin{center} 
\resizebox{145pt}{120pt}{\includegraphics{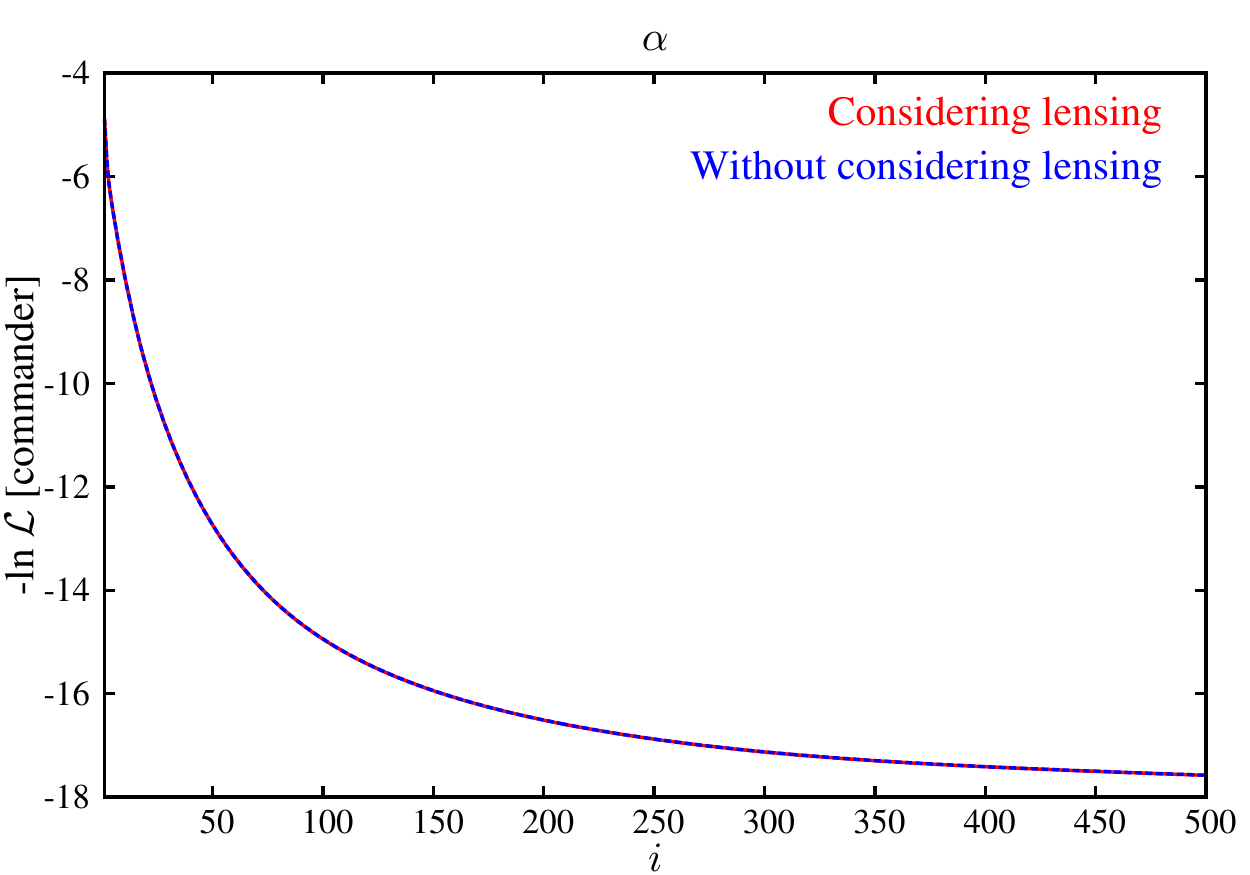}} 
\hskip -7pt\resizebox{145pt}{120pt}{\includegraphics{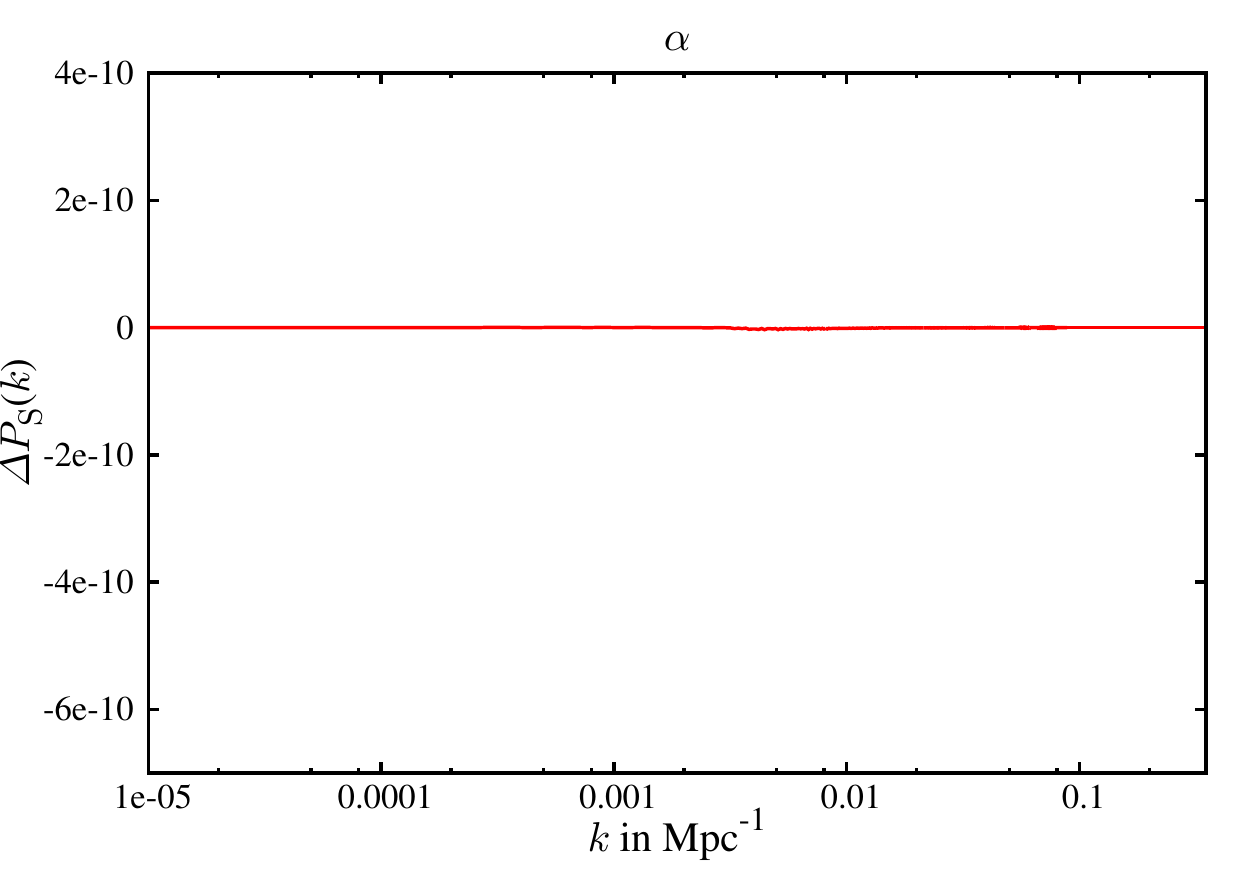}} 
\hskip -7pt\resizebox{145pt}{120pt}{\includegraphics{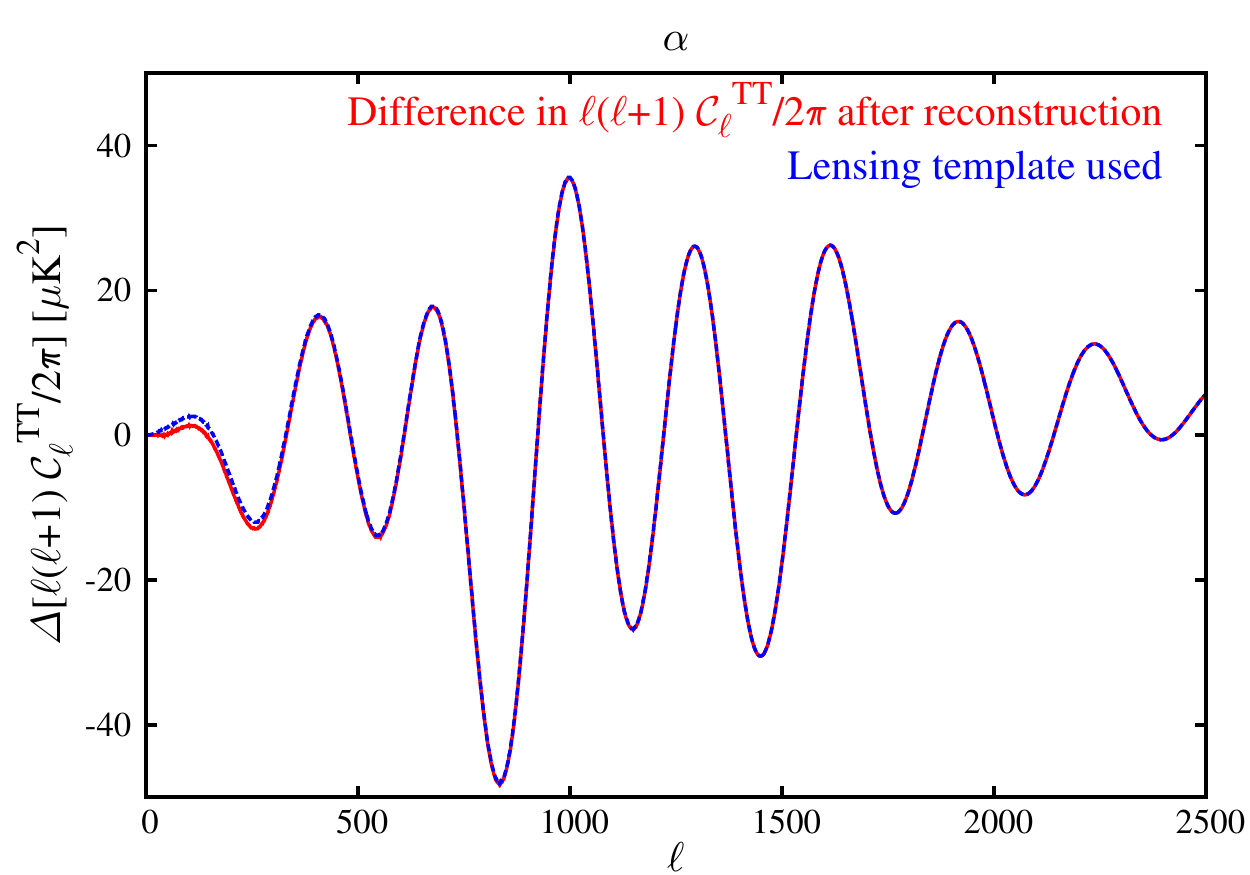}} 

\resizebox{145pt}{120pt}{\includegraphics{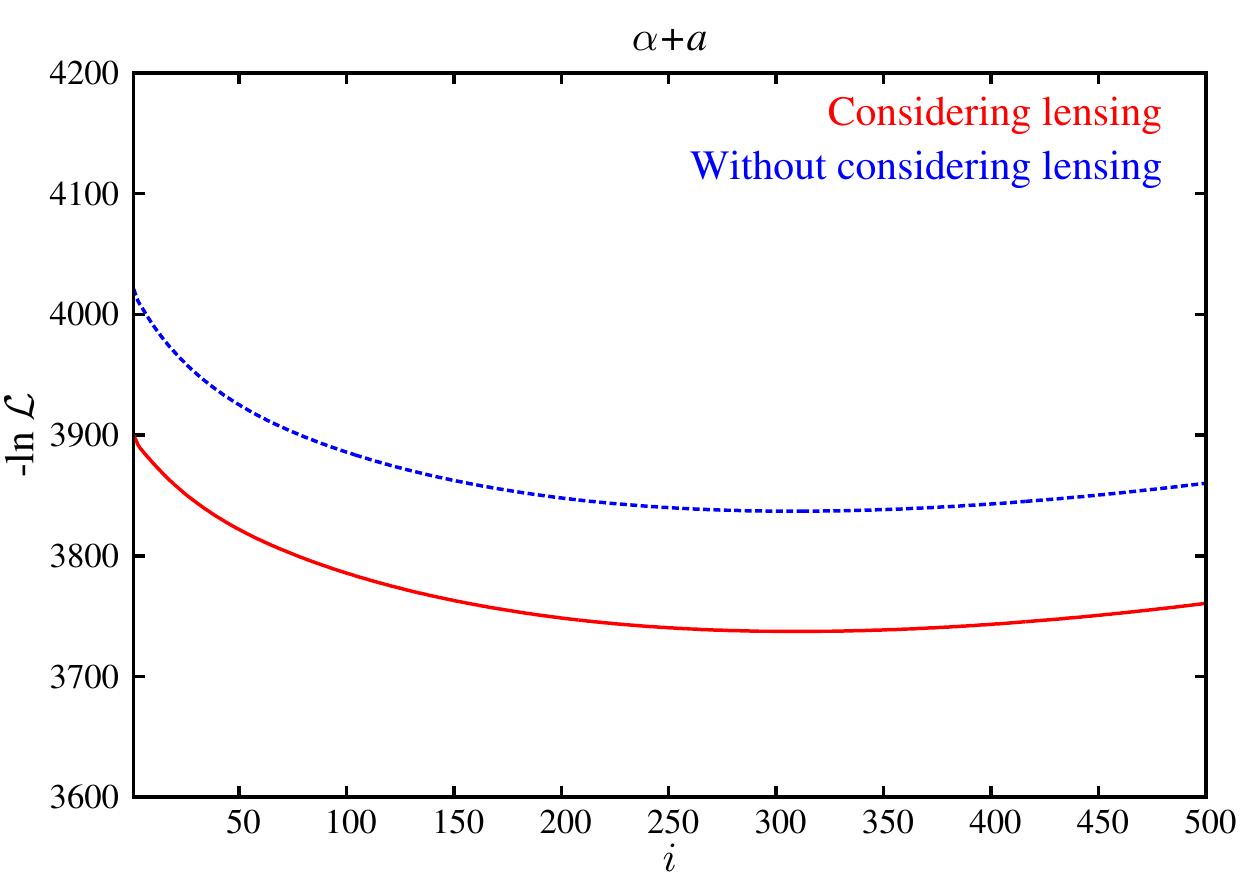}} 
\hskip -7pt\resizebox{145pt}{120pt}{\includegraphics{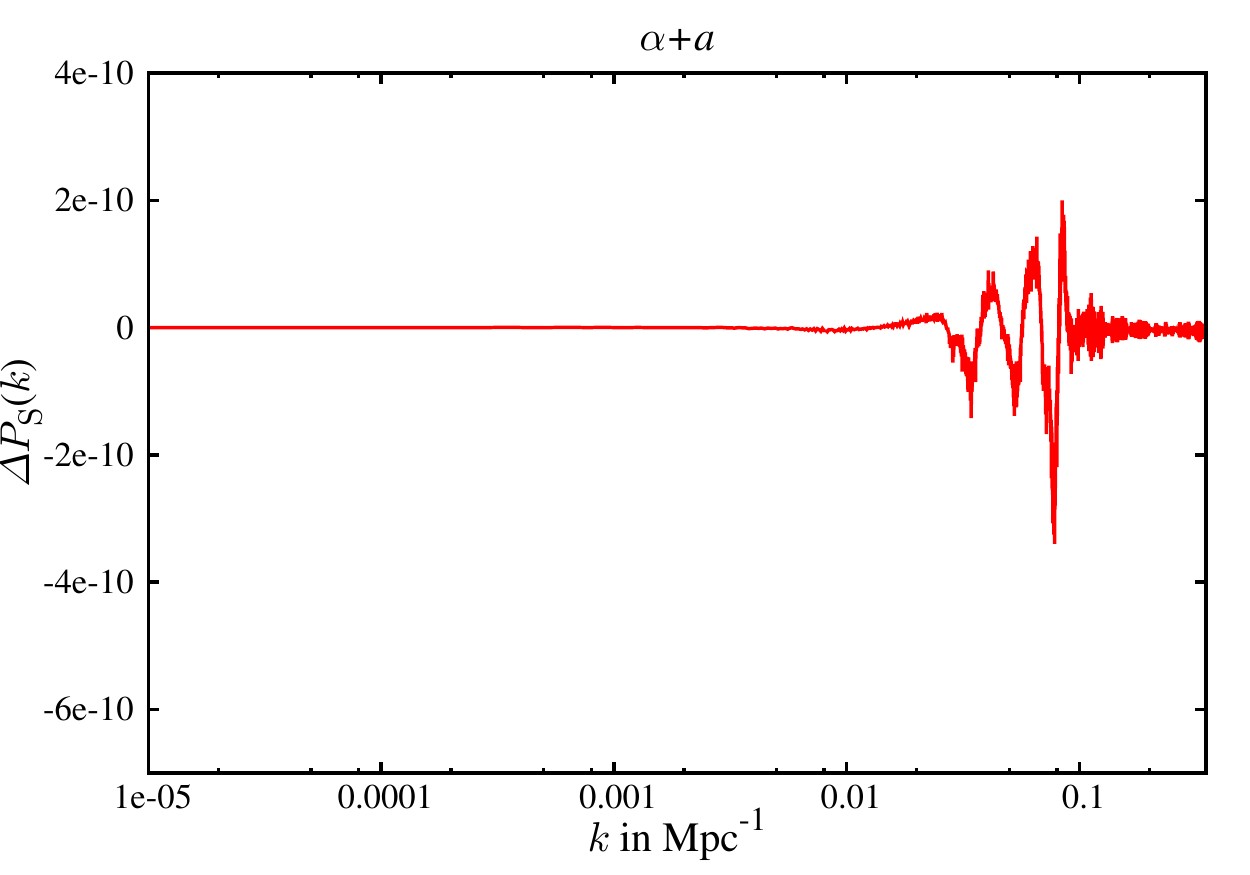}} 
\hskip -7pt\resizebox{145pt}{120pt}{\includegraphics{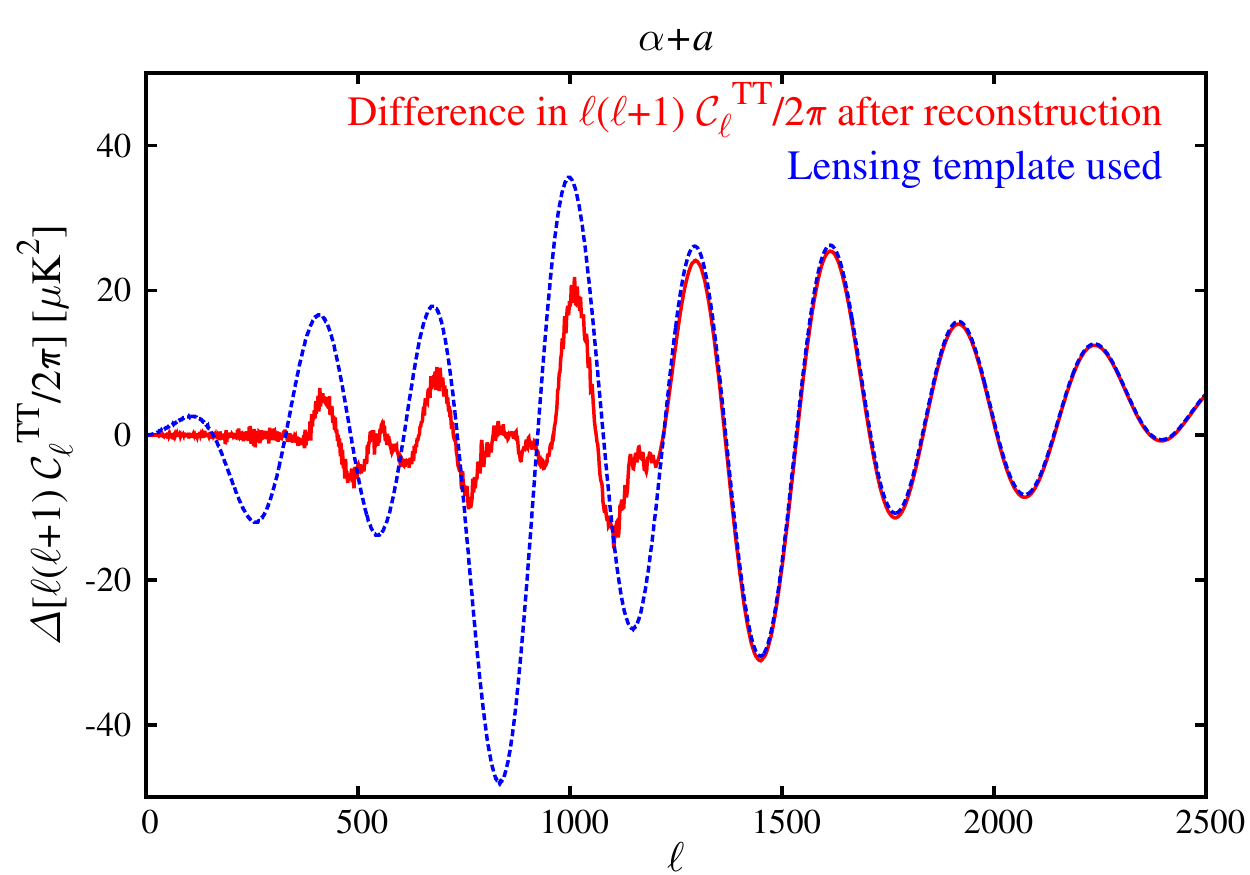}} 

\resizebox{145pt}{120pt}{\includegraphics{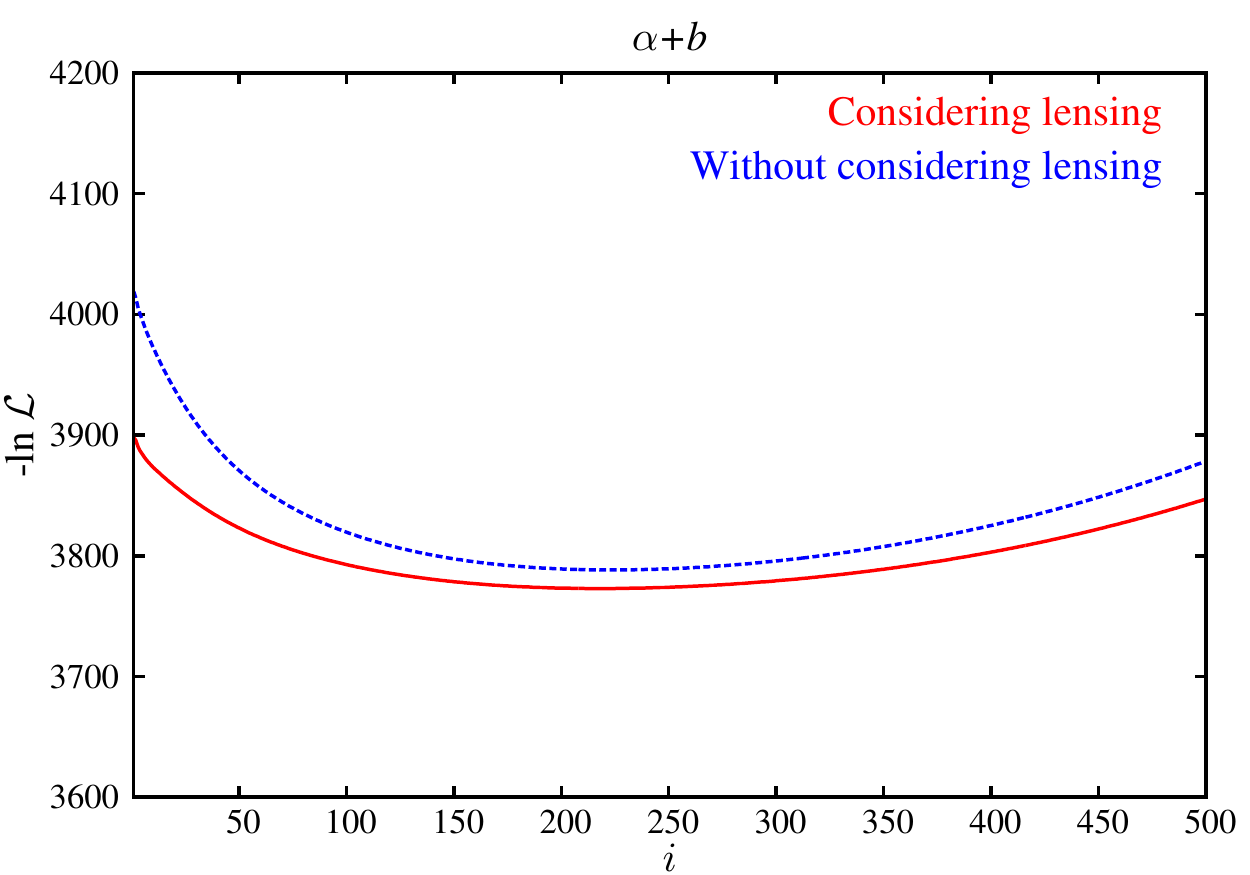}} 
\hskip -7pt\resizebox{145pt}{120pt}{\includegraphics{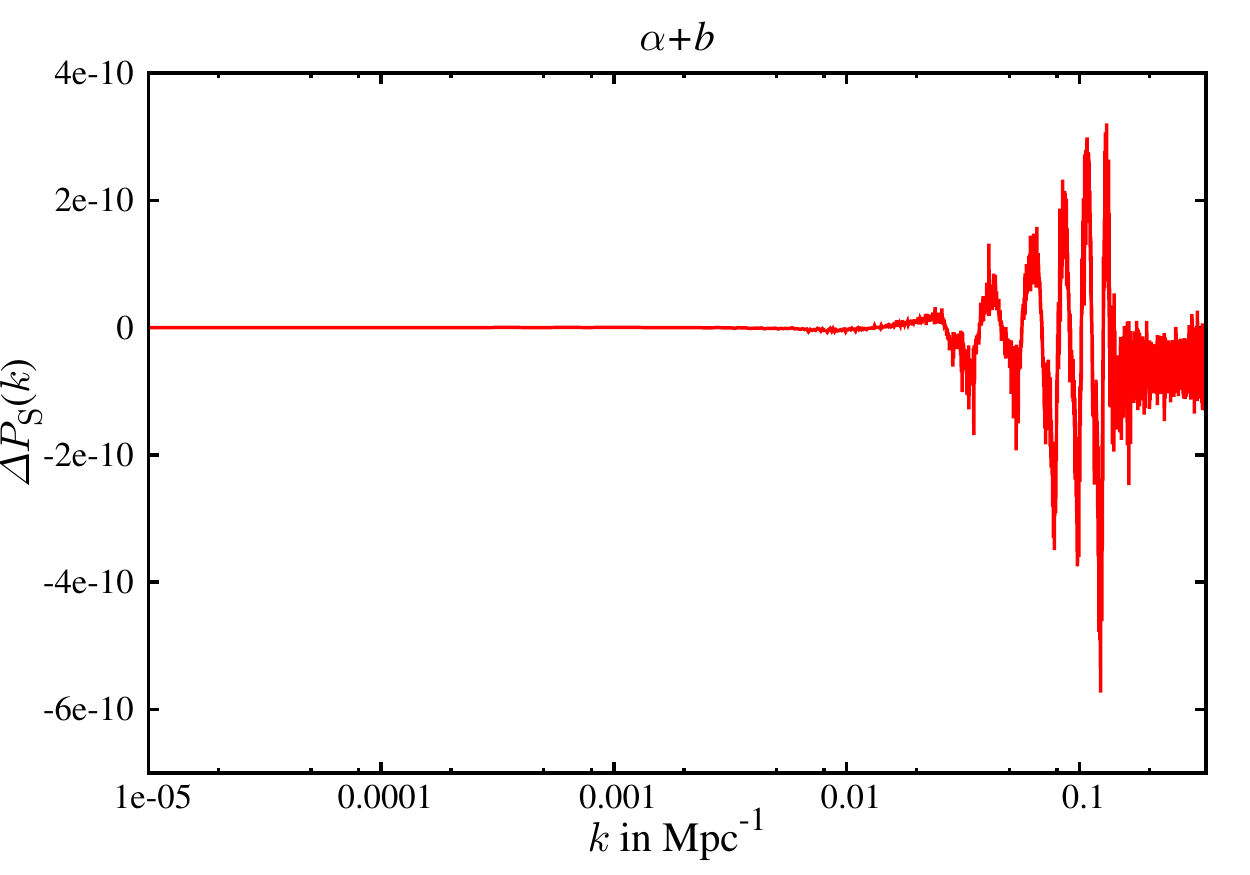}} 
\hskip -7pt\resizebox{145pt}{120pt}{\includegraphics{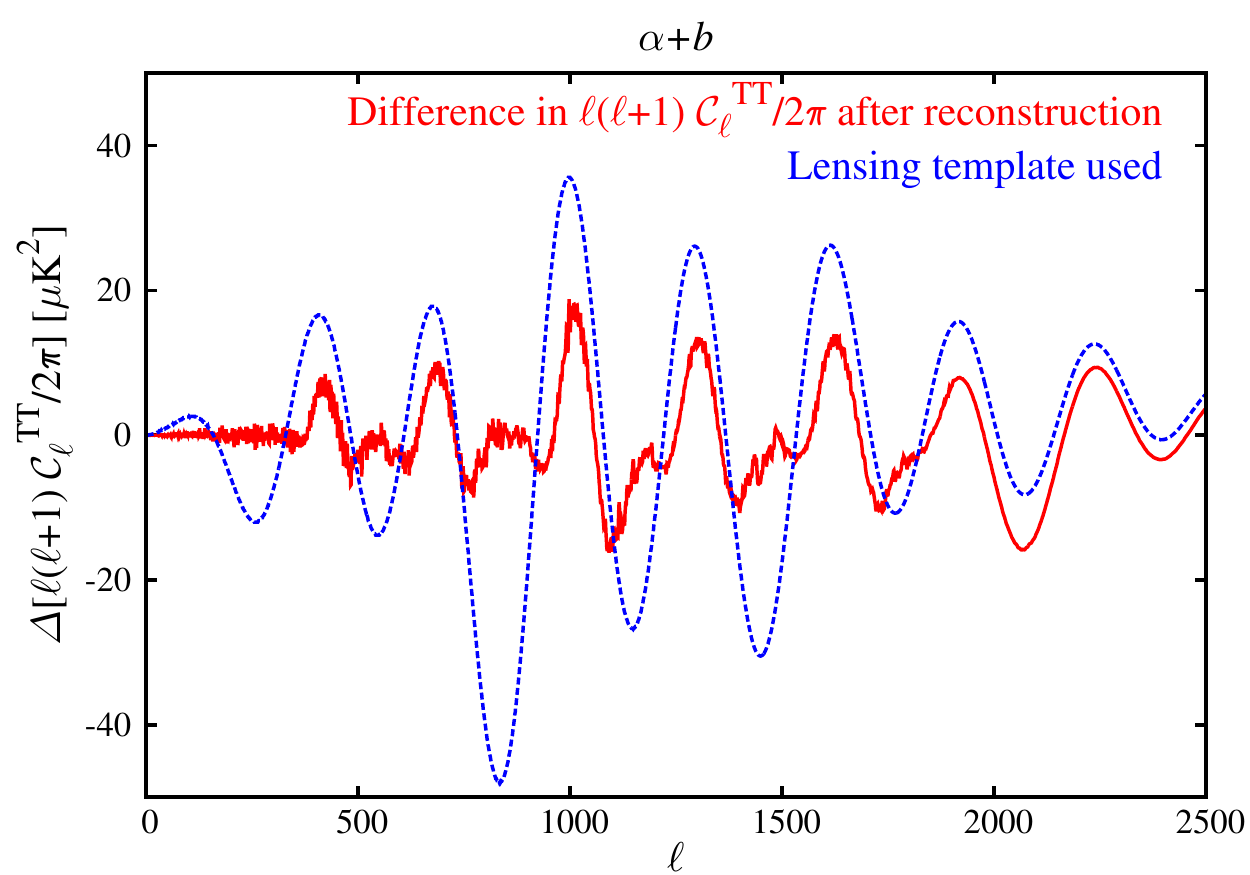}} 

\resizebox{145pt}{120pt}{\includegraphics{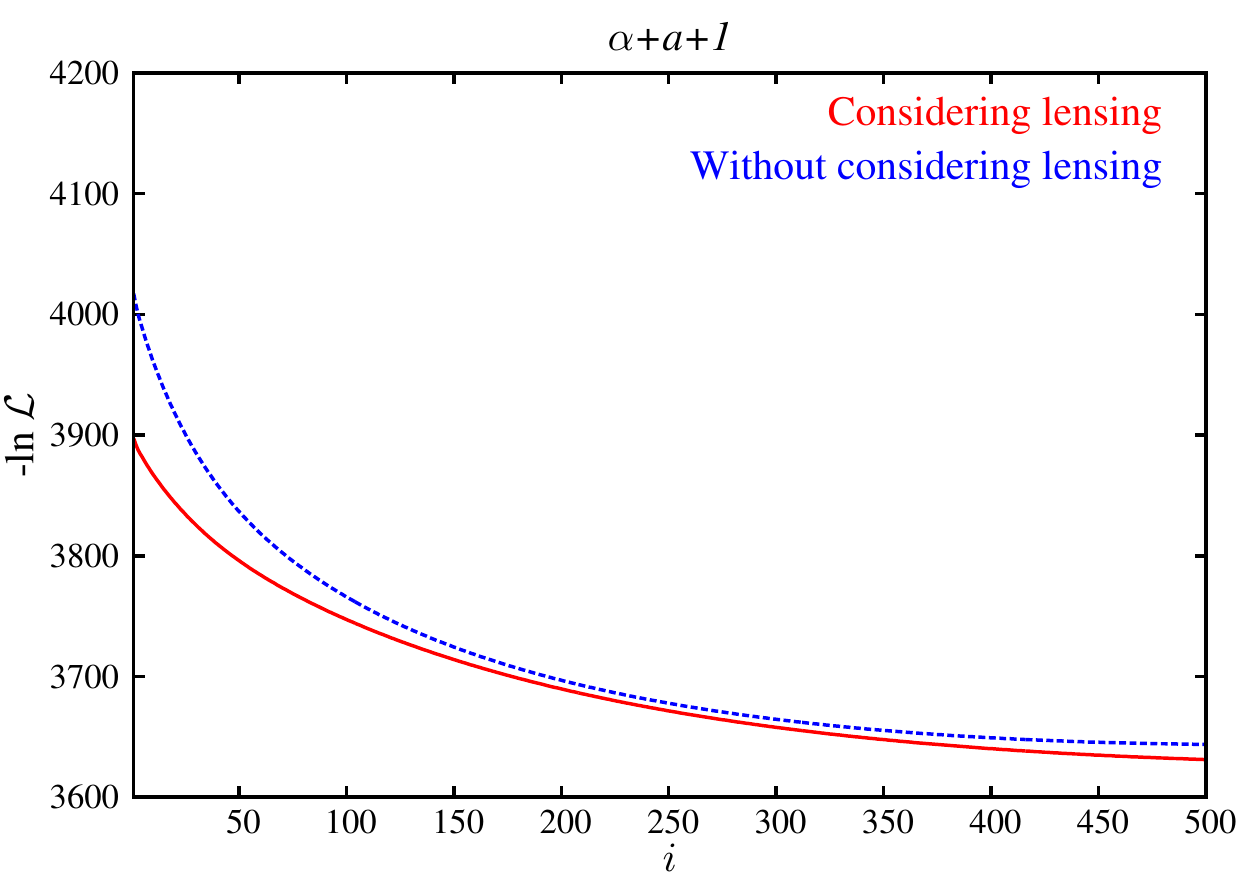}} 
\hskip -7pt\resizebox{145pt}{120pt}{\includegraphics{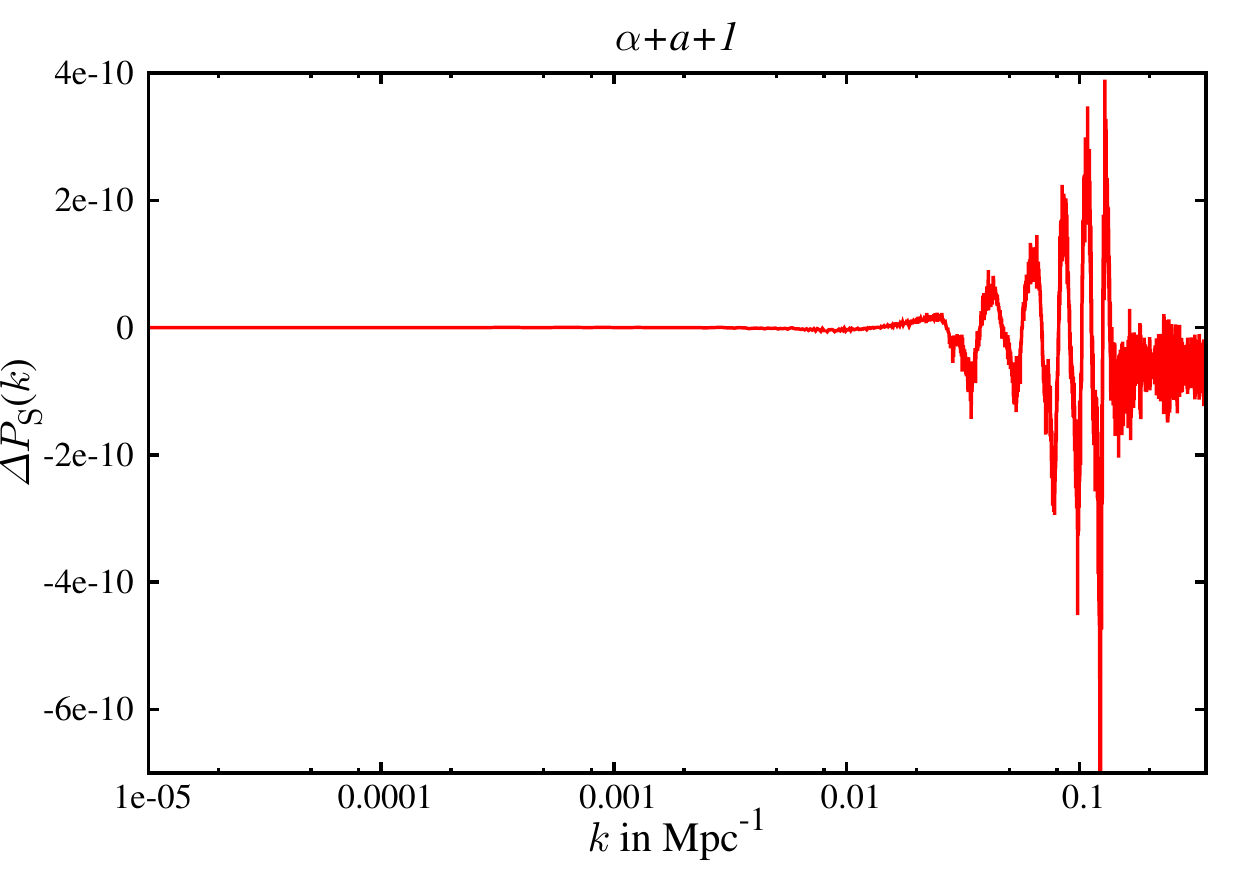}} 
\hskip -7pt\resizebox{145pt}{120pt}{\includegraphics{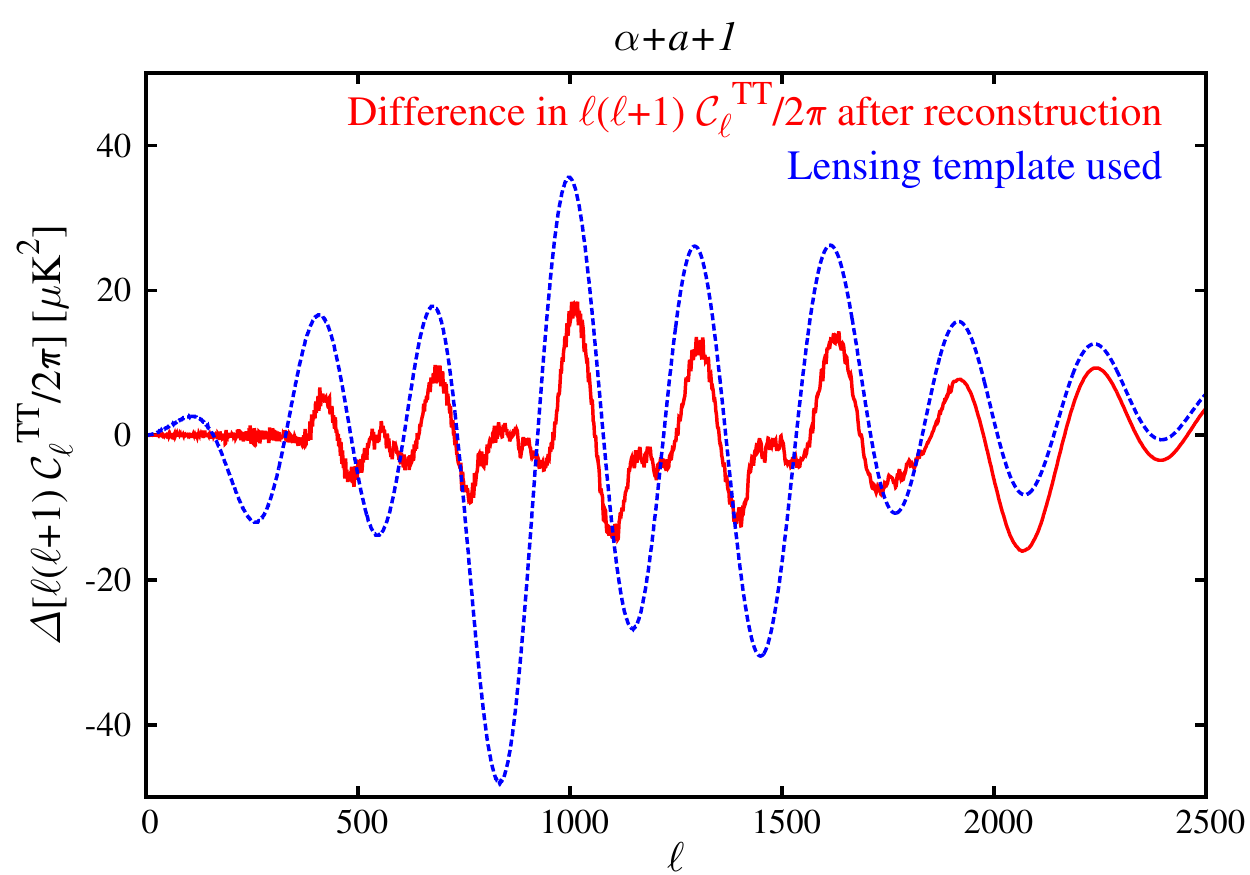}}

\end{center}
\caption{\footnotesize\label{fig:lens-1}[Left] $-\ln {\cal L}$ from the reconstructed PPS with(red) and without(blue) considering the lensing effect. 
[Middle] The difference in PPS obtained with and without considering lensing after 100 iterations. [Right] The difference in $\cl$'s obtained from reconstruction 
(for 100 iterations) with and without considering lensing effect (red) and the reference lensing template used in the analysis (blue-dashed).}
\end{figure*}

\begin{figure*}[!htb]
\begin{center}

\resizebox{145pt}{120pt}{\includegraphics{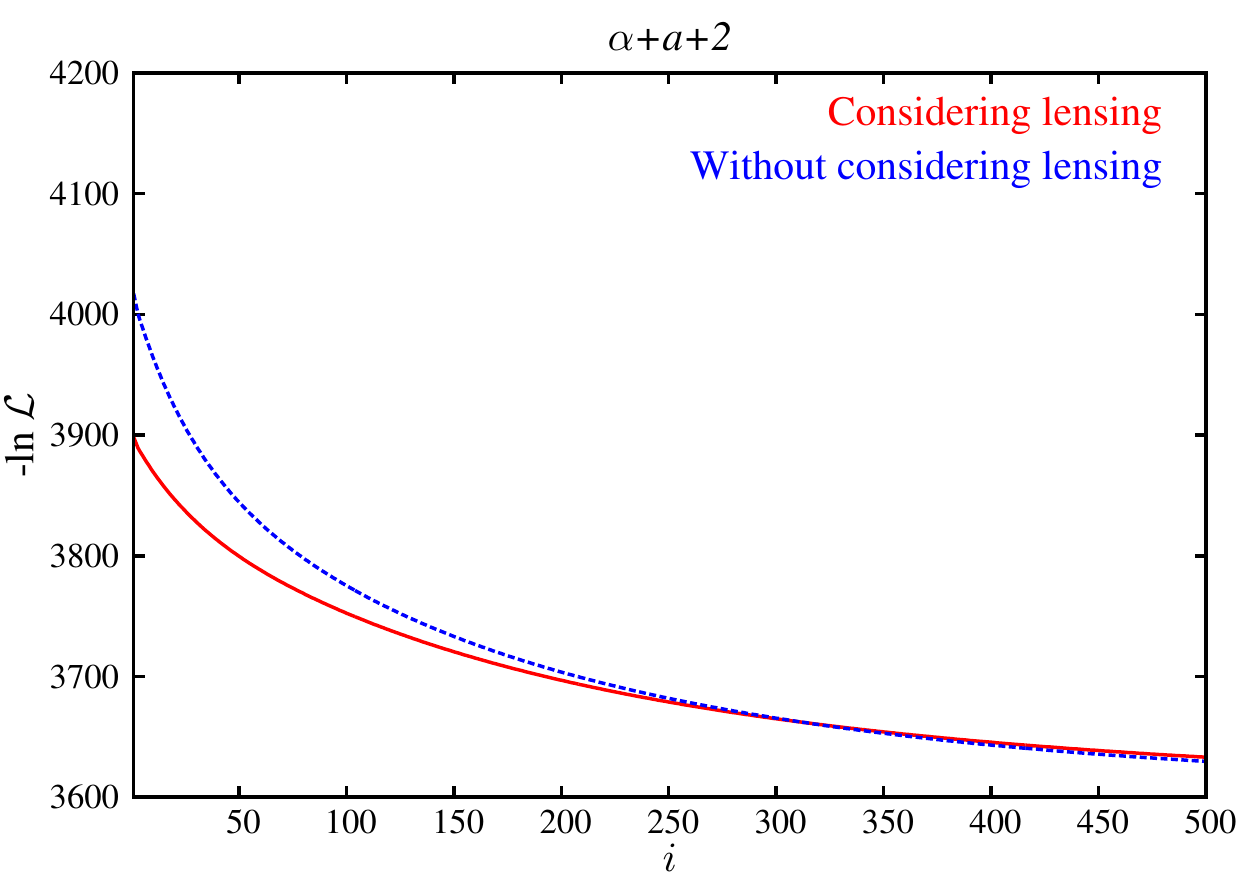}} 
\hskip -7pt\resizebox{145pt}{120pt}{\includegraphics{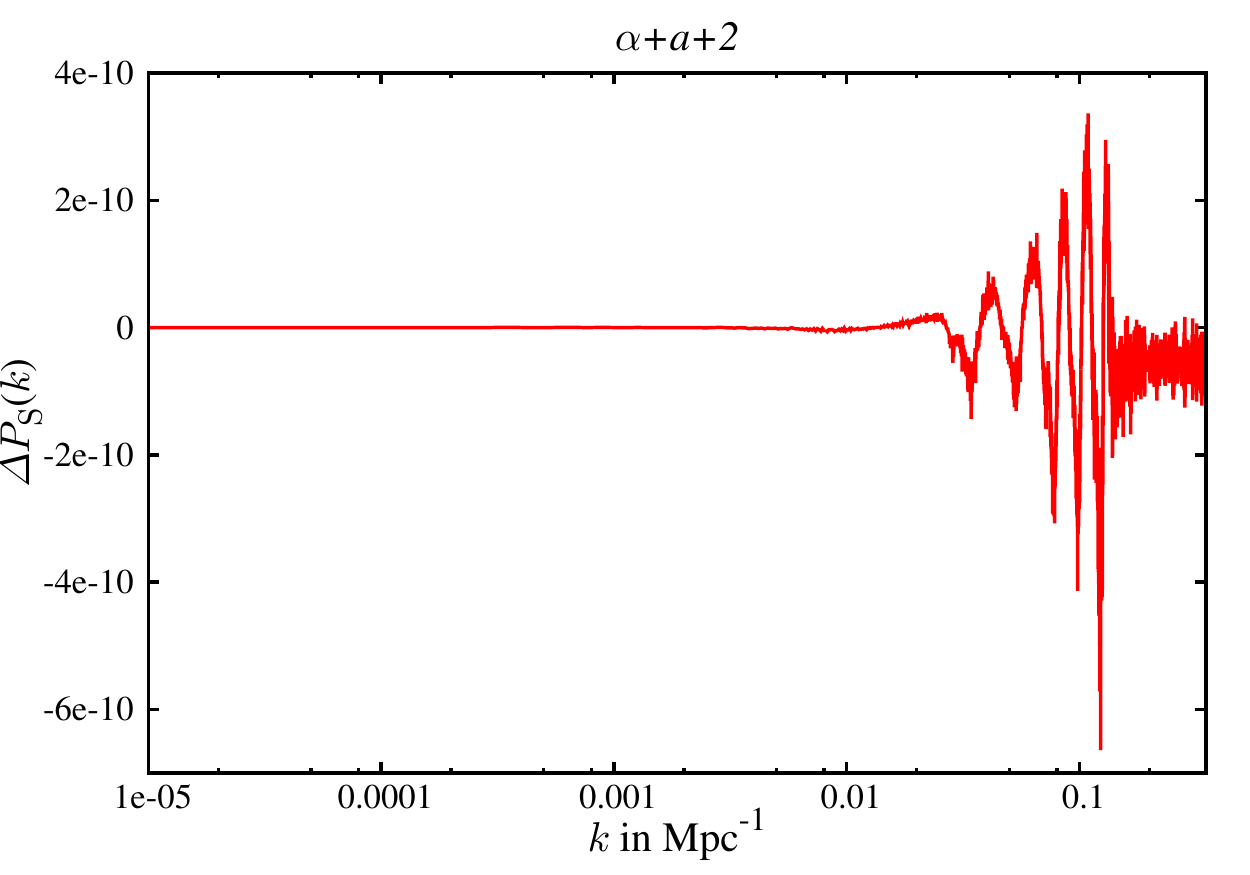}} 
\hskip -7pt\resizebox{145pt}{120pt}{\includegraphics{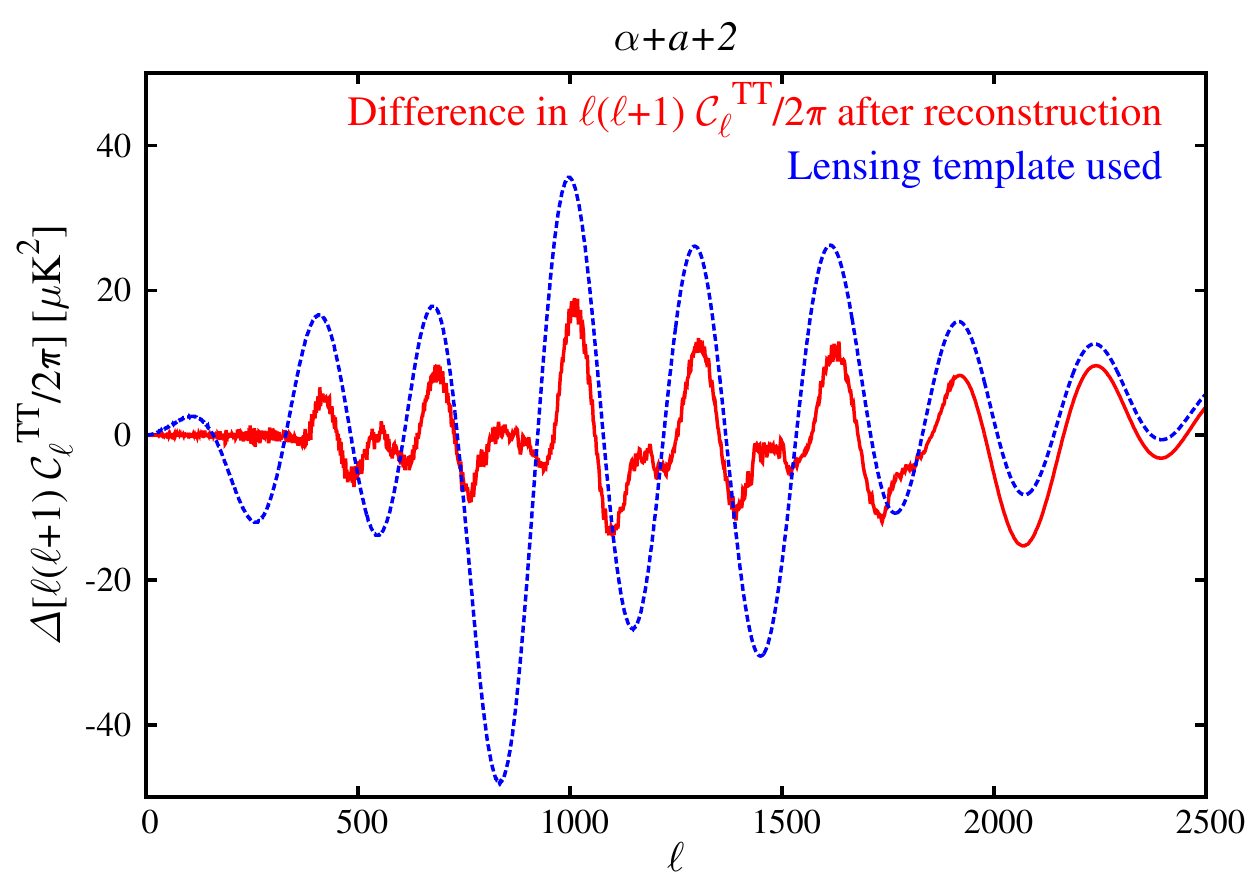}} 

\resizebox{145pt}{120pt}{\includegraphics{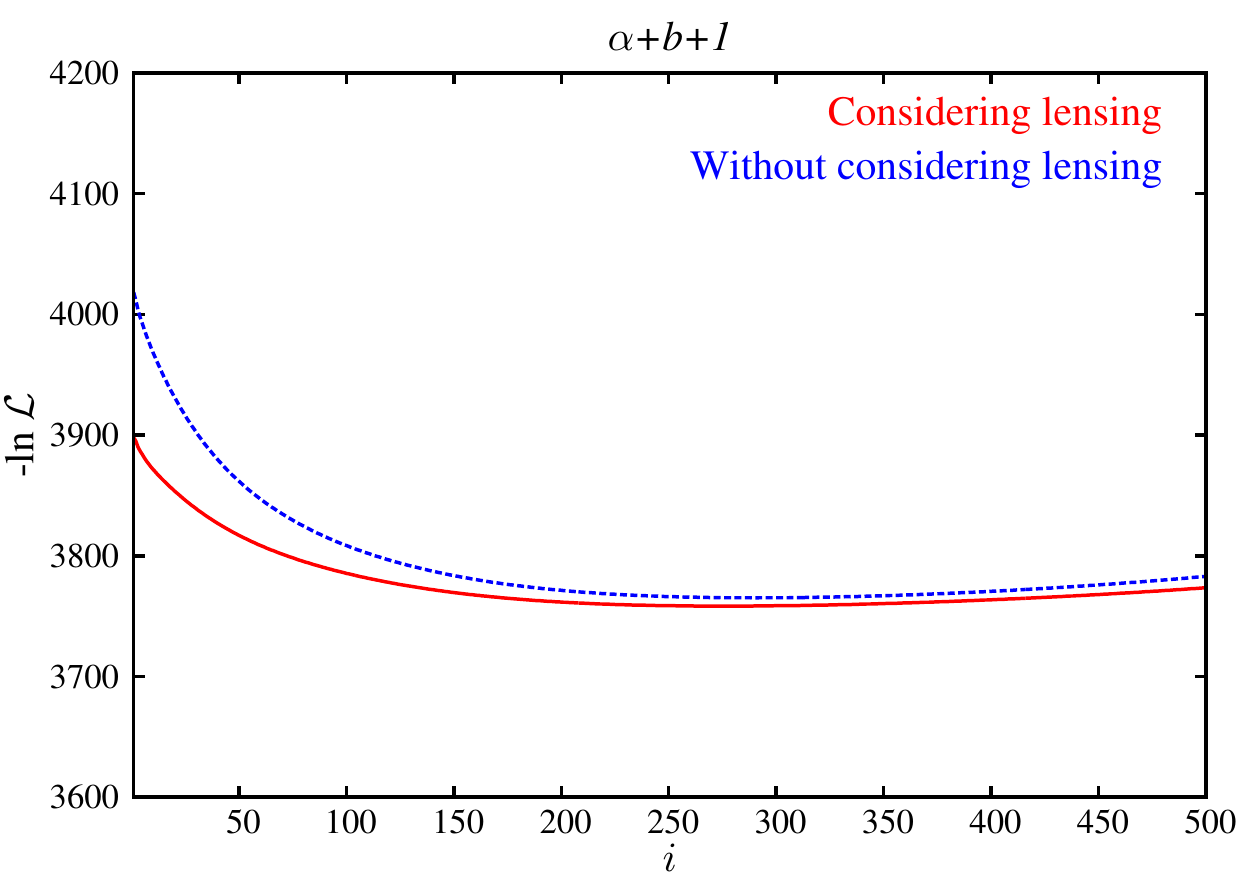}} 
\hskip -7pt\resizebox{145pt}{120pt}{\includegraphics{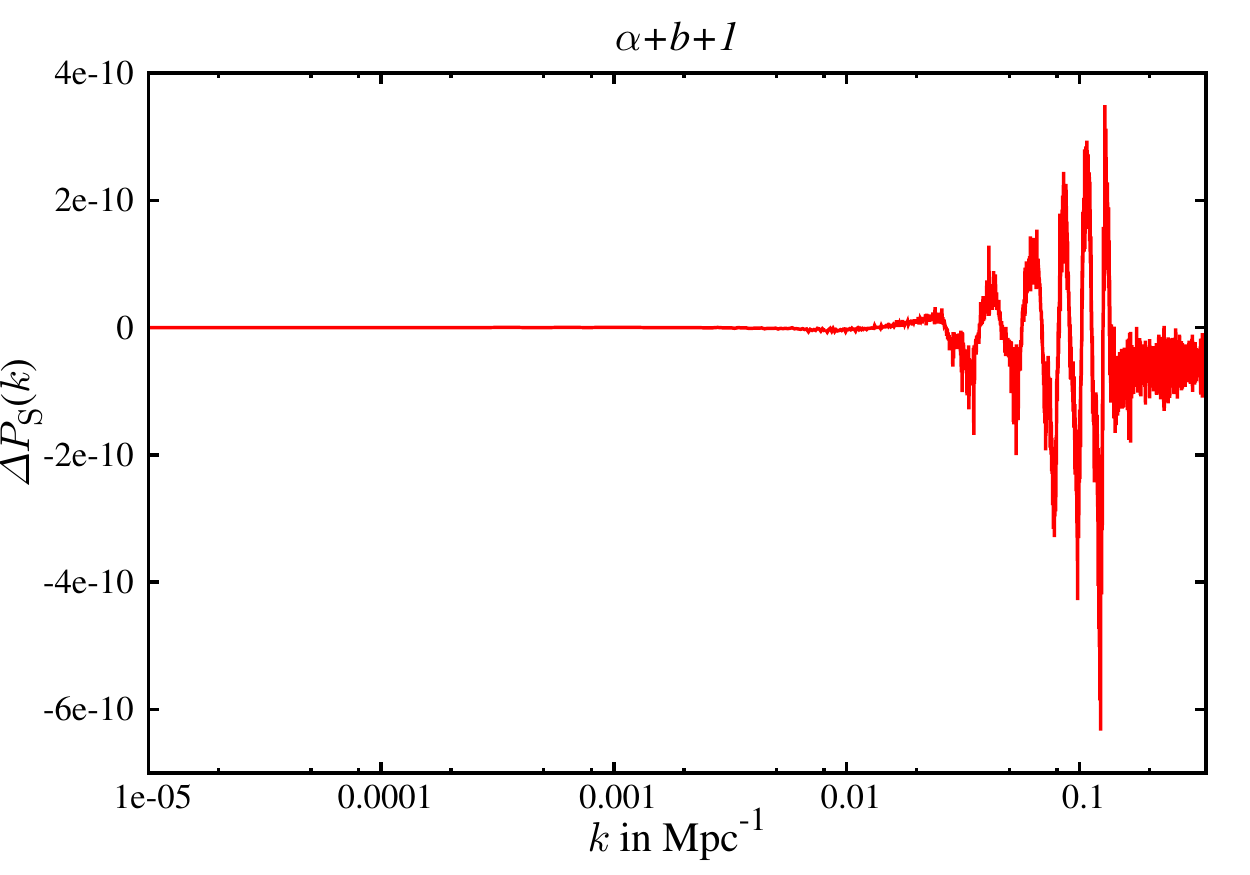}} 
\hskip -7pt\resizebox{145pt}{120pt}{\includegraphics{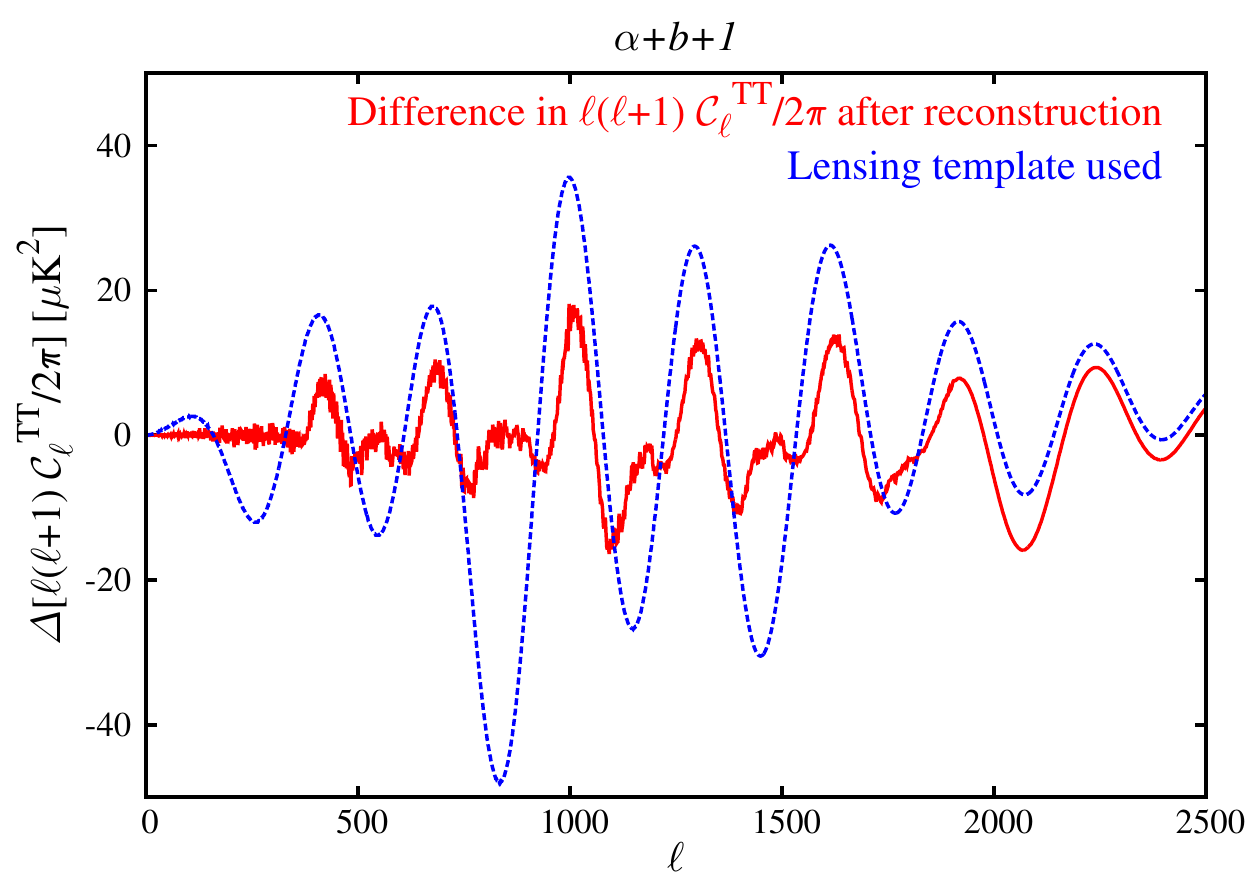}} 

\resizebox{145pt}{120pt}{\includegraphics{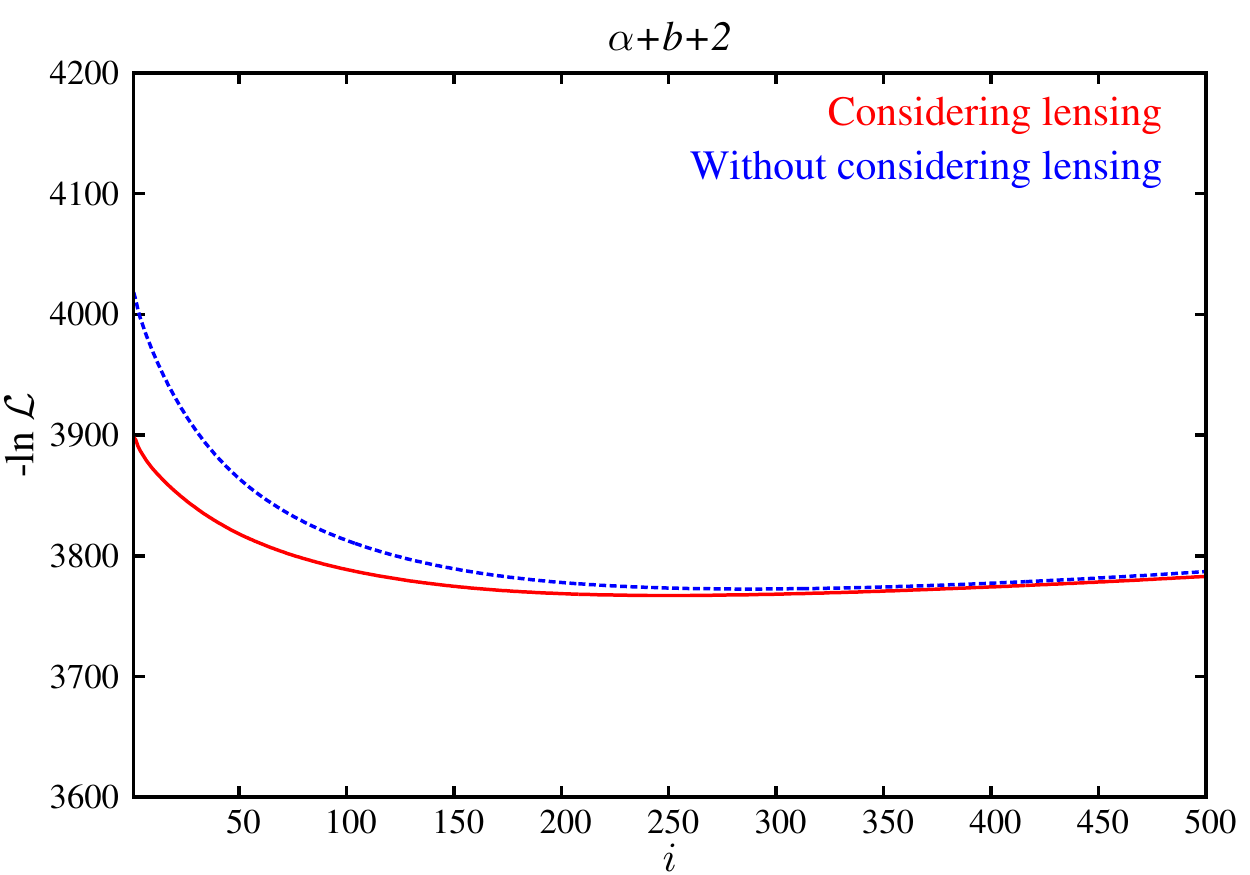}} 
\hskip -7pt\resizebox{145pt}{120pt}{\includegraphics{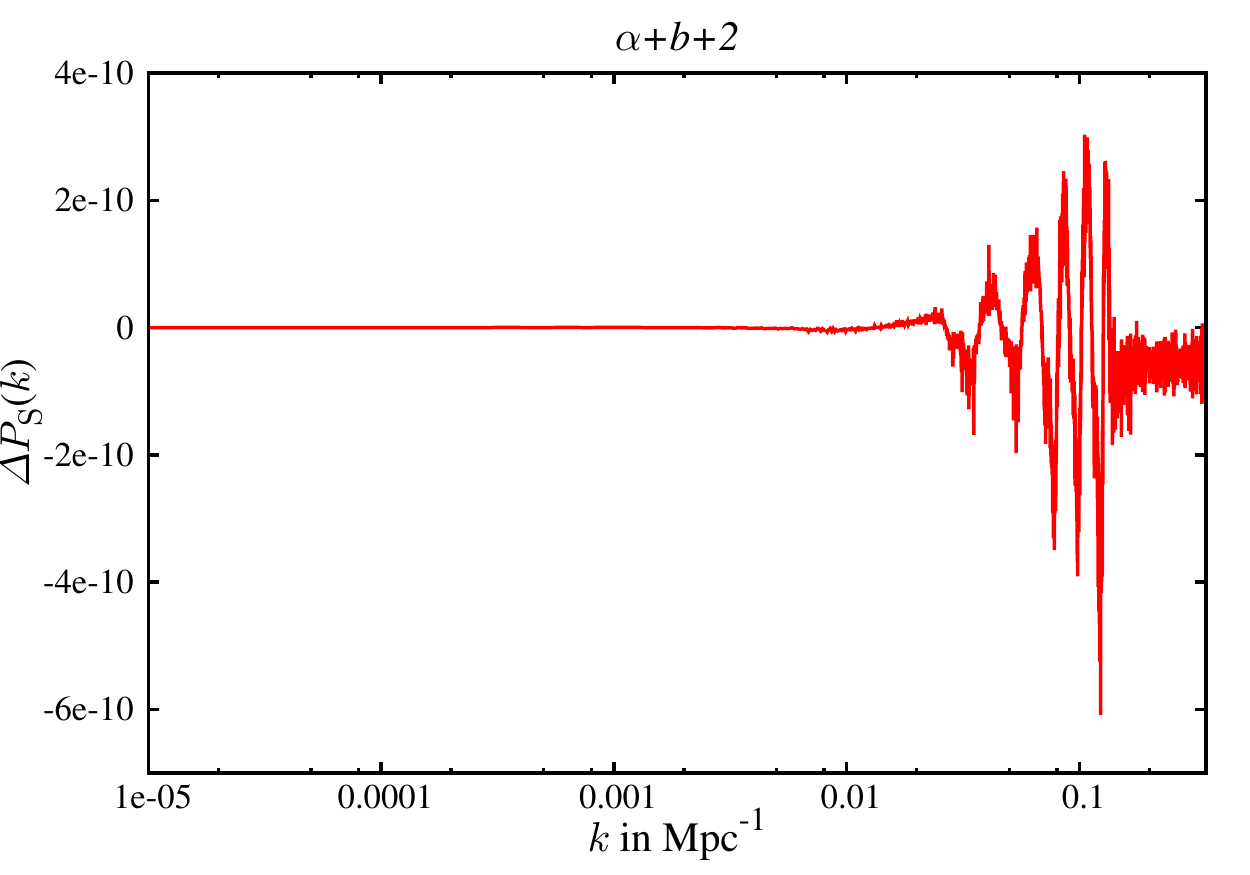}} 
\hskip -7pt\resizebox{145pt}{120pt}{\includegraphics{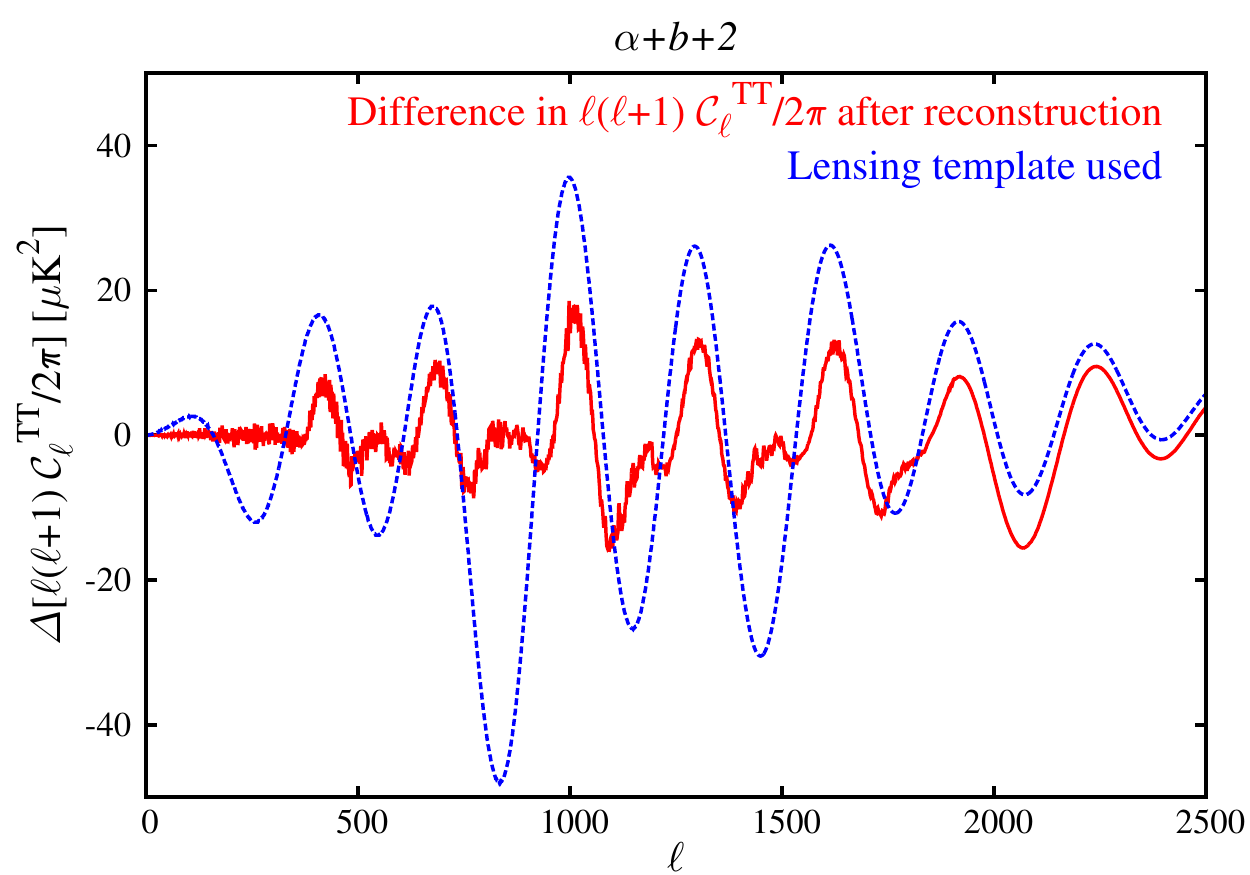}} 

\resizebox{145pt}{120pt}{\includegraphics{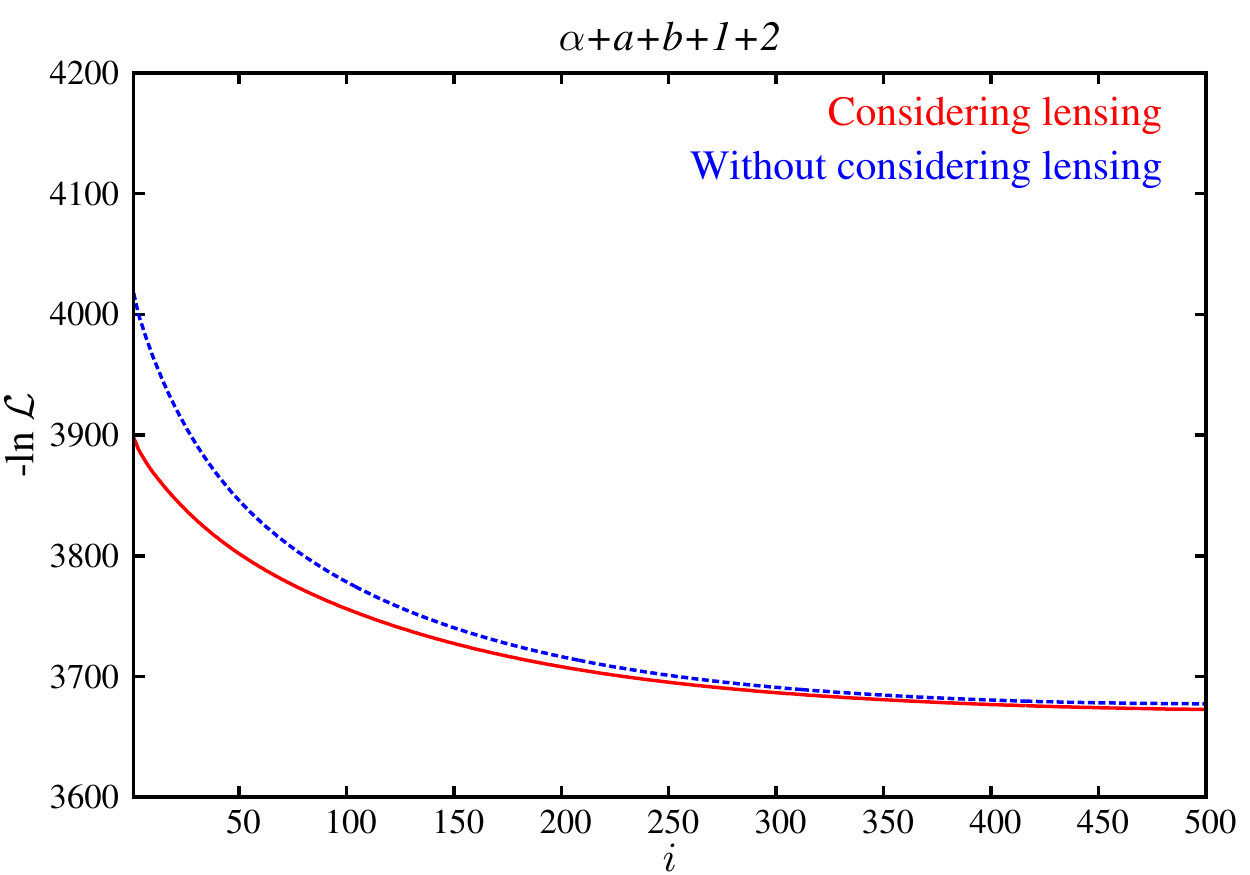}} 
\hskip -7pt\resizebox{145pt}{120pt}{\includegraphics{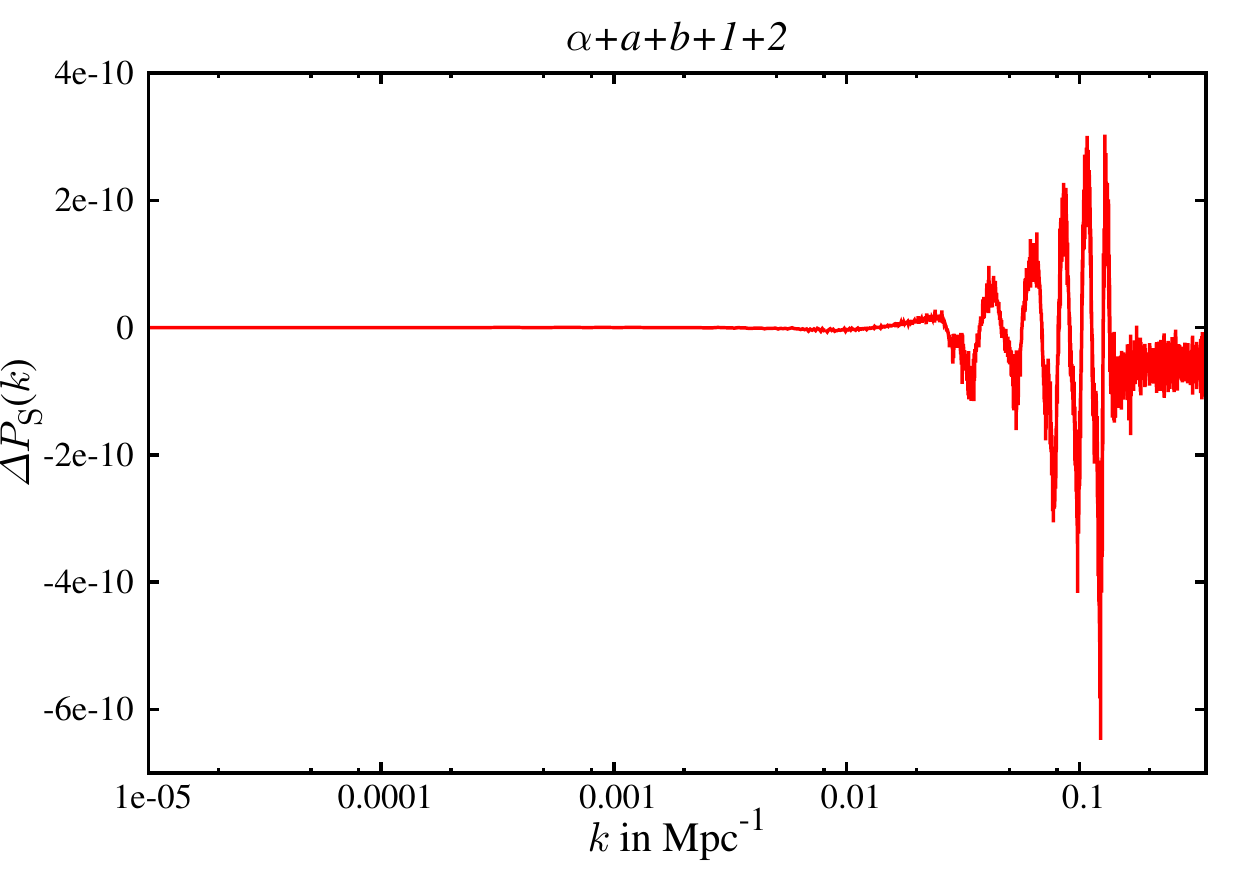}} 
\hskip -7pt\resizebox{145pt}{120pt}{\includegraphics{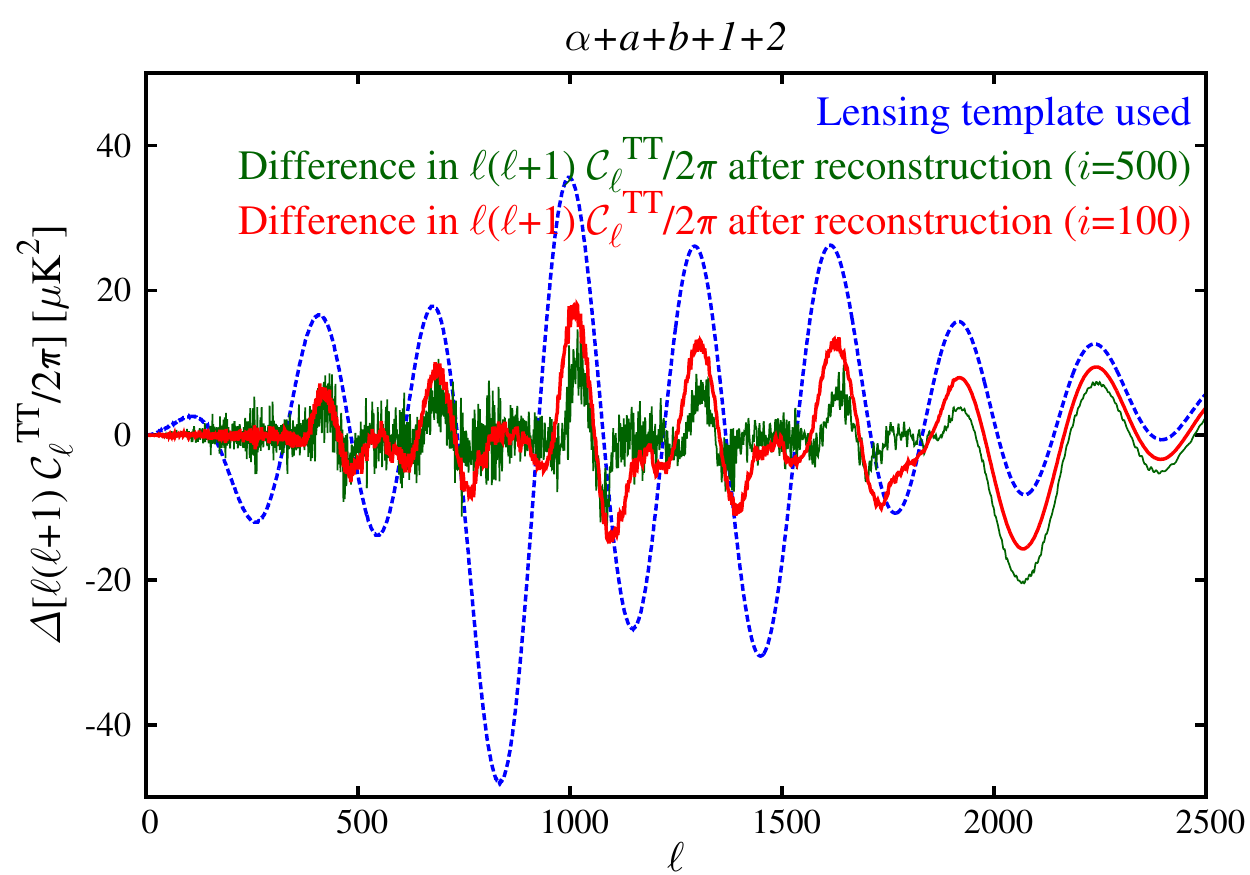}} 
\end{center}
\caption{\footnotesize\label{fig:lens-2}[Left] $-\ln {\cal L}$ from the reconstructed PPS with(red) and without(blue) considering the lensing effect. [Middle] The difference 
in PPS obtained with and without considering lensing after 100 iterations. [Right] The difference in $\cl$'s obtained from reconstruction (for 100 iterations) with and without 
considering lensing effect (red) and the reference lensing template used in the analysis (blue-dashed). Green curve in $\alpha+a+b+1+2$ represents same difference in $\cl$'s for 500 iterations.}
\end{figure*}

 In Fig.~\ref{fig:lens-1} and in Fig.~\ref{fig:lens-2} we plot our results. The left panels contain the $-\ln {\cal L}$ as a function of MRL iteration~\footnote{Note that
 only for the first case $\alpha$ we calculate the likelihood from the {\tt commander}. For any other spectra combinations we calculate
 the likelihood from both {\tt commander} and {\tt CAMspec}.}
 for the PPS obtained with (red) and without (blue) considering the lensing effect. Apart from the reconstruction from $\alpha$ in all other cases the $-\ln {\cal L}$ is plotted for the complete likelihood
 (using {\tt commander} and {\tt CAMspec}). For $\alpha$ we have plotted the results from {\tt commander} only. Middle panels contain $\Delta\psk$ obtained after 100 iterations. 
 Right panels contain $\Delta\cl^{\rm TT}$ (red) obtained after 100 iterations and the reference lensing template (blue-dashed) used in the analysis.

 Below we highlight the results obtained in different combinations of spectra, following Fig.~\ref{fig:lens-1} and~\ref{fig:lens-2}.
 
 \begin{itemize}
  \item $\alpha$ : Since $\alpha$ contains angular power spectrum on largest scales only, where we expect negligible lensing, the $-\ln {\cal L}$ is essentially 
  equivalent (the red and the blue curves overlap). $\Delta\psk$ is {\it zero} at all scales due to the following reasons. At largest scales lensing is not significant and 
  apart from large scales in both the cases the PPS is unaltered and equal to the initial choice of the PPS which we assumed to be power law. $\Delta\cl^{\rm TT}$ simply 
  follows the subtracted lensing template used, as expected. We get back what we have subtracted in the first case.  
  
  \item $\alpha+a$ : For this combination we find a significant improvement in fit ($2\Delta\ln {\cal L}\simeq200$) when we consider lensing effect. Hence the lensing effect becomes evident when we 
  include data from intermediate scales. $\Delta\psk$ shows oscillations in the PPS corresponding to the acoustic peak positions in $\ell$-space. 
  $\Delta\cl^{\rm TT}$ is suppressed at scales, $\ell\le1200$ compared to the lensing template that highlights that a part of lensing effect is 
  addressed by the oscillations in the PPS (prominent in $\Delta\psk$ plot). {\it Roughly} after $\ell=1200$, $\Delta\cl^{\rm TT}$
  follows the template since $a$ does not provide the data beyond that scale. 
  
  \item $\alpha+b$ : Similar to $\alpha+a$, we observe lensing effect is significant in this spectra combination. However, the extent of 
  improvement in fit has decreased compared to $\alpha+a$ since we are using 800 more data 
  points in $b$ for the reconstruction. The use of more data increase the degeneracy between PPS features and 
  lensing effect. $\Delta\psk$ contains even more oscillations corresponding to smaller scale 
  lensing effects. $\Delta\cl^{\rm TT}$ shows similar suppression till smaller scales. {Note here the $\Delta\cl^{\rm TT}$ obtained 
  differs from the lensing template even after $\ell=2000$. This is due to the fact that the transport kernel has significant 
  overlap in wavenumbers for $\ell=2000$ and the few hundred multipoles beyond that. Hence, when we reconstruct the PPS
  from $\alpha+b$ it changes the PPS at some wavenumbers which also
    contribute to the convolution integral for $\ell=2500$ and hence it affects the $\cl$'s beyond 
    $\ell=2000$ to some extent.}  
  
  \item $\alpha+a+1$, $\alpha+a+2$, $\alpha+b+1$, $\alpha+b+2$ and $\alpha+a+b+1+2$ :  When we add the smaller scale data we find that 
  till $i\simeq250$ the PPS obtained considering the lensing effect provides better likelihood. $\Delta\psk$ shows even higher oscillations 
  in all the combinations and $\Delta\cl^{\rm TT}$'s 
   continue to show the suppression compared to the lensing template used. Since the $\Delta\cl^{\rm TT}$ is plotted for 100'th iteration 
   we can see that it is not completely equal to {\it zero}, since till 100 iterations
   lensing is providing better likelihood in all combinations that can not be mimicked by the features in the PPS. 
   However, beyond $i\simeq250$ the lensing effect becomes completely degenerate with PPS features. Around 500 iteration we see 
   there is no difference in likelihood (say for $\alpha+a+b+1+2$). Hence we can expect no difference in $\cl$'s obtained in the two cases. 
   We find the $\Delta\cl^{\rm TT}$ obtained at $i=500$ (green curve in $\alpha+a+b+1+2$) is {\it nearly zero} in a broad range of multipoles 
   before $\ell=1900$. Beyond $\ell=1900$ since we are using binned data, the flexibility of the second PPS towards mimicking lensing effect 
   decreases.
   
  \end{itemize}

  From the results above we can certainly argue that the lensing of the CMB are reflected prominently through the reconstruction in all combinations of spectra. However, given substantially large freedom to the PPS (that can fit
the noise and fluctuations significantly) it is indeed possible to
mimick the lensing effect up to a very high extent (say for large
iterations and complete data used in  $\alpha+a+b+1+2$). However consideration
of lensing always provide better likelihood if we restrict ourselves
to low number of iterations where there will be less fluctuations in
the reconstructed PPS. Keeping this degeneracy in mind, a proper 
    reconstruction of the PPS must be carried out including lensing effects appropriately, otherwise the 
    lensing can be mistakenly treated 
  as features in the PPS.

% \clearpage

%%%%%%%%%%%%%%%%%%%%%%%%%%%%%%%%%%%%%%%%%%%%%%%%%%%%%%%%%%%%%%%%%%%%%%%%%%%%%%%

\subsection{Features : Where are they?}

Since WMAP, the features in the PPS and their importance have been discussed widely in literature. 
% Departures of the  
% concordance model angular power spectrum from the corresponding data can reveal the locations of probable features~\footnote{The 
% stand of concordance model in the light of the Planck data has been discussed in~\cite{Hazra:2014hma}}. 
Introduction of some particular features in the PPS can result in significant improvement in the 
likelihood~\footnote{The stand of concordance model in the light of the Planck data has been discussed in~\cite{Hazra:2014hma}
where we indicated a particular damping in the small scale CMB angular power spectrum is significantly favored by Planck data}. 
However their evidence as physical effects should be addressed through proper error analysis. 
Hence, our first job is to hunt down these features and obtain the shape of the PPS which contain possible physical and 
statistical features. In~\cite{Hazra:2013xva} we have shown MRL is 
an excellent method to locate the features and directly provides a better likelihood to the data.   

\begin{figure*}[!htb]
\begin{center} 
\vskip -30 pt
\resizebox{215pt}{160pt}{\includegraphics{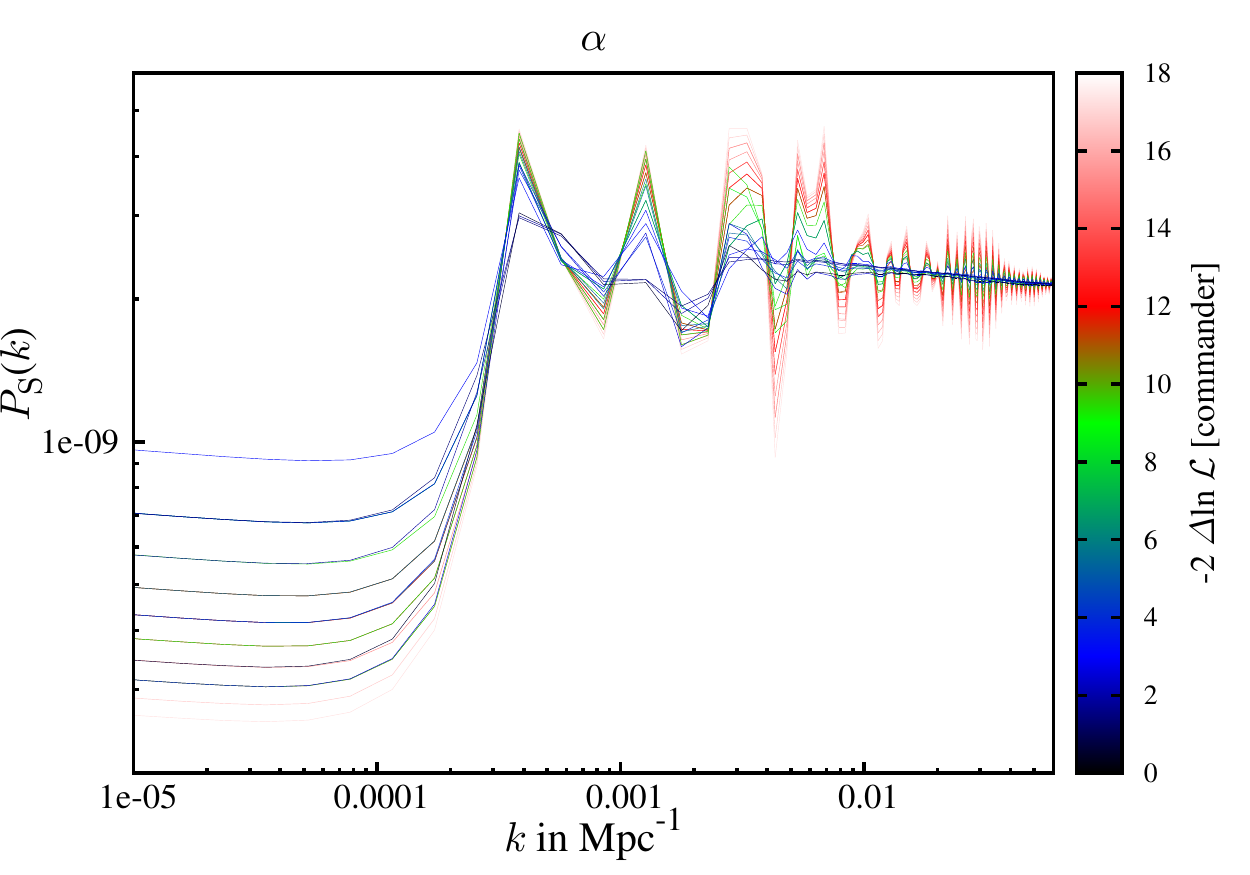}} 
\resizebox{215pt}{160pt}{\includegraphics{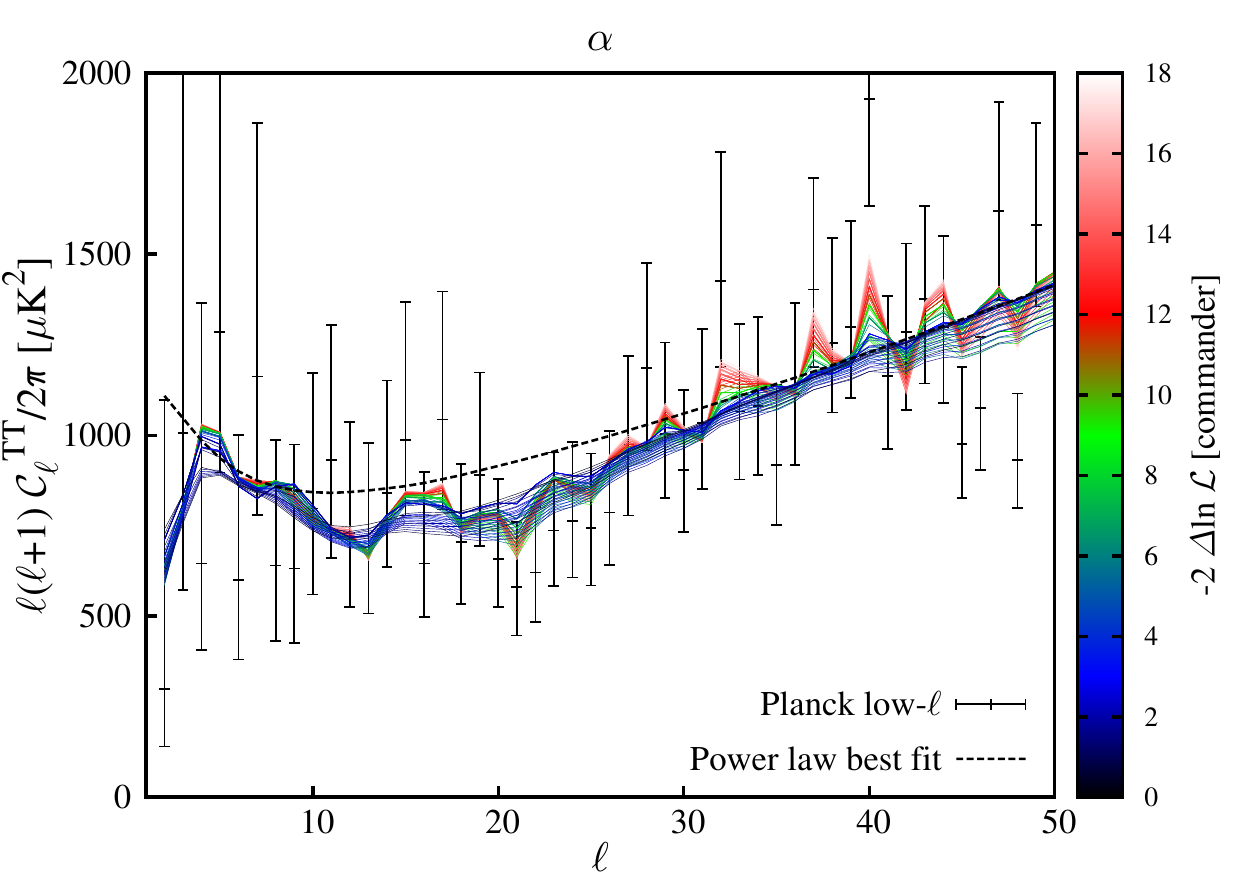}} 
\resizebox{215pt}{160pt}{\includegraphics{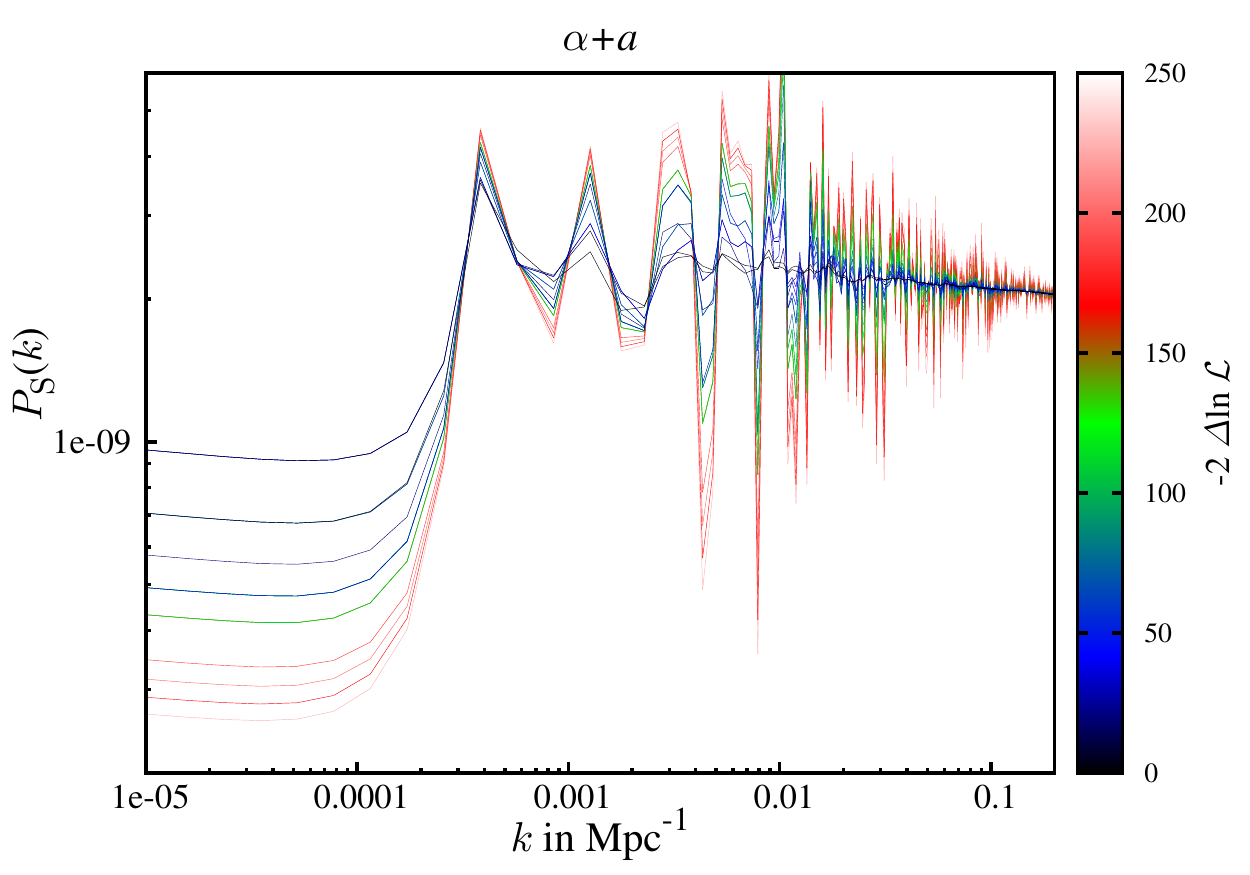}} 
\resizebox{215pt}{160pt}{\includegraphics{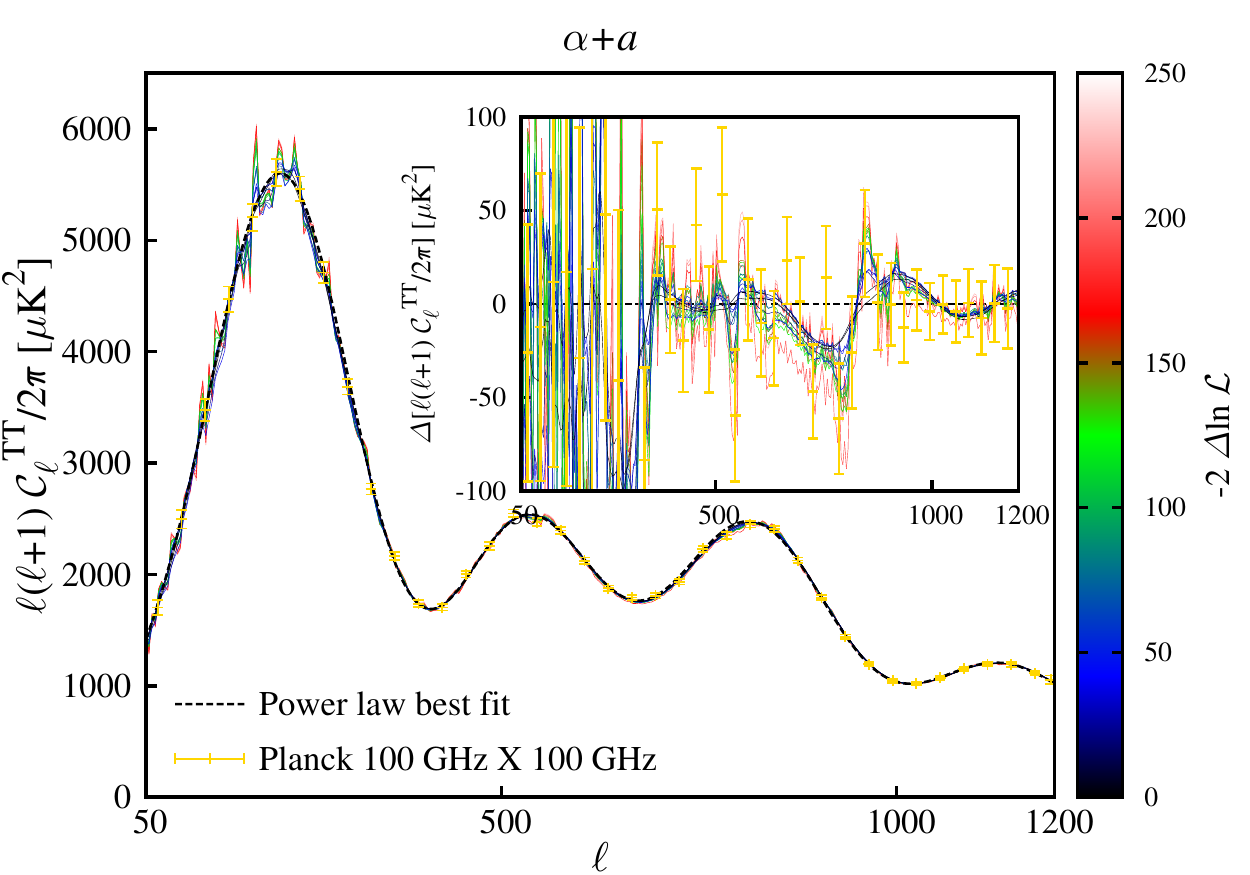}} 
\resizebox{215pt}{160pt}{\includegraphics{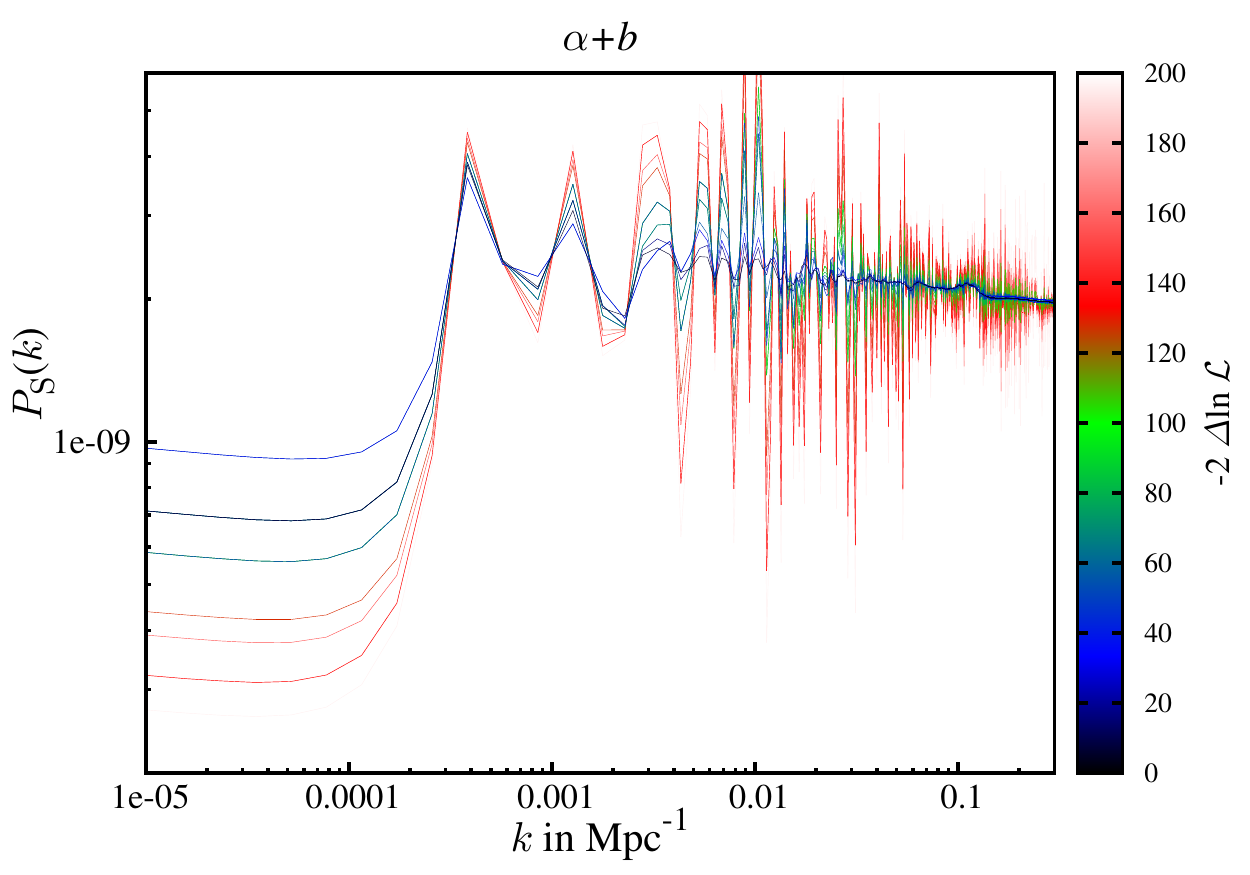}} 
\resizebox{215pt}{160pt}{\includegraphics{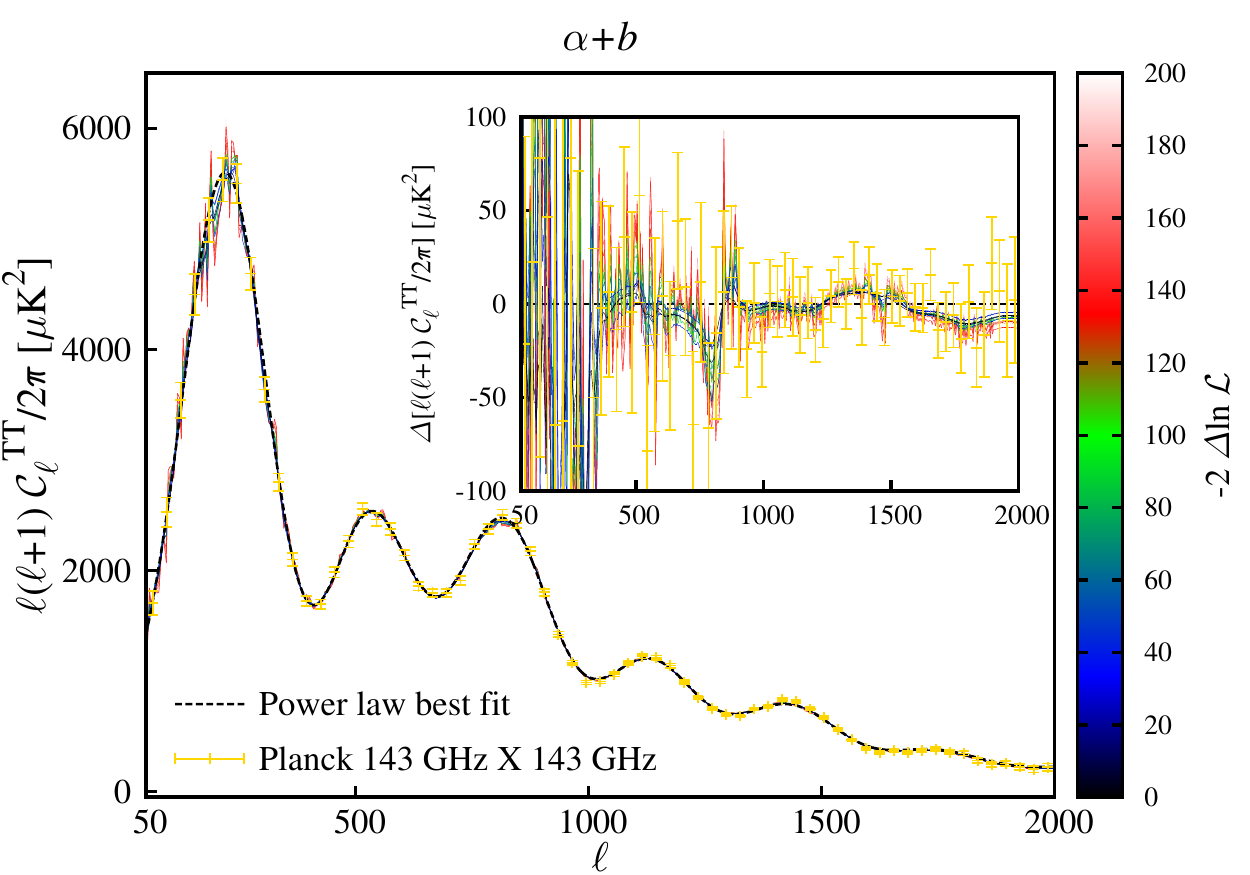}} 
\resizebox{215pt}{160pt}{\includegraphics{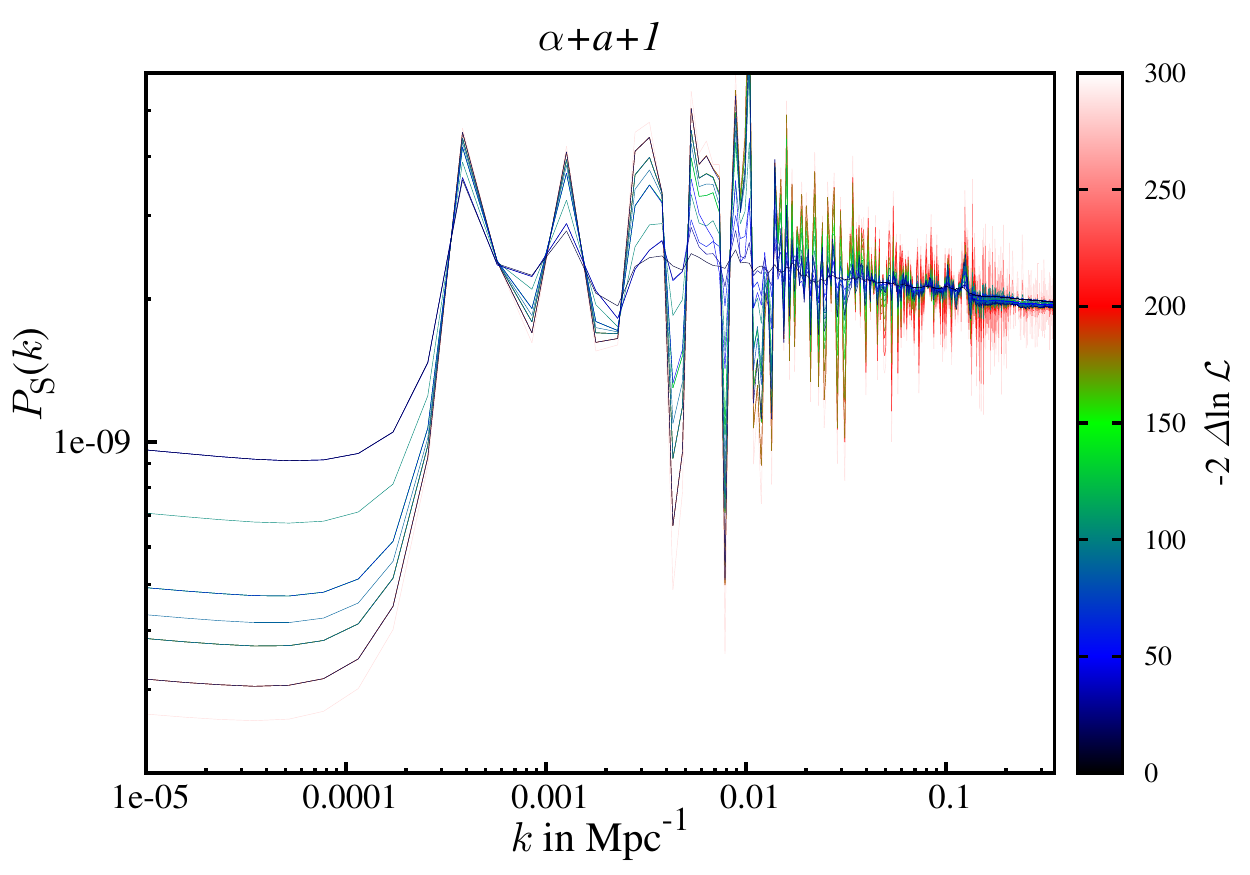}} 
\resizebox{215pt}{160pt}{\includegraphics{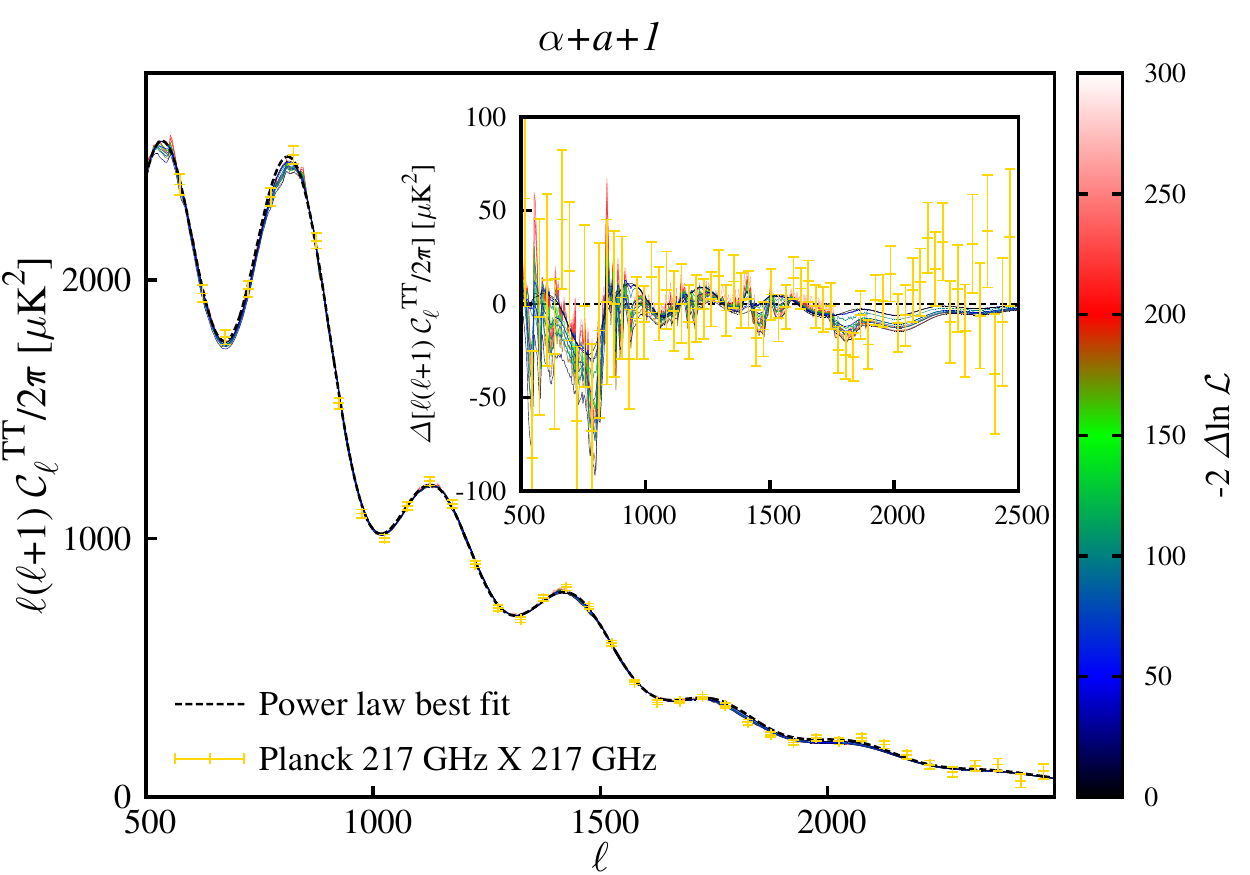}}

\end{center}
\caption{\footnotesize\label{fig:samples-1}[Left] PPS obtained for different iterations and different smoothing and the improvement to the Planck likelihood 
(in $\Delta\chi^2=-2\Delta\ln{\cal L}$)(indicated through colorbar). [Right] Angular power spectra corresponding to the same color PPS appearing in the left plots. [Inset]
Residual data with error-bars and the residual reconstructed $\cl$'s with reference to best fit Planck baseline $\cl$.} 
\end{figure*}

\begin{figure*}[!htb]
\begin{center} \vskip -30 pt
\resizebox{215pt}{160pt}{\includegraphics{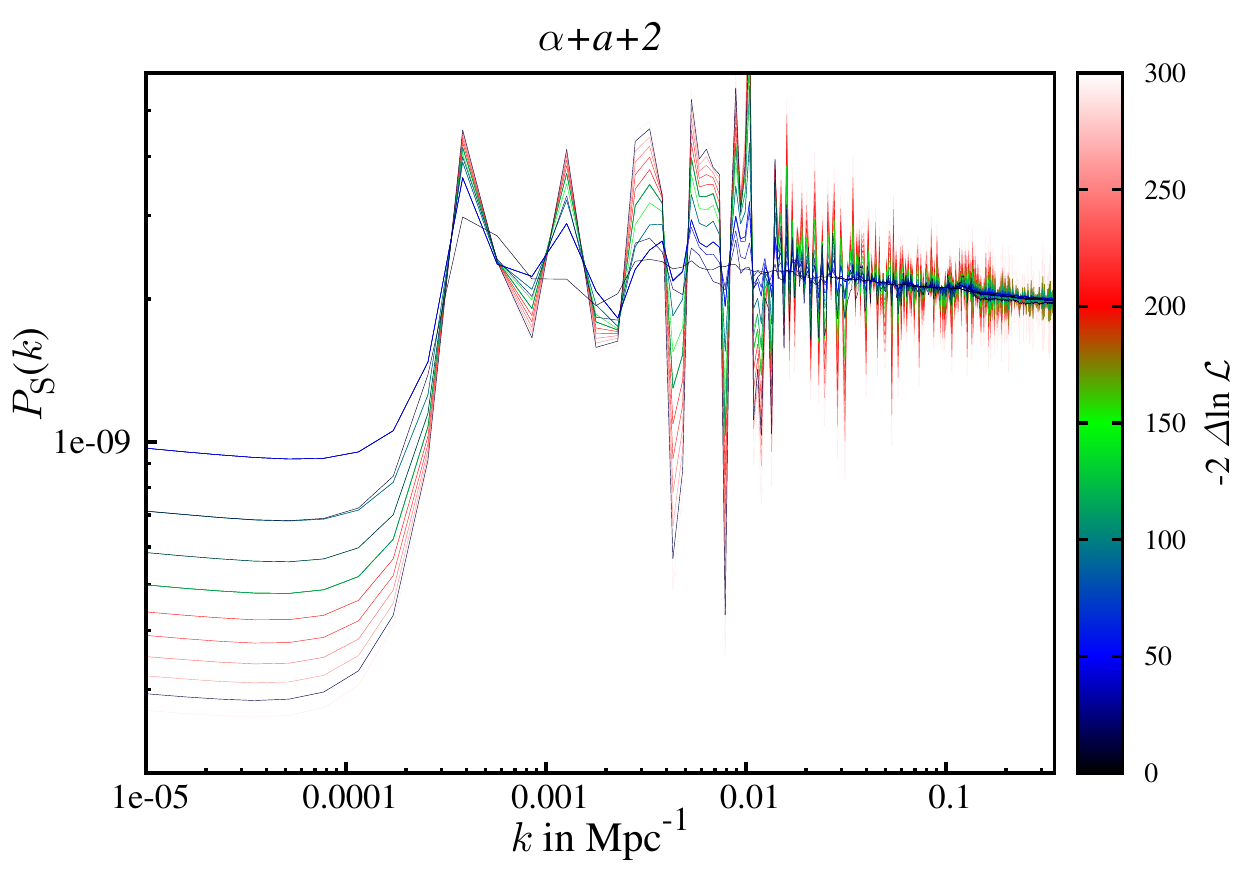}} 
\resizebox{215pt}{160pt}{\includegraphics{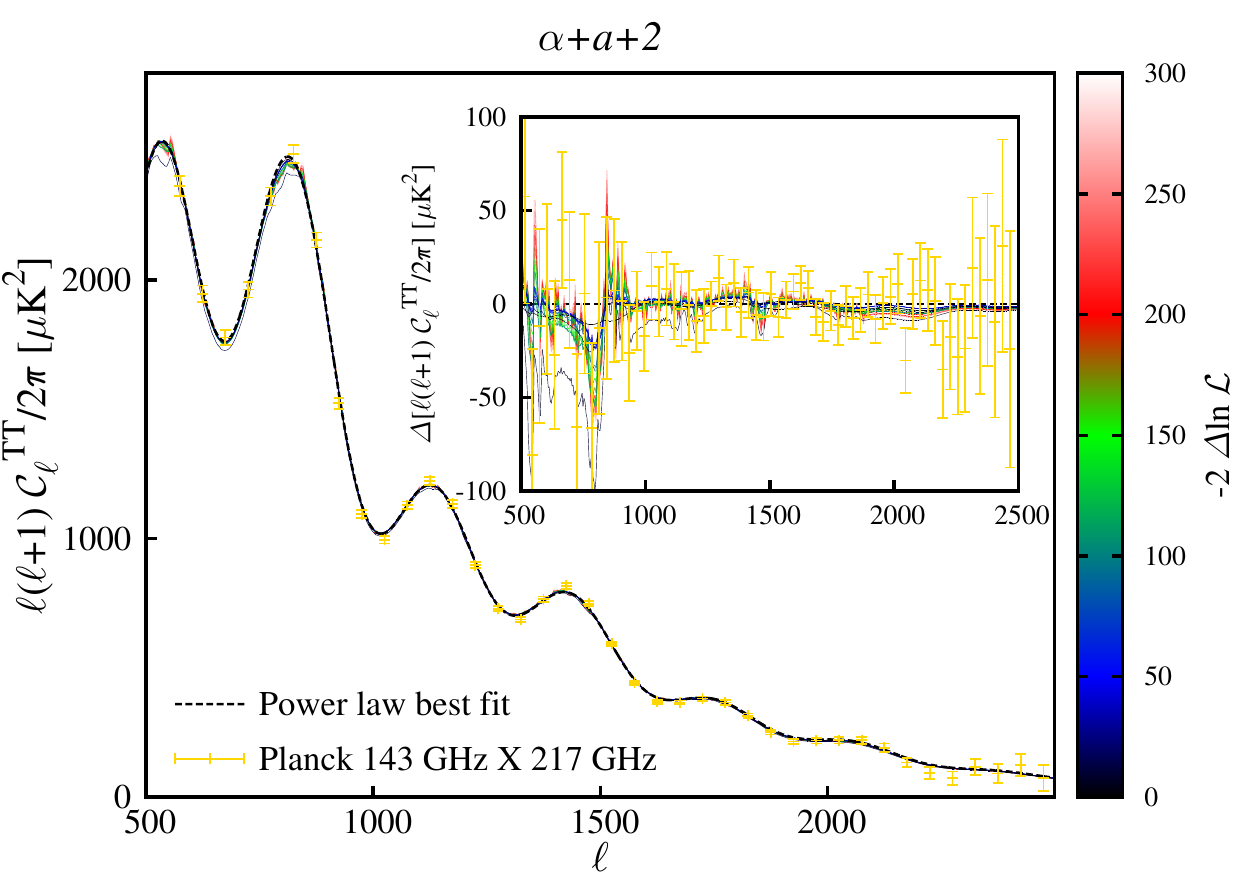}} 
\resizebox{215pt}{160pt}{\includegraphics{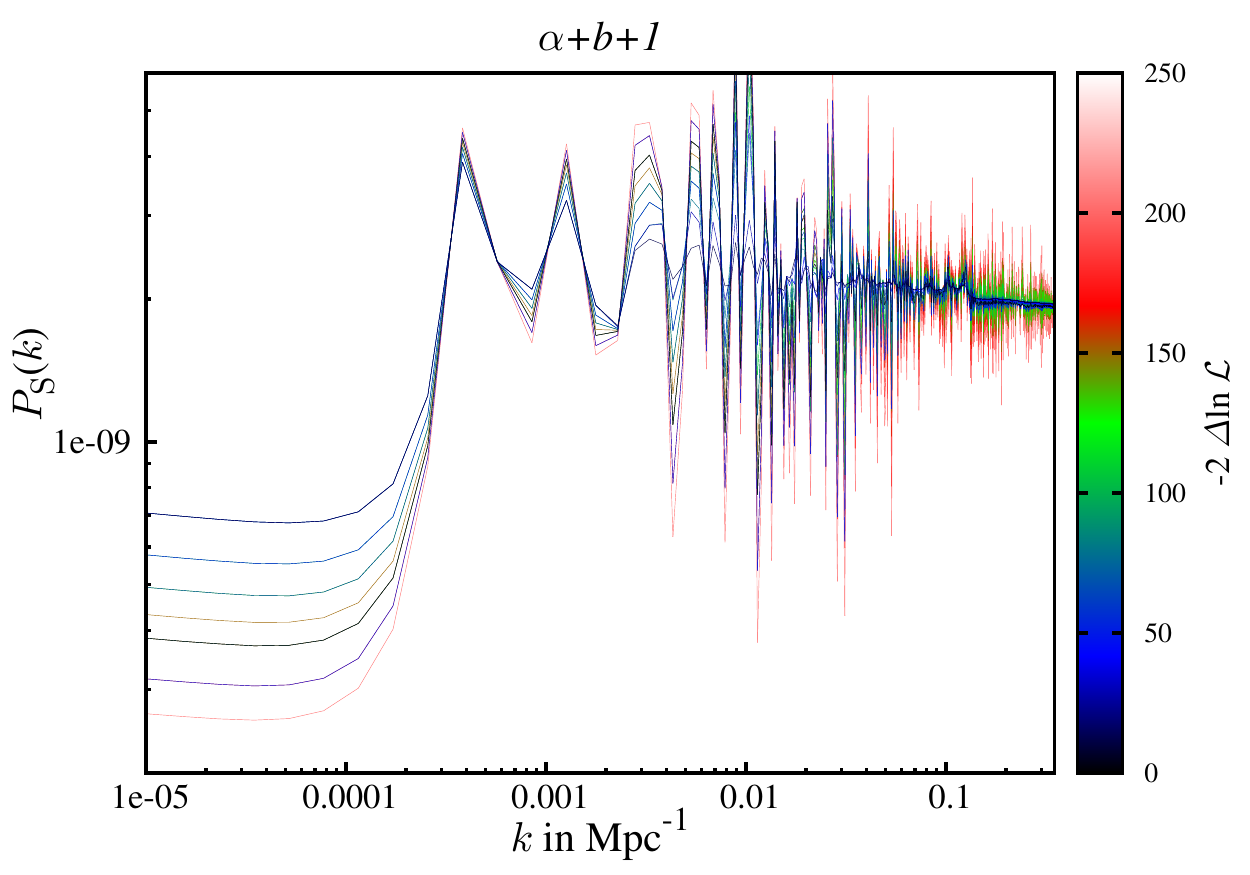}} 
\resizebox{215pt}{160pt}{\includegraphics{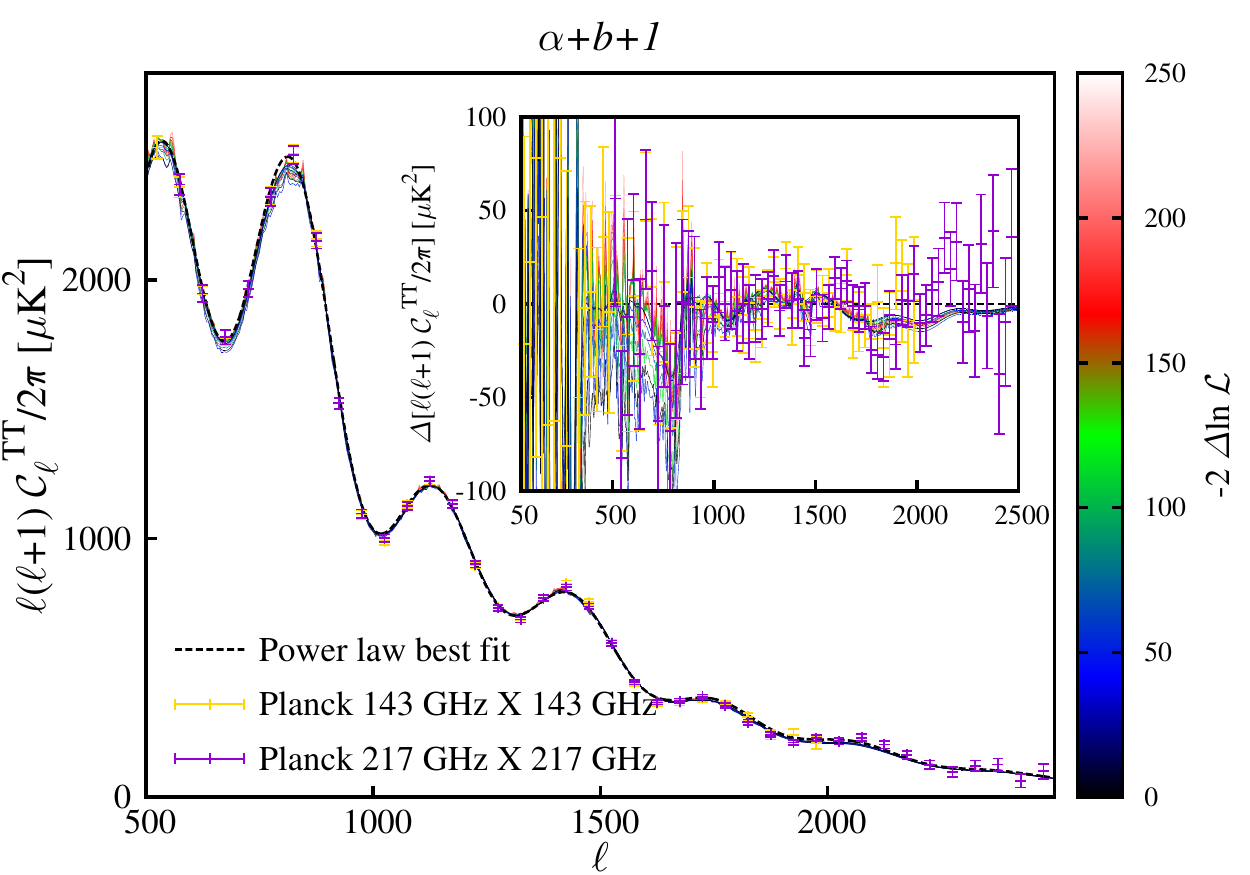}} 
\resizebox{215pt}{160pt}{\includegraphics{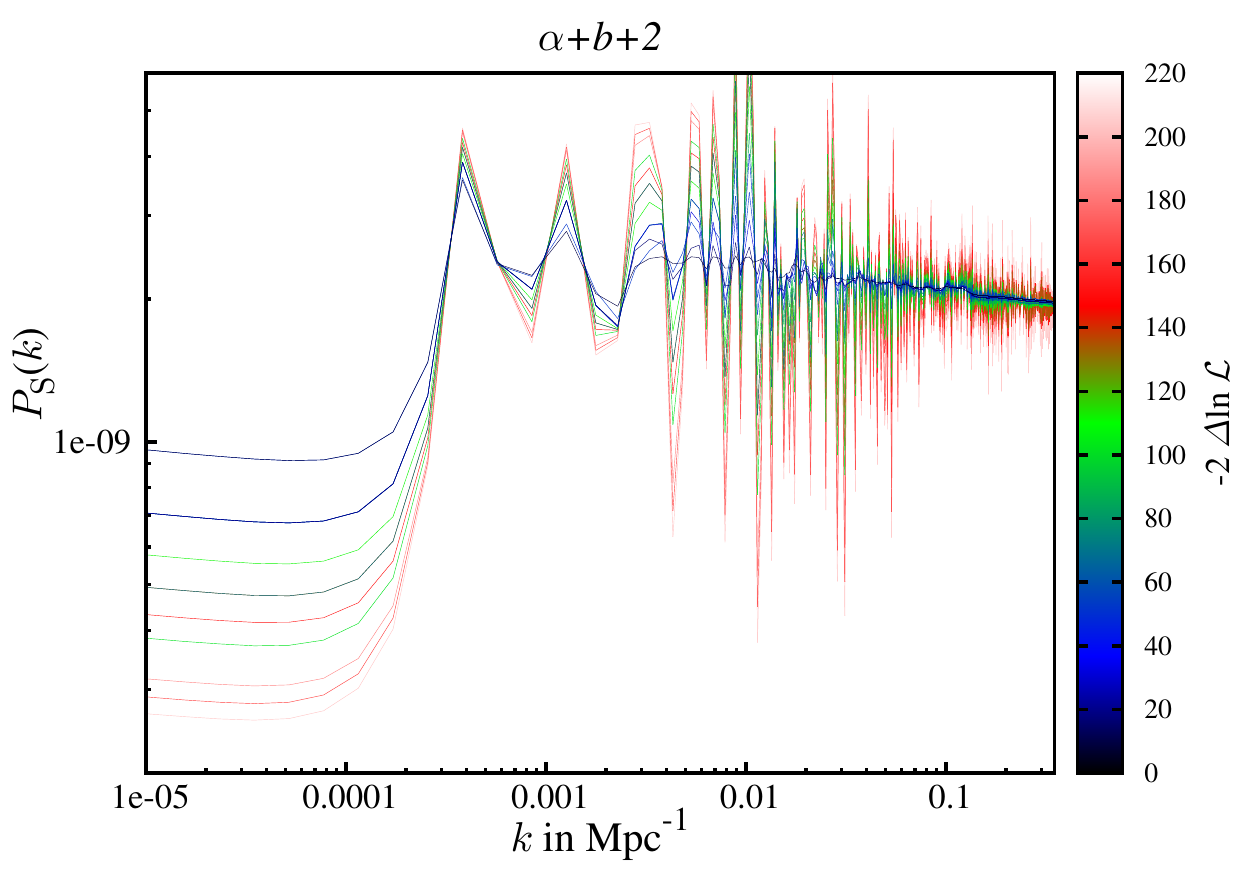}} 
\resizebox{215pt}{160pt}{\includegraphics{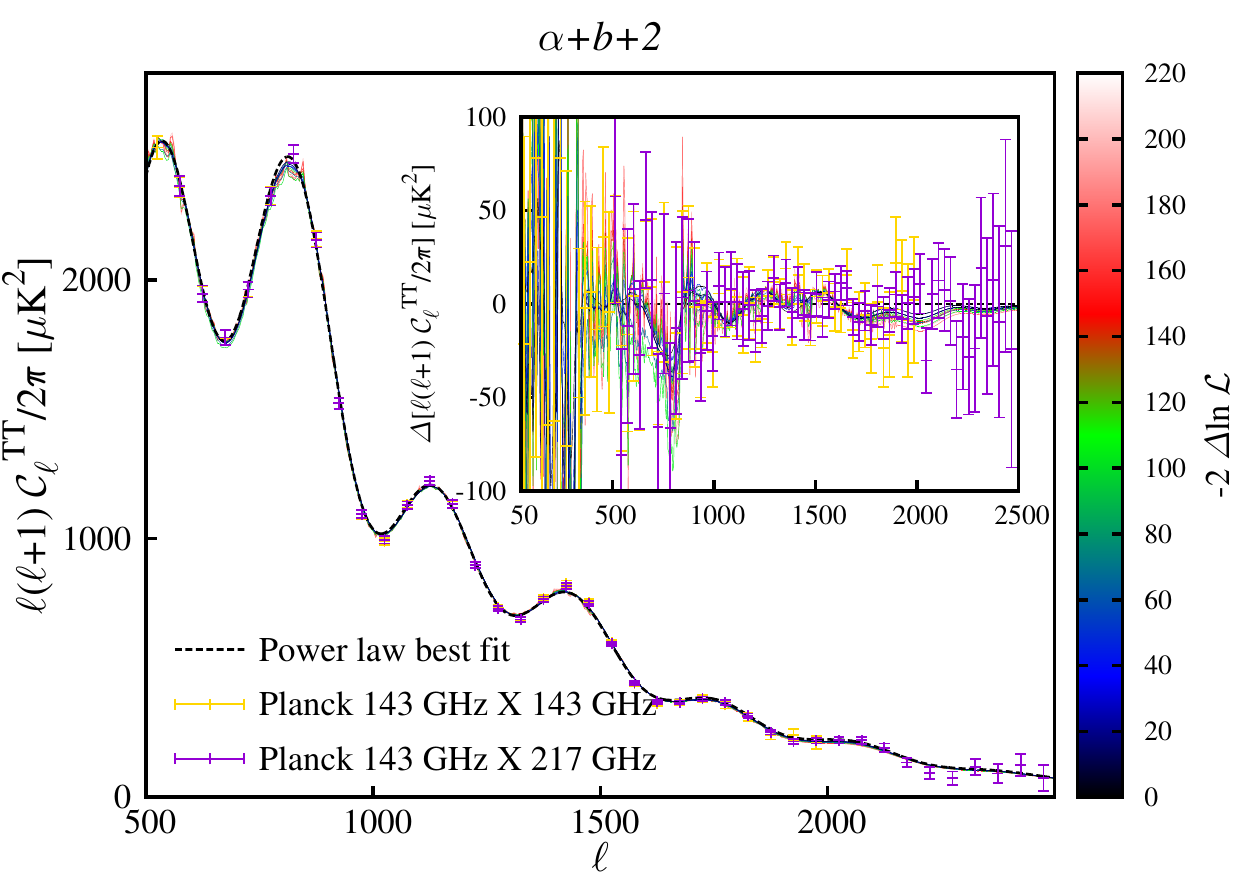}} 
\resizebox{215pt}{160pt}{\includegraphics{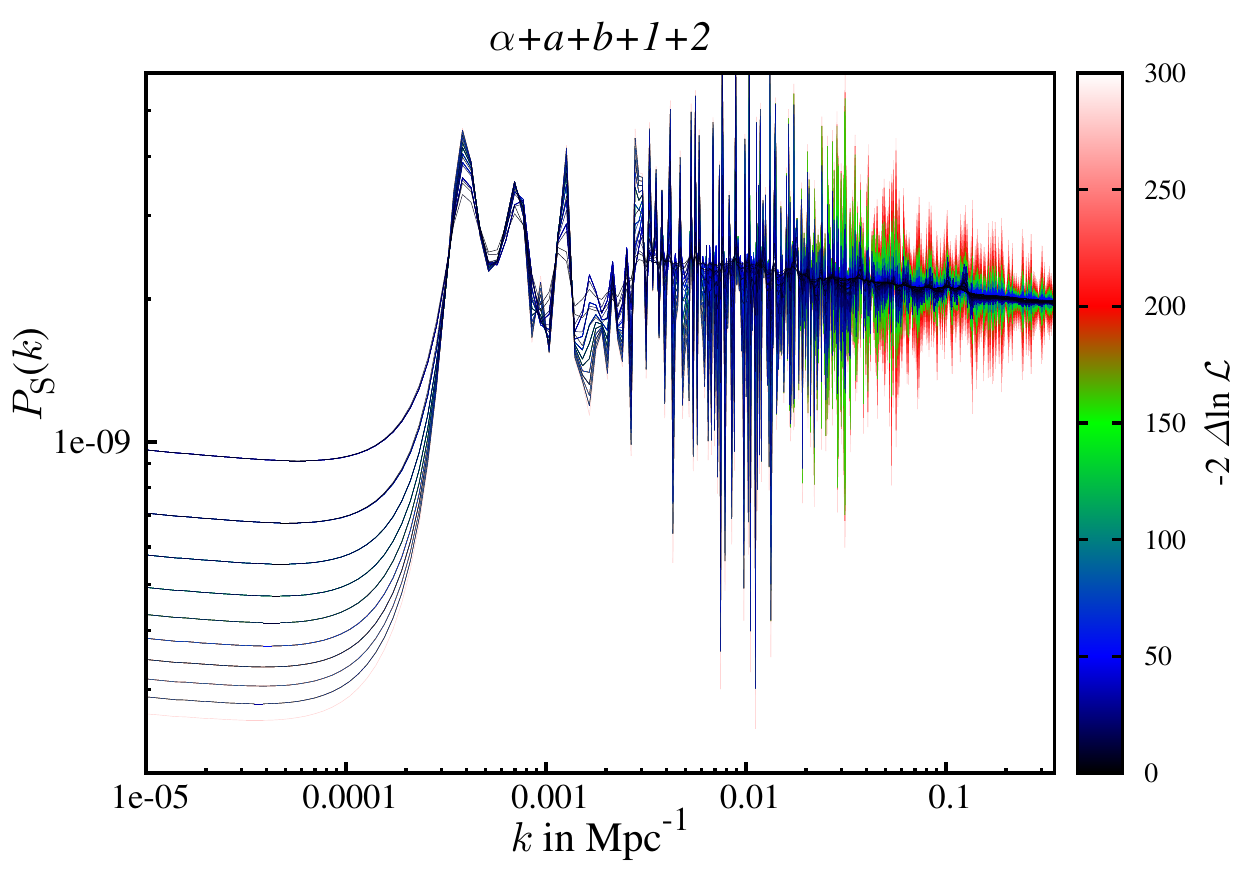}} 
\resizebox{215pt}{160pt}{\includegraphics{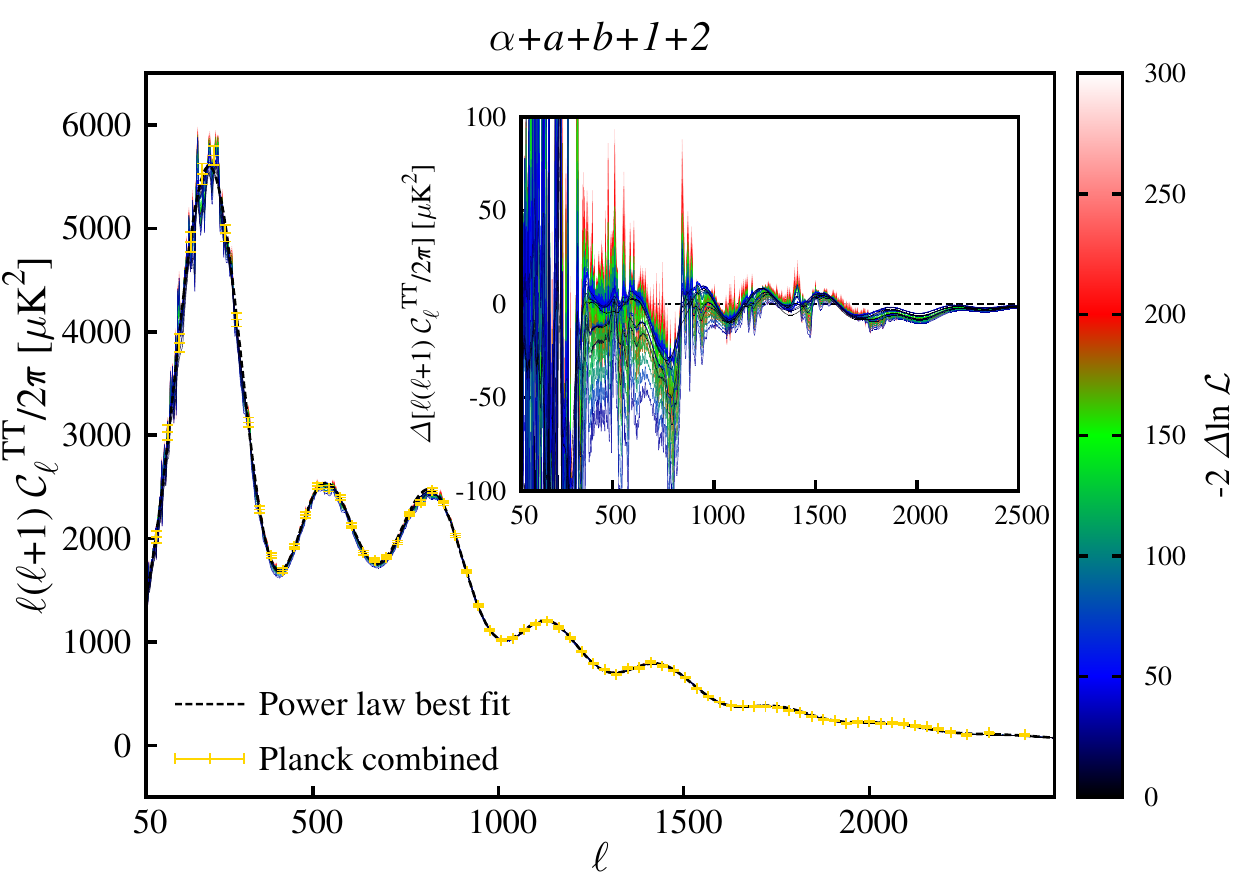}} 

\end{center}
\caption{\footnotesize\label{fig:samples-2}[Left] PPS obtained for different iterations and different smoothing and the improvement to the Planck likelihood 
(in $\Delta\chi^2=-2\Delta\ln{\cal L}$)(indicated through colorbar). [Right] Angular power spectra corresponding to the same color PPS appearing in the left plots. [Inset]
Residual data with error-bars and the residual reconstructed $\cl$'s with reference to best fit Planck baseline $\cl$.}
\end{figure*}

As a first step, we use MRL for different iterations ($i=1-50$) and use different smoothing width $\Delta$, chosen randomly, (following Eq.~\ref{eq:mrl} and~\ref{eq:gauss}) 
to generate a sample of PPS and $\cl$'s as plotted in Fig.~\ref{fig:samples-1} and in Fig.~\ref{fig:samples-2}. In both the figures the left panels contain 
the samples of PPS (in different colors) obtained using the method. The colorbars at the right represent the improvement in $\chi^2$ from Planck likelihood code compared to best 
fit power law baseline model. Note that the PPS in dark blue colors are extremely smooth at small scales due to large $\Delta$ and low iterations 
and still provide a better fit. Right panels contain the angular power spectra 
corresponding to the same PPS from the left plots in same color. The data from different spectra are plotted too. The inset of the 
plots contain the residual angular spectra data, {\it i.e.} $\cl^{\rm {D'_\nu}}-\cl^{\rm Planck~best~fit}$ and residual reconstructed 
angular power spectra ($\cl^{{\rm T}(i)}-\cl^{\rm Planck~best~fit}$)~\footnote{Note that
$\cl^{{\rm T}(i)}$'s here correspond to the reconstructed angular power spectra obtained after smoothing and adding the lensing template}. 
The dashed black lines appearing in the plots at the right correspond to the Planck best fit baseline power spectrum. 

Below we highlight the features in multipole and wavenumber space obtained from different combinations of Planck spectra.

\begin{itemize}
 \item $\alpha$ : For reconstruction obtained only from $\alpha$, we plot the colorbars that represent $-2\Delta\ln{\cal L}=\Delta\chi^2$ only from {\tt commander}. Note that the 
 features shown in the plots can improve the likelihood by 18 compared to power law PPS.
 Due to quadrupole suppression ($\ell=2$) at largest scales (smallest $k\sim 2 \times 10^{-4} {\rm Mpc}^{-1}$) we observe a cut-off in  
 power compared to the best fit baseline model. The dip in power around $\ell\sim 15-30$ (specifically, near $\ell\simeq22$)
 imprints a broad dip in the PPS around $k\simeq 0.002 {\rm Mpc}^{-1}$. Around $\ell\simeq 40$ we see the data indicate slight 
 excess in power compared to the black dashed spectrum obtained from power law PPS. The power enhancement in the PPS around 
 $k\simeq 0.0035{\rm Mpc}^{-1}$ signifies this excess. However, the smooth dark blue line only picks up the cut-off around 
 $k\simeq 2\times 10^{-4} {\rm Mpc}^{-1}$ and the dip near $k\simeq 0.002 {\rm Mpc}^{-1}$ which indicate that compared to other 
 features in low-$\ell$ these two features are more significant. More generally, one can say that 
 the low-$\ell$ data indicate an overall suppression in PPS (similar to what we have found in~\cite{Hazra:2013nca}).
 
 \item $\alpha+a$ : Note that with more data from $\alpha+a$ we have started to fit the Planck angular power spectrum 
 better ($-2\Delta\ln{\cal L}\simeq250$) than the best fit power law PPS. The residual $\cl$'s and residual data points (in yellow) 
 help to locate the possible features. 
 We observe oscillations in the angular power spectrum with a dip around $\ell\sim250-300$  in the power which correspond to a 
 dip in PPS near $k\simeq 0.02 {\rm Mpc}^{-1}$. Followed by this 
 we find a dip and a bump around $\ell\simeq750-850$ which correspond to similar dip and bump in the PPS near 
 $k\sim (0.055-0.065) {\rm Mpc}^{-1}$.
 
 \item $\alpha+b$ : Use of the spectrum from $b$ allows us to locate the features even smaller scales. Similar to the last case, 
 we find $\sim 200$ improvement in fit from the reconstructed PPS. The features near $\ell\sim250-300$ and $\ell\sim750-850$ are confirmed 
 by $b$ spectrum too. Moreover near $\ell\sim 1800-2000$ we find a dip in the reconstructed angular power spectra~\footnote{We mention
 again that this particular feature is related to a systematic effect discussed in the revised version of Planck papers~\cite{Planck:inflation}}. 
 The residual data points are below {\it zero} around that region. Corresponding to that dip we find similar dip in PPS near $k\sim (0.12-0.14){\rm Mpc}^{-1}$.
 
 \item $\alpha+a+1$ : The dip in the angular power spectrum around $\ell\sim750-850$ is visible in spectrum 1 too. Near $\ell\sim1800-2000$ we 
 observe the feature again. Note that the residual data points from 1 indicate a more prominent dip compared to residual data from spectrum $b$
 due to smaller uncertainties. This particular feature was indicated in Planck analysis~\cite{Planck:inflation}.

 \item $\alpha+a+2$ : Feature around $\ell\sim750-850$ is still visible in spectrum 2. 
 However, we find $\ell\sim1800-2000$ feature is not prominent in this spectrum compared to spectra $b$ and 1 (obtained from the reconstruction 
 using angular power spectra data $\alpha+b$ and $\alpha+a+1$).
 
 \item $\alpha+b+1$ and $\alpha+b+2$ : In these two spectra combinations we find both the features 
 around $\ell\sim750-850$ and $\ell\sim1800-2000$. 
 
 \item $\alpha+a+b+1+2$ : Results from reconstructions using the complete Planck spectrum are plotted in the last panels in Fig.~\ref{fig:samples-2}.
 Combining all spectra, we find low-$\ell$ suppressions near $\ell=2$ ($k\simeq 2\times 10^{-4} {\rm Mpc}^{-1}$), 
 and dip and bump at $\ell\simeq22$ ($k\simeq 0.002 {\rm Mpc}^{-1}$) and $\ell\simeq40$ ($k\simeq 0.0035{\rm Mpc}^{-1}$) respectively.  
 Apart from low-$\ell$ we also find features in the angular power spectrum near $\ell\sim 300,750-850,1800-2000$.
   
\end{itemize}

%  Through all the results listed above we have identified the locations and the nature of the features in the PPS. Next section is dedicated for
%  checking the statistical significance of these features.

\clearpage
%%%%%%%%%%%%%%%%%%%%%%%%%%%%%%%%%%%%%%%%%%%%%%%%%%%%%%%%%%%%%%%%%%%%%%%%%%%%%%%
\subsection{Error estimation : Hunting down the probable physical features}

It is always expected that a standard model theory can not address all the data points and we can always expect features. However, finding a feature
does not necessitates the venture to building models to explain the feature since the feature can simply be statistical rather than being a physical artifact.
Moreover it can be easily understood from the reconstructed PPS in the previous section that we are getting huge improvement in fit by fitting the 
noise in the data. To hunt down the possible physical features we need a proper error estimation.
In our analysis with WMAP-9 data~\cite{Hazra:2013xva} we had found the standard power law model is completely consistent with the data at all cosmological 
scales within the constraints on the PPS. With Planck data we get tighter constraints on the PPS compared to WMAP since the precision of Planck data is 
significantly higher than WMAP-9. Using similar formalism as in~\cite{Hazra:2013xva} we address the significance of the features and check the 
consistency of power law PPS with the data.

For the 8 different combinations of spectra we generate 1000 realizations of the data considering Gaussian random fluctuations. The variance of the fluctuations are 
equated to the diagonal term of the error covariance matrix. Since we have asymmetric errors at low-$\ell$, for $\alpha$ we use the upper-error when the random fluctuations
are above the mean value and we use lower-error for the opposite. For 1000 data in each combination MRL generates 1000 PPS. In each $k$ from the reconstructed PPS we extract the most densely populated 
$68.3\%$ and $95.5\%$ region which will act as $1\sigma$ and $2\sigma$ constraints on the PPS.  It should be kept in mind that since we are considering only one set of
foreground parameters, small scale errors on the PPS are underestimated in our analysis. We expect increase in errors with the addition of  
foreground parameter marginalization, which is beyond the scope of this paper. To compare with the PPS reconstructions discussed in Planck papers, 
see~\cite{Planck:inflation}.

\begin{figure*}[!htb]
\vskip -30 pt
\begin{center} 
\resizebox{210pt}{160pt}{\includegraphics{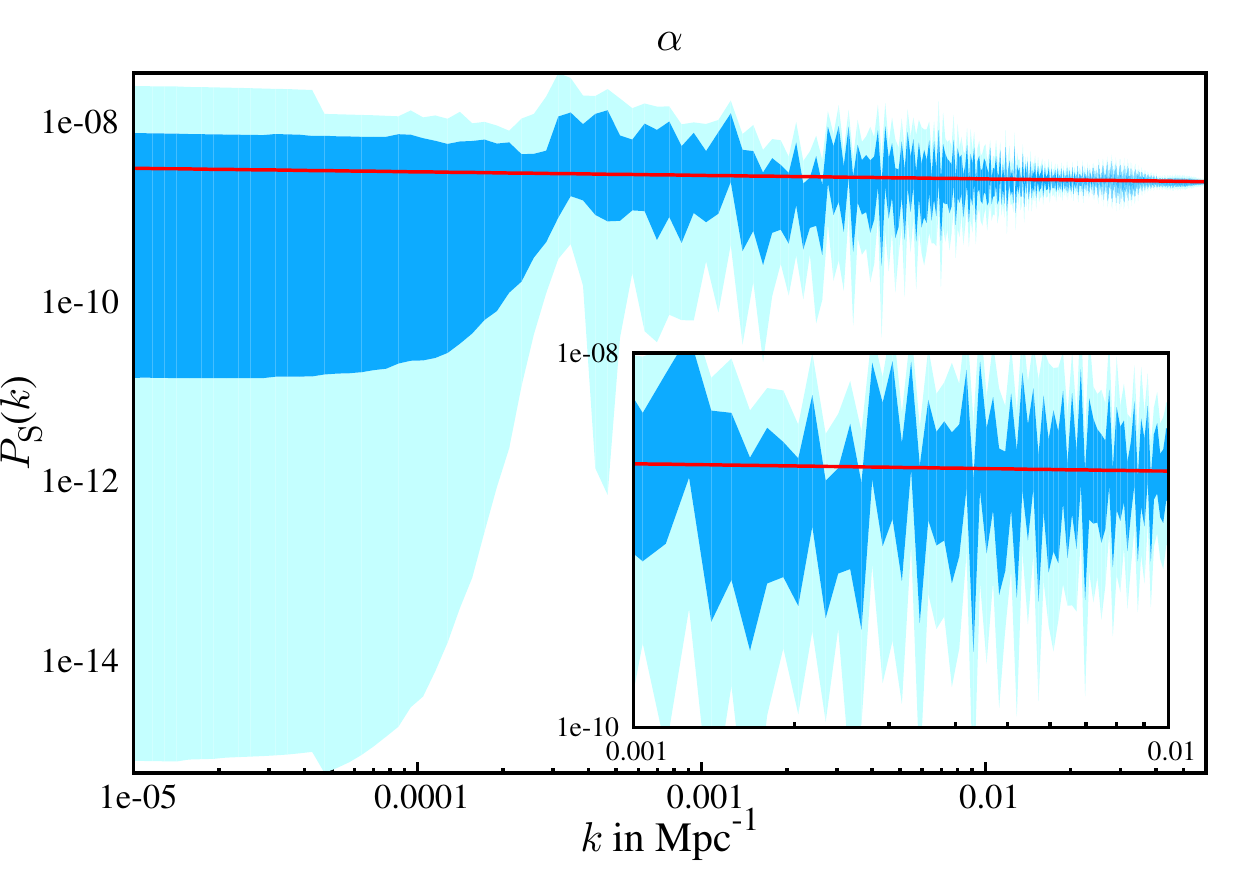}} 
\resizebox{210pt}{160pt}{\includegraphics{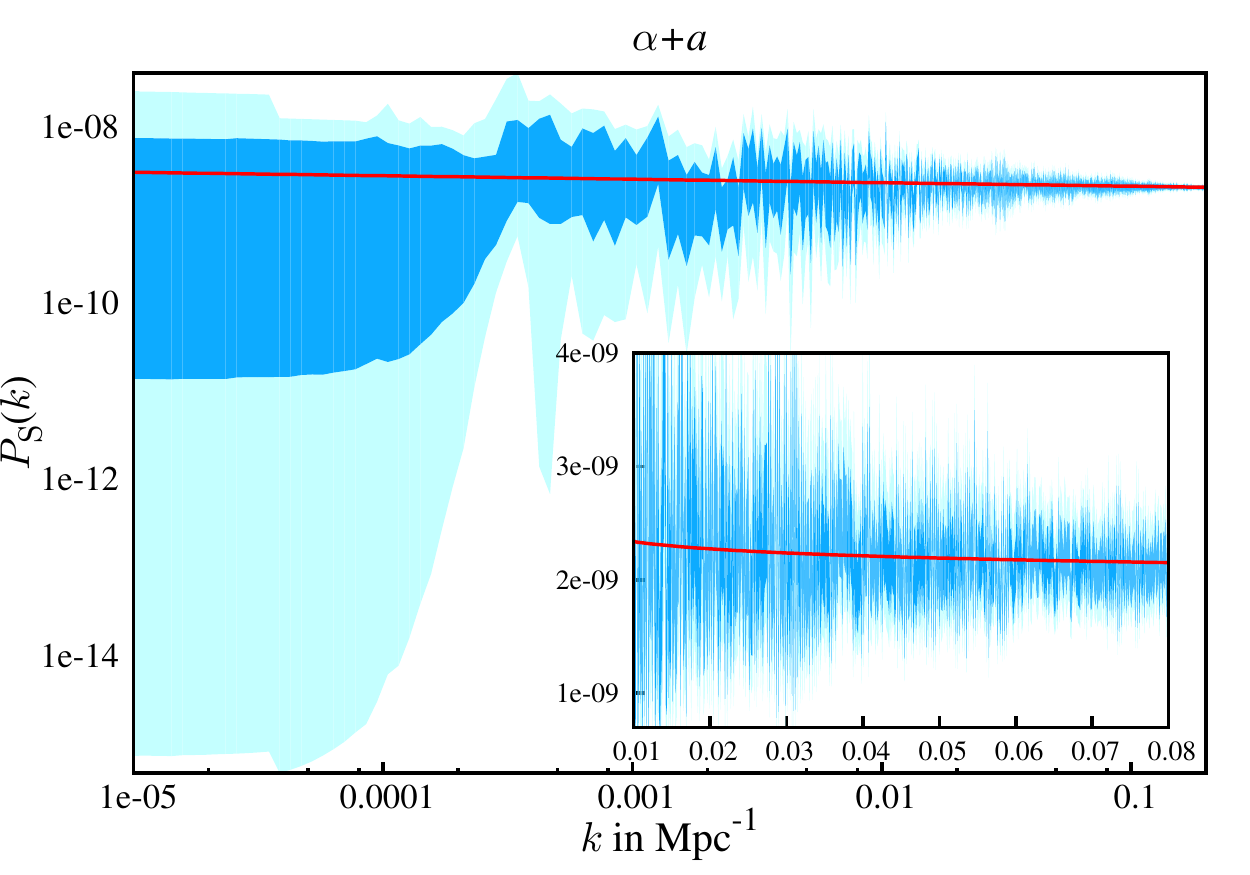}} 
\resizebox{210pt}{160pt}{\includegraphics{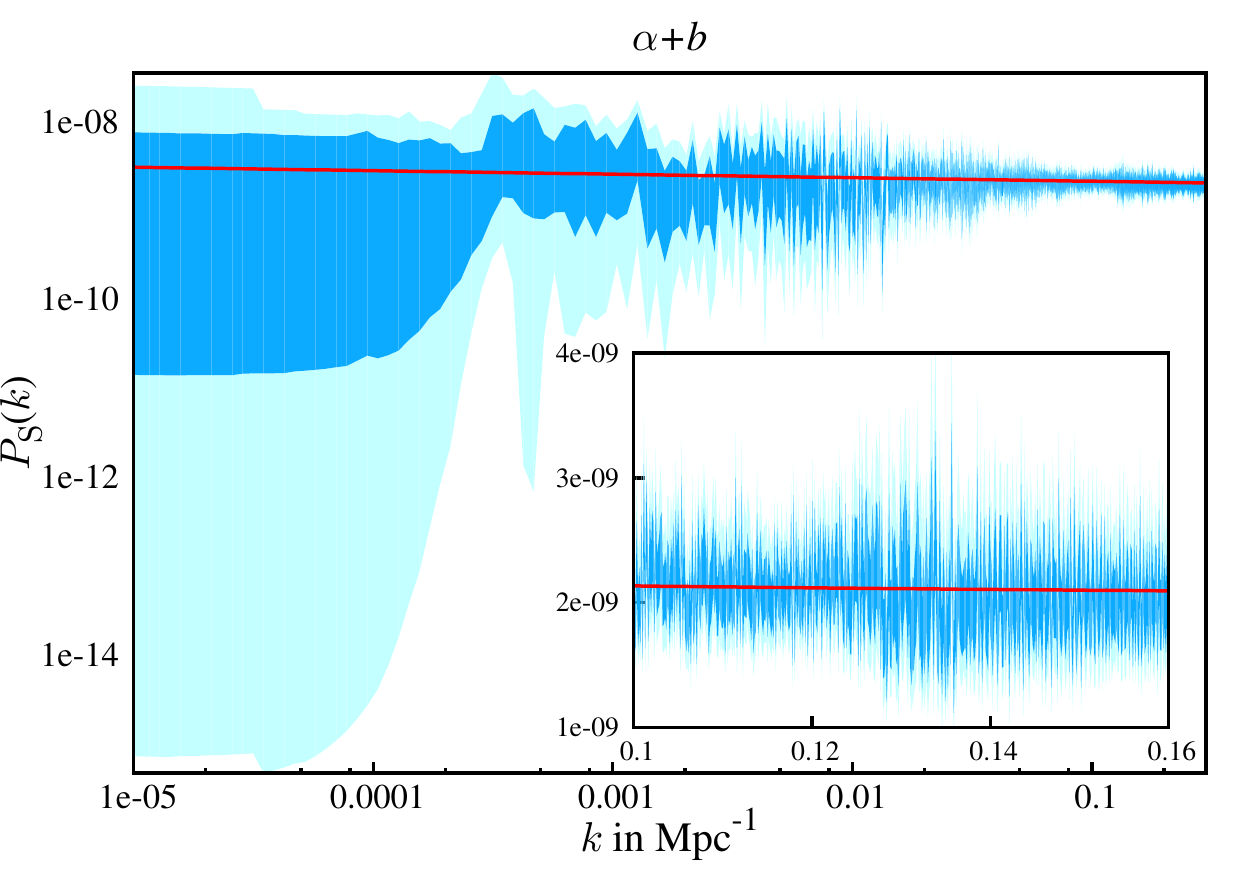}} 
\resizebox{210pt}{160pt}{\includegraphics{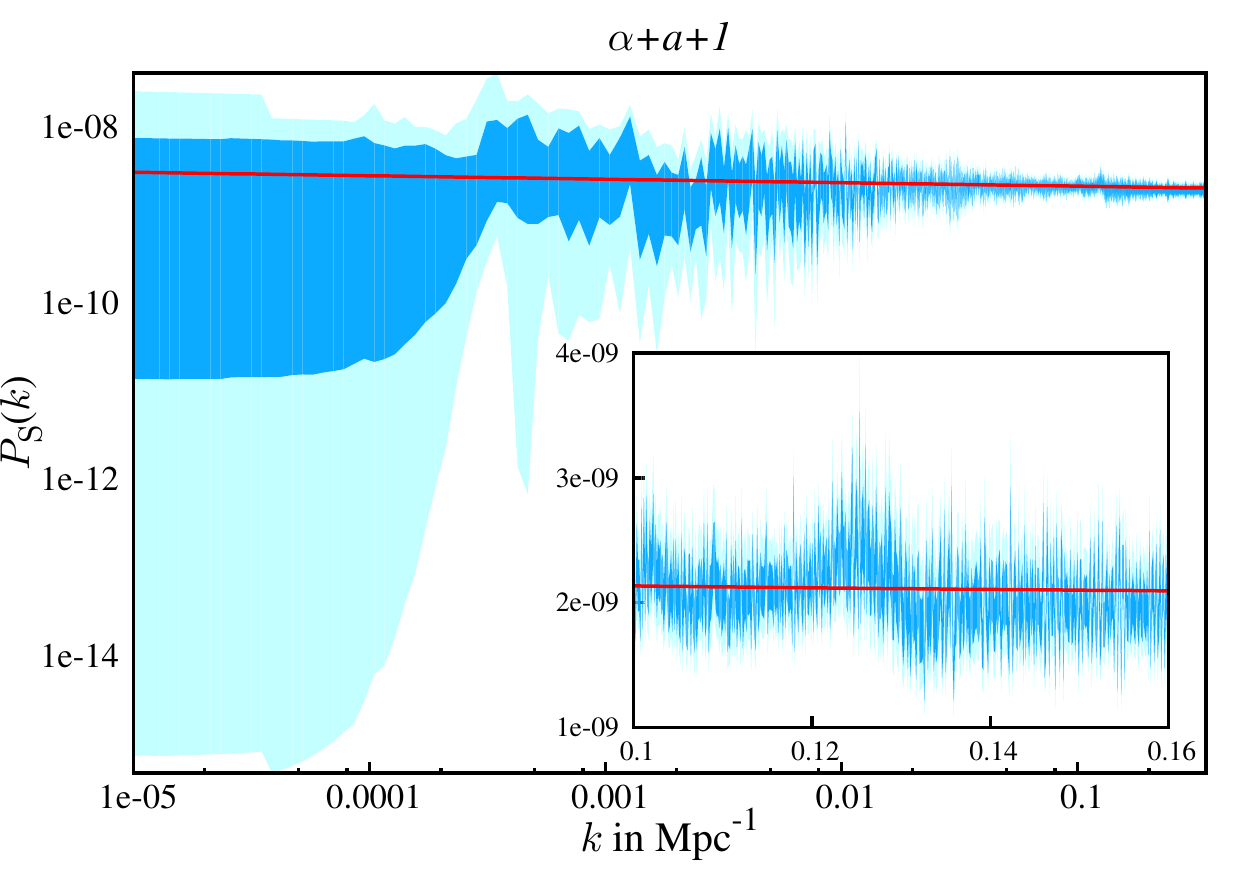}} 
\resizebox{210pt}{160pt}{\includegraphics{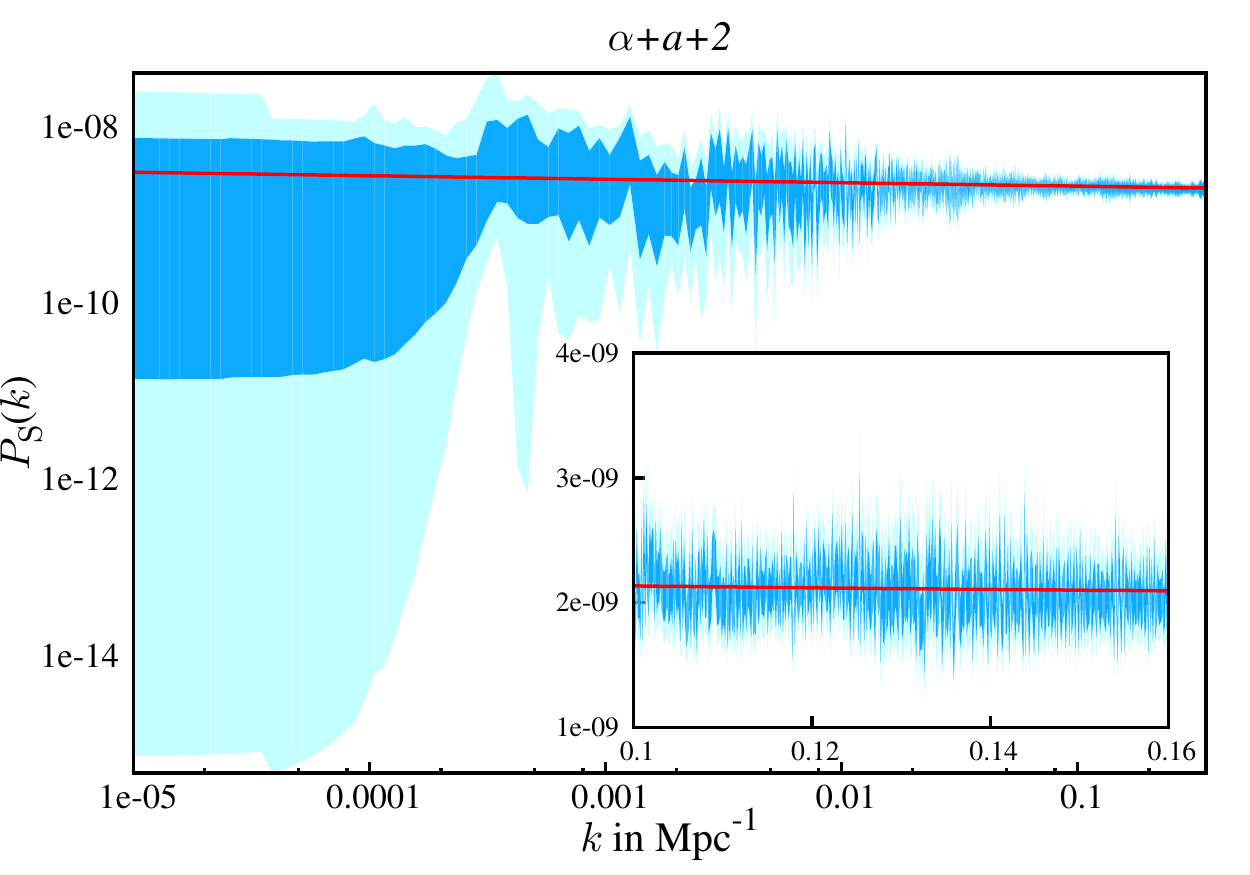}} 
\resizebox{210pt}{160pt}{\includegraphics{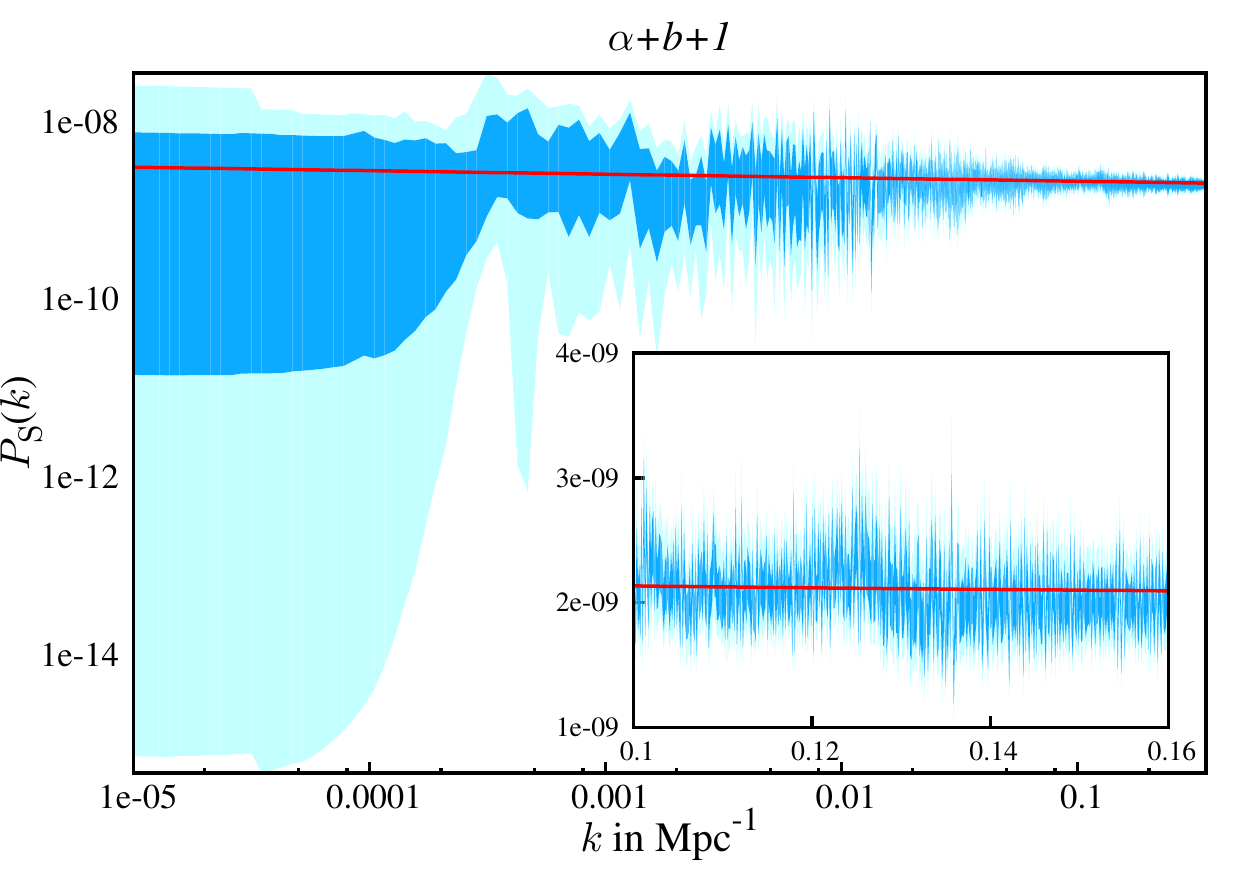}} 
\resizebox{210pt}{160pt}{\includegraphics{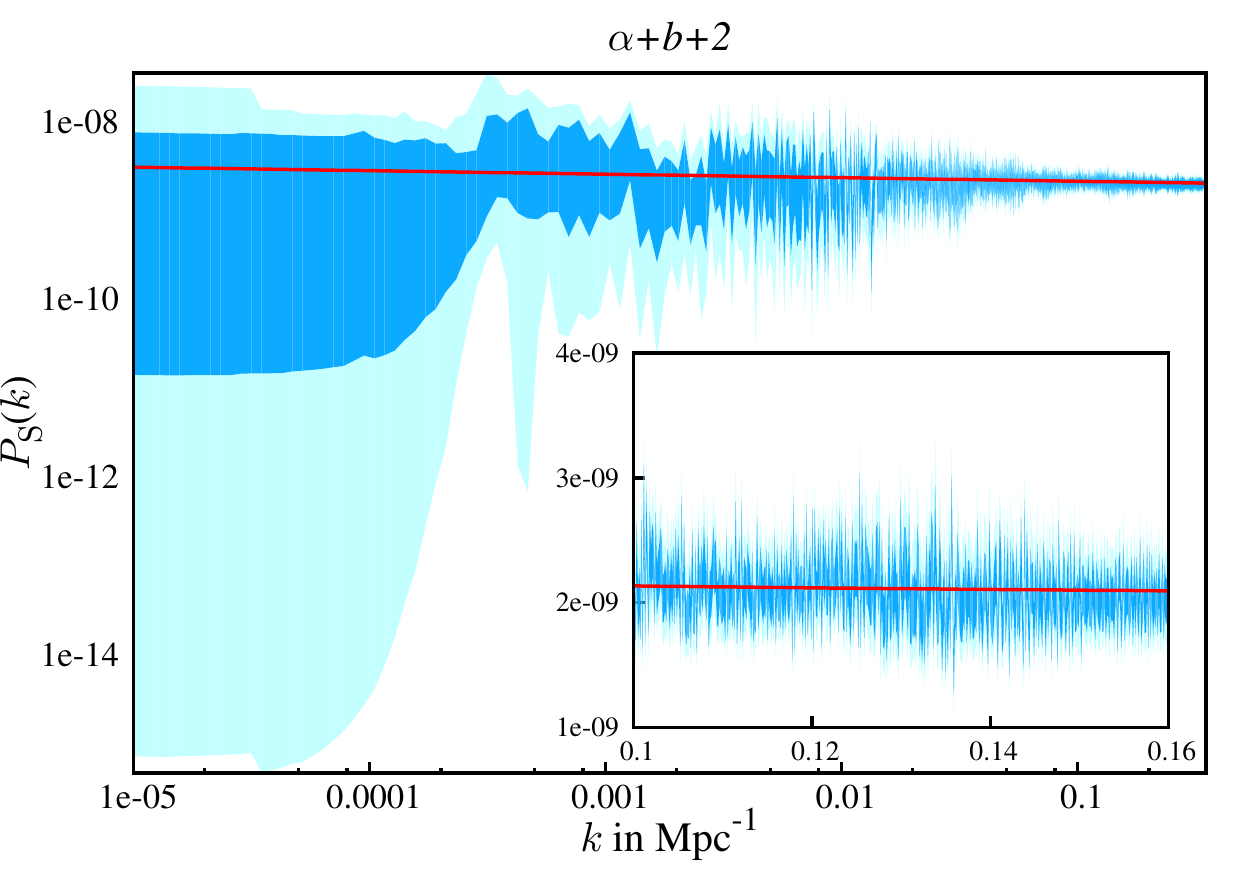}} 
\resizebox{210pt}{160pt}{\includegraphics{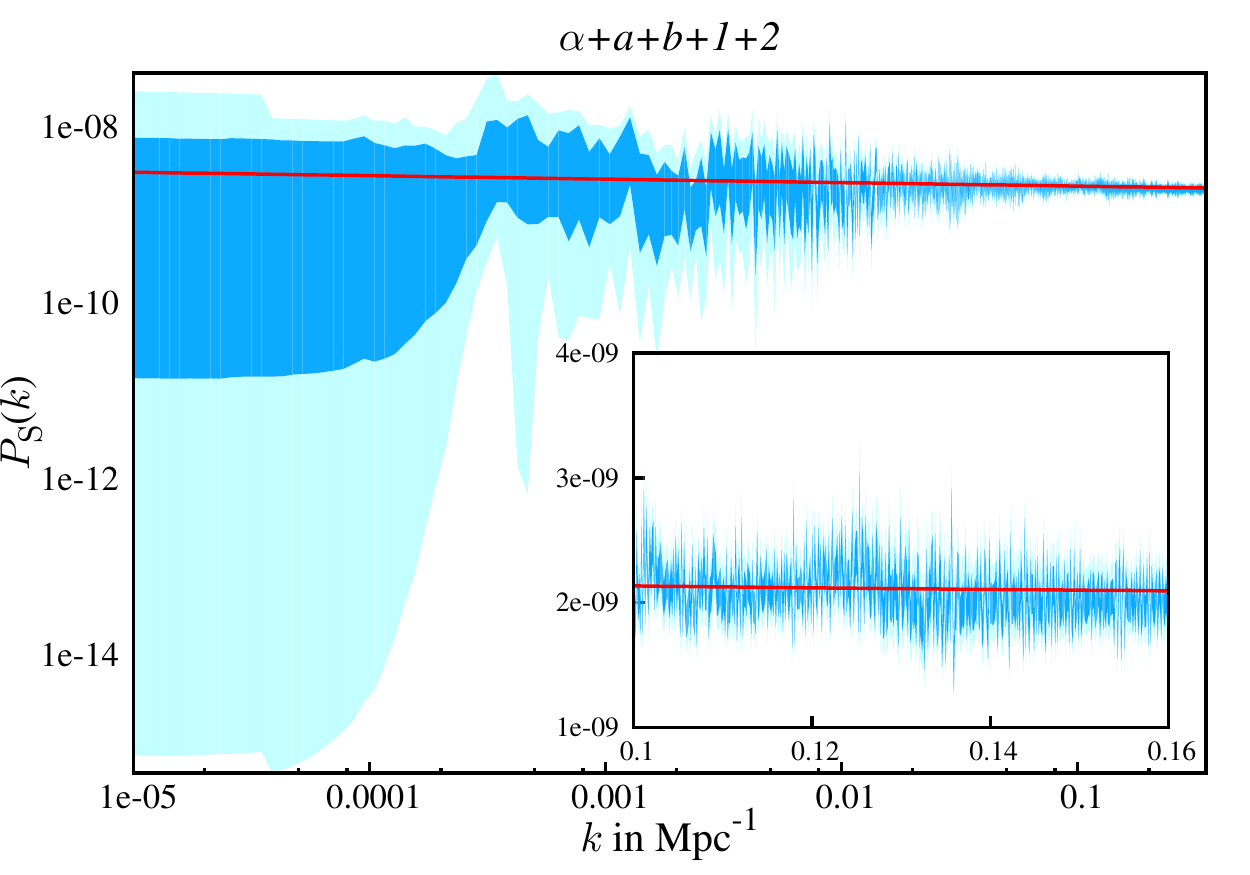}} 
\end{center}
\caption{\footnotesize\label{fig:error}The 1$\sigma$ (blue) and 2$\sigma$ (cyan) error on the PPS for different 
combinations of Planck spectra and the best fit power law PPS (in red).}
\end{figure*}

In Fig.~\ref{fig:error} we plot the $1\sigma$ (blue) and $2\sigma$ (cyan) errors on the PPS for different combinations of Planck spectra as a 
function of cosmological scales. We have used 50 MRL iterations. In red we plot the best fit power law PPS obtained from Planck baseline model. 
From the figure we can directly spot the features at $k\simeq 0.002 {\rm Mpc}^{-1}$ ($\ell\simeq22$) and around $k\sim (0.12-0.14){\rm Mpc}^{-1}$ 
($\ell\sim 1800 - 2000$) since for both the cases power law stands outside $1\sigma$ bounds. Note that in $\alpha+a+1$ and $\alpha+b+1$ the $k\sim (0.12-0.14){\rm Mpc}^{-1}$ feature is prominent, 
however for $\alpha+a+2,\alpha+b,\alpha+b+2$ we find power law is within 
1$\sigma$ errorband which point towards the high significance of this feature only in spectra 1 (217 GHz $\times$ 217 GHz). 
Combining all the spectra (for $\alpha+a+b+1+2$) we find features around $k\simeq 0.002 {\rm Mpc}^{-1}$ and $k\sim (0.12-0.14){\rm Mpc}^{-1}$ remain prominent where the constraints 
on the power spectrum pushes away the power law best fit spectrum by more than 1$\sigma$. Though not visible clearly in the plots we 
found noticeable departures from power law PPS in few other scales for which we extend our error analysis further. 

%%%%%%%%%%%%%%%%%%%%%%%%%%%%%%%%%%%%%%%%%%%%%%%%%%%%%%%%%%%%%%%%%%%%%%%%%%%%%%%

\clearpage

\begin{figure*}[!htb]
\begin{center} 

\resizebox{400pt}{300pt}{\includegraphics{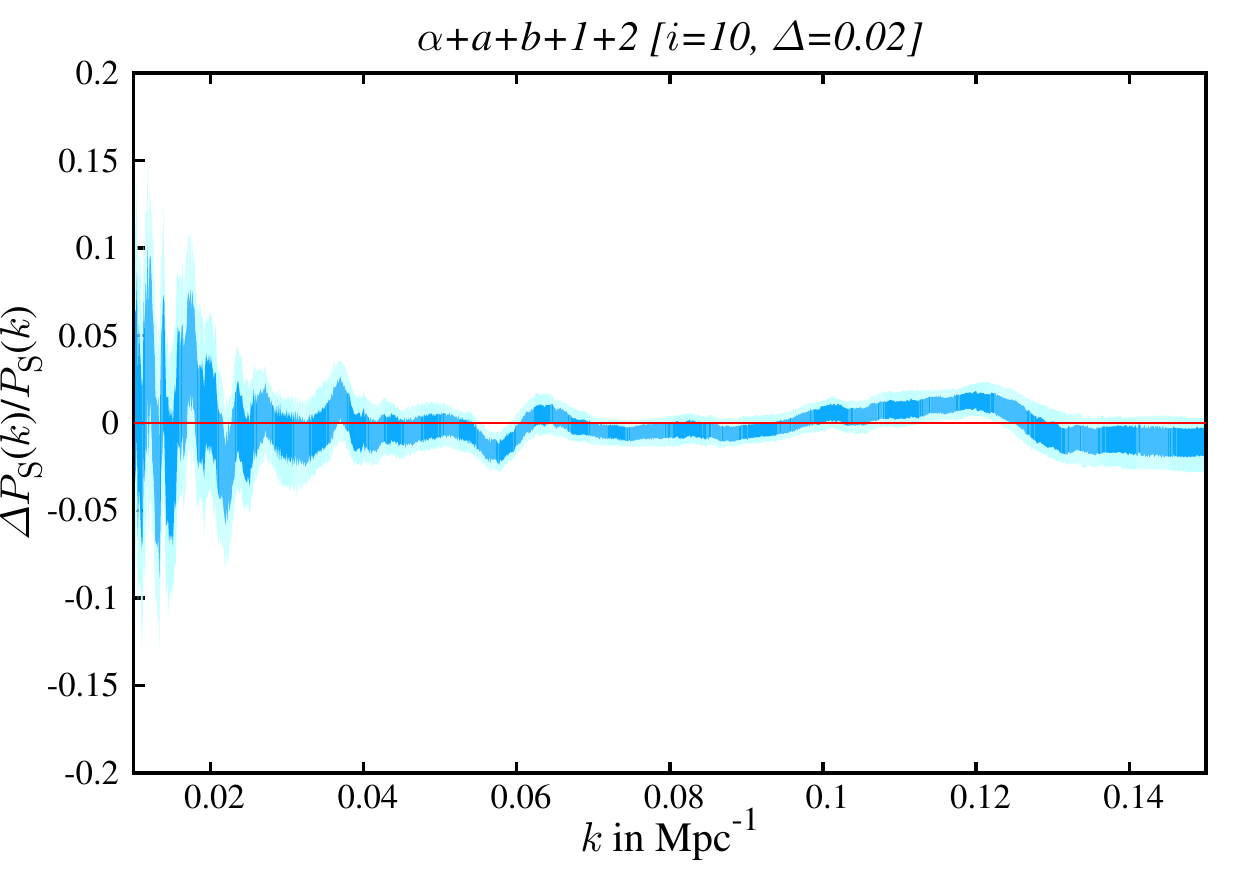}} 
 
\end{center}
\caption{\footnotesize\label{fig:error-smooth} Errors on fractional deviations from the best fit PPS obtained using a particular combination
of iterations and smoothing width, provided in the title of the plot.}
\end{figure*}

Apart from the 2 features discussed above, samples in Fig.~\ref{fig:samples-1} and Fig.~\ref{fig:samples-2} also highlight oscillations near 
$\ell\simeq250-300$ and a broad dip near $\ell\simeq750-850$. These are features correspond to oscillations near $k\simeq0.02 ~{\rm Mpc}^{-1}$ and $k\sim0.055-0.065~{\rm Mpc}^{-1}$ in the PPS respectively. 
However, since MRL works with the un-binned and correlated data, the errors obtained are correlated too.
To reduce the correlation and to highlight broad and dominant features in the data we 
repeat our error analysis with different iterations ($i$) and different smoothing width ($\Delta$), for $\alpha+a+b+1+2$. 
Out of few analysis we only present here one result that can show 
dominant features in the data within $k\sim (0.01-0.15) ~{\rm Mpc}^{-1}$ in Fig.~\ref{fig:error-smooth}. The plot in the figure
represents the fractional deviations ($(\psk|_{\rm Reconstructed}-\psk|_{\rm Power~law~best~fit})/\psk|_{\rm Power~law~best~fit}$) 
from best fit power law model (for $\alpha+a+b+1+2$) and the 
significance (blue and cyan bands correspond to 1 and 2$\sigma$ errors respectively) of these deviations for $i=10$ and $\Delta=0.02$. 
This plot indicates $k\sim {0.055-0.065}~ {\rm Mpc}^{-1}$ ($\ell\sim750-850$) and $k\sim0.12-0.14~ {\rm Mpc}^{-1}$ features can be significant. 

We should note that the estimation of the significance of the individual reconstructed features in the PPS can be complicated due to correlations in the PPS. One can use approaches such as discrete wavelet decomposition~\cite{Shafieloo:2006hs} 
to separate features from each other and estimate the significance and importance of the reconstructed individual features. This is beyond the scope of this work and requires further investigation~\cite{Hazra-Shafieloo-Hope}.  

%Feature around $k\simeq0.02 ~{\rm Mpc}^{-1}$ ($\ell\simeq250-300$) have $<2\sigma$ significance. 
%Moreover we should mention that for unless we smooth the PPS with $\Delta\sim0.1$, 
%the dip around $k\sim {0.055-0.065}~ {\rm Mpc}^{-1}$ always remain more than 2$\sigma$
%significance. We have discussed in last section that this dip is present in all combinations of spectra which also increases 
%its significance.} 

% In this context, particularly, we 
% highlight an important result from Planck lensing analysis~\cite{Planck:lensing}. In Fig.~10 and Fig.~11 
% of paper~\cite{Planck:lensing} it has been demonstrated that lensing 
% potential power spectrum estimates from 143 GHz and 217 GHz have a drop in power around $\ell\sim750-850$ compared to best fit 
% $\Lambda$CDM model. 217 GHz spectrum also shows a dip in the potential power spectrum around $\ell\sim1800-2000$ {\color{red}}. Since, the information of 
% the lensing potential is imprinted in the lensed temperature power spectrum, these features (more importantly $\ell\sim750-850$ feature), 
% seen in the lensing potential power spectrum, also support the primordial features indicated in our analysis.

The small scale features (one near $\ell\sim1800$), appears due to imperfectly subtracted 
electromagnetic interference and has been mentioned in revised versions of Planck 2013 papers~\cite{Planck:likelihood,Planck:cparam,Planck:inflation}. 
% Apart from being physical this feature can appear from unaddressed foregrounds or systematics in the data~\cite{Hazra:2014hma}. 
Having said that, we should mention that we observe the dip near $\ell\sim1800-2000$ in spectra $b$ and $1$, however it is 
only spectra $1$ ($217~{\rm GHz}\times217~{\rm GHz}$) where we find the effect more than 1$\sigma$ significant. It is interesting to notice that spectrum $2$
($143~{\rm GHz}\times217~{\rm GHz}$) does not indicate the feature. In this context, compare the reconstructed angular power spectra residuals and 
residual data points in the plot of $\alpha+a+2$ in Fig.~\ref{fig:samples-2}. We can see that residual data points of $143~{\rm GHz}\times217~{\rm GHz}$ 
in Fig.~16 of~\cite{Planck:inflation} also indicate similar behavior.

While we find the power law primordial power spectrum to be consistent within 2$\sigma$ at almost all cosmological scales, we also report 
certain localized feature around $\ell\sim 750-850$ which has the highest significance. 
Since compared to WMAP-9~\cite{Hazra:2013xva}, with Planck the features are found to be more significant, 
our analysis can encourage models of inflation with features.  For example, the large scale cut-off and dip are 
usually addressed through a punctuation in inflation (with an inflection point in the 
inflaton potential) and a step in the inflationary potential. The oscillations on the other hand are very well explained by 
axion monodomy model. For detailed discussion on these models, see~\cite{features-all}. 

\subsection{Filtering out the noise: towards a smooth primordial power spectrum}

\begin{figure*}[!htb]
\begin{center} 
\resizebox{420pt}{250pt}{\includegraphics{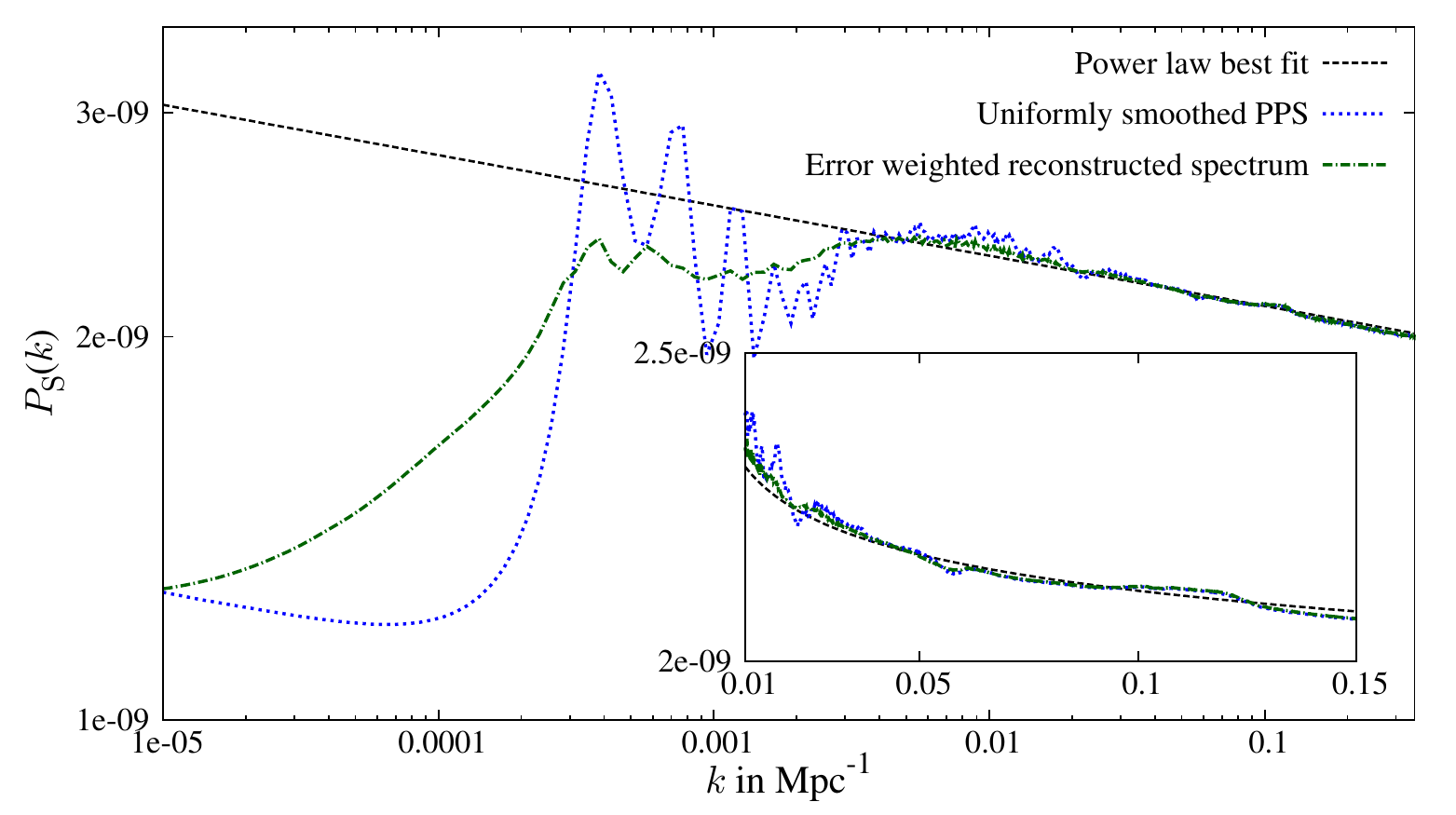}} 
\end{center}
\caption{\footnotesize\label{fig:bestpsk}Smooth primordial power spectra obtained using uniform smoothing (blue dashed) and error-weighted 
smoothing (green dashed) in logarithmic scale. The best fit power law in dashed black is plotted as a reference. The blue and the green PPS 
fit the Planck angular power spectrum data better compared to power law PPS with improvement in $\Delta\chi^2$ of 16 and 12 respectively. 
The inset highlight small scale feature in linear scales.}
\end{figure*}

Towards the end of the feature hunt, it is important that we present a shape of PPS with only significant features included 
providing a significant improvement in fit to the Planck data over power law. While reconstructing an inflationary model from a PPS with 
too many oscillations is difficult and certainly unrealistic, a smooth PPS can provide the necessary information towards 
modification of a slow-roll inflationary potential to achieve similar PPS.
In Fig.~\ref{fig:bestpsk} we plot 2 smooth PPS obtained using the smoothing following Eq.~\ref{eq:gauss}. Using a constant smoothing width 
$\Delta$ we optimize the iterations and the smoothing width in order 
to get a significantly better likelihood than power law PPS with minimal variations in the PPS. 
Such a PPS is given in blue dashed line which provides 16 improvement in $\chi^2$ compared to the power law PPS. 
Moreover, the obtained errorband in our analysis provided in last subsection 
opens up a possibility of having error weighted PPS. The errors in the PPS capture the signal-to-noise ratio in the data. Hence a 
smooth PPS weighted with the signal-to-noise ratio of the corresponding data contain important features in the data. 
In order to achieve that we smooth the PPS assuming $\Delta$ to be a proportional to the $1\sigma$ error ($\ln\psk|{\rm 1\sigma~up}-\ln\psk|{\rm 1\sigma~low}$)~\footnote{$\ln\psk|{\rm 1\sigma~up/low}$ represent
the logarithmic value of the upper and lower 1$\sigma$ error on $\psk$.} obtained for $\alpha+a+b+1+2$. 
The smooth PPS is plotted as green dashed line in Fig.~\ref{fig:bestpsk}. The green line provides 12 better fit over power law PPS.
The best fit power law PPS is plotted black dashed line. 

Comparing the 2 PPS with the power law model we can understand that 
the error weighted PPS has much less oscillations at large scales compared to the uniformly smoothed PPS and can be described by 
simple forms. This PPS can come in handy to be attempted from an inflationary potential with few parameters. 
Note that the green PPS indicates a blue tilt up-to $k\simeq0.01~{\rm Mpc}^{-1}$ reflecting that till this scale the data does not constrain
the PPS to have a necessary red tilt. This fact re-establishes our result obtained in~\cite{Hazra:2013nca} for a PPS broken in 2 bins.
Oscillations around $0.02~{\rm Mpc}^{-1}$ is visible in blue curve. Both the green and blue PPS agrees on small scales and indicate a broad dip at $k \sim 0.055-0.065~{\rm Mpc}^{-1}$ 
(corresponding to $\ell\sim 750-850$) and a broad oscillation at $k \sim 0.12-0.14~{\rm Mpc}^{-1}$ (corresponding to $\ell\simeq 1800-2000$).
However, due to probable systematics in the data mentioned in Planck paper in~\cite{Planck:inflation} immediate priority should not be 
given to the feature at $\ell\simeq 1800-2000$.

\begin{figure*}[!htb]
\begin{center} 
\resizebox{420pt}{320pt}{\includegraphics{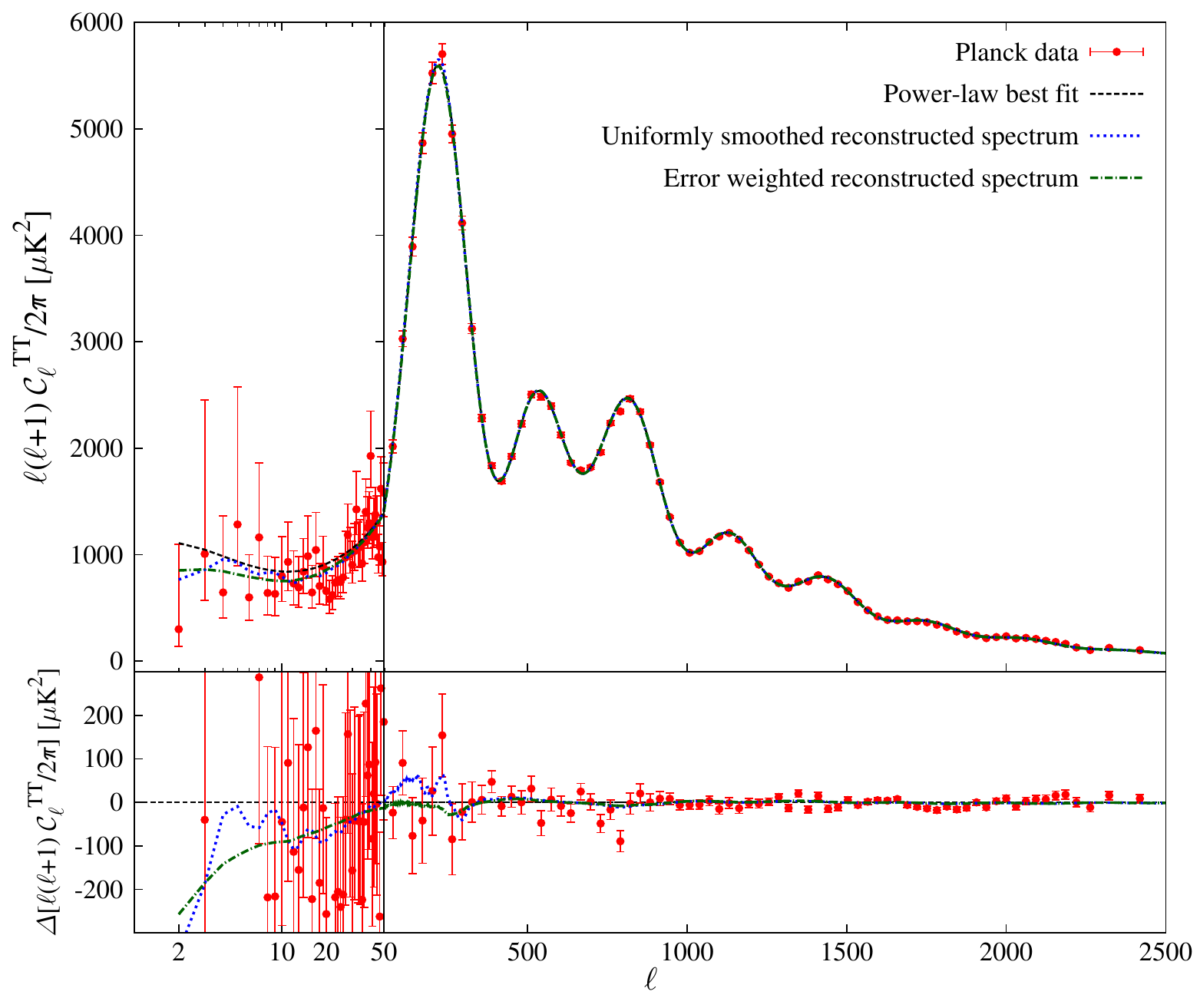}} 
\end{center}
\caption{\footnotesize\label{fig:bestcl} Binned Planck data (red) (combining all the spectrum) and the angular power spectrum obtained from 
power law PPS (dashed black line), uniformly smoothed reconstructed PPS (blue dashed) and error-weighted PPS (green dashed) as plotted 
in Fig.~\ref{fig:bestpsk}. Angular power spectrum and data from low-$\ell$ (2-49) are plotted in logarithmic scales and high-$\ell$ (50-2500) are plotted in linear scale.
Residual data and the angular power spectra (with reference model as Planck best fit baseline model) for the same models are piloted at the bottom.}
\end{figure*}

For the same reconstructed PPS and the power law PPS appearing in Fig.~\ref{fig:bestpsk} we plot the corresponding $\cl$'s in plot~\ref{fig:bestcl}.
The data points with errors are plotted in red. Power spectra and the data from low-$\ell$ are plotted in logarithmic scale and from high-$\ell$
are plotted in linear scale. Note that in apart from the large scale cut-off, all 3 spectra look identical. To identify the difference 
between the spectra we plot the spectra in the space residual to the Planck best fit baseline angular power spectra, 
{\it i.e.} we plot $\ell(\ell+1)(\cl^{\rm TT}|_{\rm Data/reconstructed~spectrum}-\cl^{\rm TT}|_{\rm Planck~best~fit})/2\pi$.
Note that apart from the large scale cut-off, the dip near $\ell\simeq22$ is addressed in both the spectra. Both the green and blue 
spectra attempts to fit the feature in the 
data near $\ell\sim250-300$. Note that while the blue curve fits data with oscillations present around multipole 250-300 
(with a pronounced dip near 300), the error weighted green spectrum gives importance to the particular dip near 
300. Fig.~\ref{fig:bestpsk} also indicates the attempts to fit features around $\ell\sim 750-850$ 
and $\ell\sim 1800-2000$ which are not prominently visible in this figure due to large range covered by the plot.

\section{Discussion}\label{sec:discussion}

In this paper we discuss about reconstruction of the primordial power spectrum (PPS) from Planck CMB data using Modified Richardson-Lucy (MRL) algorithm. 
Planck best fit baseline model has been used as the background cosmology and we used 8 different combinations of Planck spectra for 
the purpose of reconstruction. In our analysis we attempt to fit WMAP-9 data using the reconstructed PPS from Planck spectra and 
address the consistency of the WMAP-9 and Planck data. Our results indicate that Planck low-$\ell$ and the 100 GHz spectra are 
consistent to WMAP-9 data allowing an overall amplitude shift. Combining all Planck spectra we show that the reconstructed PPS is 
consistent to WMAP-9 allowing an overall $\simeq2.5\%$ amplitude shift, which is in agreement with the other studies along this 
line~\cite{Planck:likelihood,Hazra:2013oqa}. 

We do also discuss the reconstruction of the PPS studying the effect of lensing and report
definitive evidence of lensing in the data for all combinations of spectra except for the case of
low-$\ell$ spectrum (as expected). Our results indicate that allowing large amount of fluctuations in 
the form of PPS it is possible to mimic the lensing effect through features in the PPS. This degeneracy is an important 
issue to consider when we are looking for the features in the form of the PPS.

Using MRL algorithm we have been able to locate possible features in 
the form of the PPS by analyzing different combinations of the Planck spectra. We have performed an extensive error analysis using 
1000 realizations of the data in all 8 combinations of the spectra to obtain realistic constraints on the form of the PPS and 
estimate the statistical significance of the reconstructed features. $\ell\simeq22$ ($k\simeq0.002 ~{\rm Mpc}^{-1}$), $300$ 
($k\simeq0.02 ~{\rm Mpc}^{-1}$) and 
$750-850$ ($k\sim {0.055-0.065}~ {\rm Mpc}^{-1}$) features are found to be the most prominent ones, of them a dip around 
$\ell\sim750-850$ with highest significance. We report that feature at $\ell\sim1800-2000$ is evident with high significance only in 217 GHz spectra and apparently appears due to small systematics as has been reported by Planck~\cite{Planck:inflation}. 
In a conservative error estimation we find standard power law PPS remains consistent within 2$\sigma$ constraints in almost all scales. At the end we present two smooth form of the PPS (that can fit the data significantly better than the power law model) 
that can be described by relatively simple functional forms with limited number of parameters. This can be helpful for inflationary 
model building to address the data using alternative scenarios.  

%Though our study in this paper is detailed with using the available Planck data, the work %lacks completeness since polarization data is not 
%yet published. We hope to revisit the primordial spectra from Planck using polarization %data. We expect tighter constraints on PPS with Planck
%polarization.

\section*{Acknowledgments}
The authors would like to thank George Efstathiou for his important comments and suggestions on the manuscript.
D.K.H and A.S wish to acknowledge support from the Korea Ministry of Education, Science
and Technology, Gyeongsangbuk-Do and Pohang City for Independent Junior Research Groups at 
the Asia Pacific Center for Theoretical Physics. The authors would like to thank Simon Prunet 
for his help in understanding Planck covariance matrix. A.S. would like to acknowledge
the support of the National Research Foundation of Korea (NRF-2013R1A1A2013795). 
We acknowledge the use of WMAP-9 data and likelihood from Legacy Archive for 
Microwave Background Data Analysis (LAMBDA)~\cite{lambdasite} and Planck 
data and likelihood from Planck Legacy Archive (PLA)~\cite{PLA}. 
We also acknowledge the use of publicly available CosmoMC in our analysis.

%%%%%%%%%%%%%%%%%%%%%%%%%%%%%%%%%%%%%%%%%%%%%%%%%%%%%%%%%%%%%%%%%%%%%%%%%%%%%%%

%%%%%%%%%%%%%%%%%%%%%%%%%%%%%%%%%%%%%%%%%%%%%%%%%%%%%%%%%%%%%%%%%%%%%%%%%%%%%%%
\end{document}